\documentclass[twocolumn]{aastex631}

\usepackage{mathtools}
\usepackage{hyperref}

\begin{document}

\title{Mapping Jet-Gas Coupling and Energetic Ionized Outflows in High-Redshift Radio Galaxies with JWST/NIRSpec}

\correspondingauthor{Namrata Roy}
\email{nroy13@jhu.edu}

\author[0000-0002-4430-8846]{Namrata Roy}
\affiliation{Center for Astrophysical Sciences, Department of Physics and Astronomy, Johns Hopkins University, Baltimore, MD, 21218}

\author[0000-0001-6670-6370]{Timothy Heckman}
\affiliation{Center for Astrophysical Sciences, Department of Physics and Astronomy, Johns Hopkins University, Baltimore, MD, 21218}
\affiliation{School of Earth and Space Exploration, Arizona State University, Tempe, AZ 85287}

\author[0000-0002-6586-4446]{Alaina Henry}
\affiliation{Space Telescope Science Institute, 3700 San Martin Drive, Baltimore MD, 21218} 
\affiliation{Center for Astrophysical Sciences, Department of Physics and Astronomy, Johns Hopkins University, Baltimore, MD, 21218}

\begin{abstract}


 We present spatially resolved maps of morphology, kinematics, and energetics of warm ionized gas in six powerful radio galaxies at $z \sim$ 3.5–4, using JWST/NIRSpec IFU to quantify jet-driven feedback in the early universe.  All sources exhibit broad [OIII] emission-line profiles with $\rm W_{80}$ (line width) values of 950–2500 km s$^{-1}$ across $\sim$10s of kpc, signifying large-scale outflows. The outflowing nebulae are preferentially aligned with the radio jet axis, suggesting jet-driven origin.  On average, the regions with the broadest lines and highest velocities are co-spatial with radio lobes or cores, and exhibit the strongest kinetic power.
Ionized gas masses associated with the outflows span $\sim 1$ to $8 \times 10^{9} \ \rm M_\odot$, with total mass outflow rates of 80–950 $\rm M_\odot \ yr^{-1}$ and kinetic powers between $10^{43.2}$ and $10^{45.0} \ \rm erg \ s^{-1}$. The outflow kinetic power corresponds to 0.15\%–2\% of the AGN bolometric luminosity, sufficient to impact galaxy evolution. However, only $\lesssim 1$\% of the jet mechanical energy couples to the warm ionized gas via outflows, consistent with predictions from hydrodynamic simulations. A large fraction of the jet energy may instead reside in shock-heated hot gas, supported by X-ray detection, or used to thermalize the gas and produce the observed emission-line nebulae. Our results demonstrate that radio jets in massive, gas-rich systems at high-redshift can inject significant kinetic and thermal energy to the surroundings, providing direct evidence for jet-driven feedback operating during the peak epoch of galaxy formation.

\end{abstract}



\section{Introduction} \label{sec:intro}


Active Galactic Nuclei (AGNs) are among the most luminous and energetic sources in the universe, and are a crucial component in the evolution of galaxies. 
AGNs can return vast amounts of mass, energy, momentum, and metals to the surrounding medium, potentially suppressing the host galaxy's star formation and influencing their evolutionary trajectory. This process of energy injection, termed as AGN feedback, can occur through multiple pathways, mainly via radiation pressure \citep{ciotti10}, relativistic jets \citep{fabian12}, and multi-phase gas outflows \citep{crenshaw03}.  
 Theoretical models predict that even a small fraction of the total feedback energy is sufficient to reheat or expel most of the interstellar medium (ISM) gas if deposited efficiently \citep{dimatteo05, croton06}. Cosmological simulations have also shown that without AGN feedback, massive galaxies would continue to form stars rather than quench,  leading to inconsistencies with the observed galaxy luminosity function and bimodal color distribution at z$\sim$ 0 \citep{dimatteo05, croton06, hopkins10}.
 However, observational constraints on the energy and momentum deposition by AGN feedback and their impact on the ISM remain scarce, particularly at high redshifts (z $>$ 2).


Multiple studies at lower redshifts (z $\sim$ 0--1) have now detected powerful, high-velocity gas outflows in AGN host galaxies  \citep{xu19, xu20, villar-martin21, cicone15, morganti21, speranza21, harrison14, mullaney13, roy21a}. 
These studies have demonstrated that AGN feedback can disturb the ISM gas to varying degrees, but the kinetic energy transport is almost always measured to be below 0.1\% of the AGN bolometric luminosity. This is much lower  than the theoretical benchmark of 0.5-5\% to have any sufficient impact on the ISM \citep{baron19, fiore17}. This discrepancy raises questions about the efficiency of energy coupling in different AGN environments and feedback modes. \citep{fiore17, baron19, speranza21}. 

Recently, \cite{heckman23} adopted a different approach and quantified the cumulative energy and momentum from various AGN classes, integrating their feedback contributions over cosmic time per unit co-moving volume. They concluded that the major  source of kinetic energy comes from relativistic jets driven by Radio AGNs. Moreover, \cite{heckman24} demonstrated that the cosmic evolution in the rate at which jets have transported energy is a near-perfect match to the rate at which quiescent massive galaxies are created, strongly suggesting that feedback from the radio jets has been responsible for quenching star-formation.

Indeed, Radio-loud AGNs are powerful candidates for efficient kinetic feedback, as large scale radio jets can extend over tens to hundreds of kiloparsecs and can pierce through the ambient ISM well into the CGM gas. High resolution hydrodynamical simulations have shown that these jets can effectively deposit energy and momentum, and often inflate large cocoons/ bubbles of jet fluid that drive shocks when they encounter ambient gas. The expanding cocoons can entrain and accelerate this ambient gas and lead to large scale outflows and turbulent gas motions along the radio jet axis \citep{begelman89, mukherjee16, mukherjee18, wagner11, wagner12, dutta24}. The mechanical power of these jets are capable of displacing, heating, or expelling the ambient gas \citep{mcnamara07, mukherjee16}. In the local universe, spatially resolved studies using integral field unit (IFU) spectroscopy have revealed fast ionized gas outflows associated with radio jets \citep{jarvis21, roy18, roy21a, roy21c, mukherjee16, kim23}. The impact of the jets are predicted to be even more prominent in higher redshift radio galaxies, since galaxies are more gas-rich and launch more powerful jets \citep{falkendal19}.  Thus, high-$z$ radio galaxies (HzRGs) can be excellent sites for constraining the cosmological evolution of feedback energetics, although observations are quite rare \citep[e.g.: ][]{nesvadba17}.

 The JWST NIRSpec IFU has recently enabled detailed mapping of multi-phase feedback in z $>$ 3 HzRGs for the first time, due to its unprecedented spatial resolution and sensitivity. Both \cite{roy24} and \cite{saxena24}  demonstrated that TNJ1338, a z $\sim$ 4 radio galaxy with extended ionized gas nebulae, show disturbed kinematics and large gas velocities co-spatial with radio lobes. The expanding jet cocoons entrain gas with a mass outflow rate $\sim$ 500 $\rm M_{\odot} \ yr^{-1}$ which indicates that a significant portion of the gas is displaced outwards, consistent with the feedback picture suggested by jet simulations \citep{mukherjee18, meenakshi22}. \cite{roy24} carefully calculated spatially resolved measurements of energy, momentum and mass along the radio jet axis and demonstrated that the jet deposits a large amount of kinetic energy and momentum, but the energy is still 0.1\% of the AGN bolometric luminosity. Thus, although the jets  at high redshift significantly alter the gas kinematics, drive large scale outflows, and expel substantial amount of gas mass, the efficiency of feedback energy transfer in this population of jets is still unclear, due to small numbers of HzRG with the requisite ionized gas observations.  This underscores the need for similar detailed kinematics and energetics analyses for more HzRGs. \cite{wang24} and \cite{wang25} presented JWST NIRSpec IFU observations of additional HzRGs at z$>$3.5, identifying widespread nebular gas and broad emission line components associated with jet structures. However, comprehensive spatially-resolved measurements of the energetics, and studies of the detailed underlying spatial correlation between the radio jet structures and the gas kinematics are required to fully assess the impact of these jets.


In this study, we leverage the JWST NIRSpec IFU observations of a sample of six HzRGs from two JWST programs (GO 1970 and GO 1964) to conduct a detailed spatially resolved analysis of ionized gas kinematics, resolved mass, energy, momentum and outflow rates of the ambient gas deposited by radio jets to demonstrate the impact of jet feedback  and radio jet-ISM interactions. Our sample consists of four HzRGs from \citet{wang24, wang25} and two sources from GO-1924 presented in \citet{roy24, saxena24}, spanning a range of radio morphologies from compact radio AGNs to large scale jet-driven systems.
All these HZRGs host dust-obscured quasars at their centers; hence the blinding
glare of light from the central AGN is blocked. So it is
straightforward to identify broad lines indicative of gas outflows in the  ISM and quantify the spatially resolved energetics associated with the outflowing component. 
We will quantify the morphological co-incidence between the radio lobes and the gas kinematics in HzRGs, and measure the kinetic energy transport of jet-driven feedback. Our work will also address how the feedback characteristics of compact jets differ from those of more extended radio jets in high redshift environments. Our analysis follows the methodology established in \cite{roy24}, to allow for a consistent comparison of feedback metrics across different HzRGs.

The paper is organized as follows: Section 2 outlines the JWST NIRSpec data used in this study and the associated data reduction steps. Section 3 discusses the kinematic analyses performed on the reduced data cubes. The subsequent results are narrated in Section 4. In Section 5, we discuss the implications of the result and end with the conclusion in Section 6.

Throughout this paper, we assume a flat cosmological model with $H_{0} = 70$ km s$^{-1}$ Mpc$^{-1}$, $\Omega_{m} = 0.30$, and  $\Omega_{\Lambda} =0.70$, and all magnitudes are given in the AB magnitude system \citep{oke83}.

\section{Data Acquisition} \label{sec:obs}


\subsection{Sample \& NIRSpec IFU Observations} \label{subsec:data_nirspec}


In this study, we utilize JWST/NIRSpec IFU observations to investigate the spatially resolved gas kinematics, energy distribution, and momentum transfer driven by radio jets in a sample of six high-redshift radio galaxies at $z > 3.5$. The data were acquired as part of two Cycle 1 GO programs: 1964 (PIs: Overzier and Saxena) and 1970 (PI: Wang). All the JWST data used in this work can be accessed via MAST doi: \href{http://dx.doi.org/10.17909/eeb5-hz26}{10.17909/eeb5-hz26}. The NIRSpec IFU delivers spatially resolved spectroscopy over a $3'' \times 3''$ field of view with $0.1'' \times 0.1''$ spatial resolution elements (spaxels). All six sources were observed with the high-resolution gratings (primarily G235H/F170LP and G395H/F290LP), yielding spectral cubes with a resolution of $R \approx 2700$.

Data for two of the targets, TNJ1338-1942 and TGSS1530+1049, were taken between UT 2023 February 22 to 2023 July 14 as part of GO-1964.  TNJ1338-1942, at $z = $ 4.1, is one of the most luminous ($\rm L_{1.4GHz} \sim 10^{28.3} \ W \ Hz^{-1}$) radio sources in the early universe. It exhibits a double-lobed radio structure with a projected separation of $\sim$ 35 kpc. This source has a rich set of ancillary multi-wavelength observations, including  X-ray  data \citep[Chandra: ][]{smail13}, rest UV spectra \citep[MUSE:][]{debreuck99, swinbank15}, far IR \citep{falkendal19}, and ground-based optical spectroscopy \citep[SINFONI: ][]{nesvadba17}. 
It was observed with JWST/NIRSpec IFU using two grating-filter combinations: G235H/F170LP and G395H/F290LP, covering 1.66 $-$ 5.27 $\mu$m in the observed frame  ($\sim$3300 $-$ 10,300 \AA\  in the rest frame). Good quality background subtraction was obtained by nodding off-scene and obtaining a separate background image. For a full description of the data acquisition
and reduction, we refer the reader to \cite{saxena24}. 
Detailed kinematic, energetic, and momentum analyses for this source have already been conducted by \cite{roy24}, and those results are incorporated in the current analysis.

TGSS1530 (z $\sim 4.0)$, the second radio galaxy observed with the same GO program, 
is a compact and ultra-steep spectrum powerful radio galaxy ($\log L_{150 \, MHz} \approx 29.1 \ W \ Hz^{-1}$) with an angular  size of $\sim 0.4''$ (physical size $\sim$ 2.5 kpc). Its high redshift coupled with relatively small radio size suggest that TGSS1530 may be a radio galaxy in an early phase of its evolution. Although initially classified as a z = 5.72 source based on tentative Lyman-$\alpha$ emission, rest-optical emission lines from the JWST/NIRSpec data revised the redshift to $z = 4.0$.  This source was observed with only  G395H/F290LP configuration, and thus provides a wavelength coverage of 2.87 $-$ 5.14 $\mu$m in the observed frame or 5700$-$10000 \AA\ in the rest frame. A 4-point nodding strategy was employed for in-scene background subtraction, given the compact nature of the source.

The remaining four HzRGs at $z\approx 3.5$ were observed by GO 1970 (PI: Wuji Wang) between UT 2022 October 30 and 2023 August 29. They were selected from a larger sample of HzRGs with existing multi-wavelength supporting datasets \citep[Hershel, Spitzer, SINFONI, MUSE, ALMA:][]{seymour07, debreuck10, nesvadba07, nesvadba17, wang24, kolwa23}. 
The additional JWST/NIRSpec observations, as presented in \cite{wang24} and \cite{wang25}, provide complementary spectroscopic information in the rest optical wavelength bands. All four galaxies were observed with the G235H/F170LP configuration, covering 1.70 -- 3.15 $\mu$m in the observed frame or $\sim$ 3700 -- 7620 \AA \  in the
rest frame. Integration times ranged from 3.7 to 4.0 hours per target, with a 4-point nodding strategy employed for background subtraction.
4C+19.71 and TNJ0205+2242 are traditional double-lobed Radio galaxies with projected lobe separations of $\sim 60$ kpc and $\sim 20$ kpc, respectively. In contrast, 4C+03.24 displays a more extended, complex, but fragmented radio structure, suggesting a recent or ongoing interaction with the surrounding ISM. Finally, TNJ0121+1320, is a compact source with an unresolved radio morphology.

Except for TGSS1530, all six HzRGs have coverage of the primary rest optical emission lines, including [OII]$\lambda \lambda$3726, 3729, H$\beta$, [OIII]$\lambda \lambda$4959, 5007, H$\alpha$, [NII]$\lambda \lambda$6548, 6584, and [SII]$\lambda \lambda$6716, 6731. For TGSS1530, only  H$\alpha$, [NII] and [SII] lines fall in the selected dispersion/grating. 
The stellar masses of the sample span $\rm 10^{10.9} - 10^{11} \ M_{\odot}$, consistent with massive host galaxies. The diversity in jet morphology, radio power, jet age, star formation rate, and gas content among the sample is expected to produce a wide range of jet-ISM interactions, and it will enable a comprehensive characterization of energy transfer mechanisms across a broad spectrum of HzRGs.



 The raw data were downloaded from Mikulski Archive for Space Telescopes (MAST). The reduced data cubes presented in this paper were obtained by running the JWST Science Calibration Pipeline \footnote{https://github.com/spacetelescope/jwst} version 1.16.1
with the calibration context file \texttt{jwst\_1303.pmap}. This latest calibration reference file
incorporates proper in-orbit flat-fielding and improved flux
calibrations. Our reduction methodology is very similar to \cite{roy24}. We executed Stage 1 and Stage 2 of the pipeline. We skipped the imprint subtraction step since it increases the overall noise level in the final data cube, and ran the Stage 3 of the pipeline. We replaced the default outlier detection step of the pipeline with our own custom module that determines outliers across the different dither positions. We used a sigma-clipping technique to remove outliers at the first order, and additionally used the python package \texttt{astroscrappy} \citep{mccully18} to remove any other residual outliers missed by the previous step. After masking and excluding the outlier pixels, we combine the cleaned dithered observations to create the final 3D data cube using the \texttt{cube\_build} step of the pipeline. We select the exponential modified Shepard method (``emsm'') of weighting when combining detector pixel fluxes to provide high signal to noise ratio (S/N) at the spaxel level. As all the HzRGs analyzed in this study are Type-2 AGNs with obscured central quasars, point spread function (PSF) subtraction is not required. The properties of our six targets are provided in Table \ref{tab:1}.

\subsection{Data Analyses} \label{subsec:data_analyses}

\begin{figure*}
    \centering
    \includegraphics[width=\textwidth]{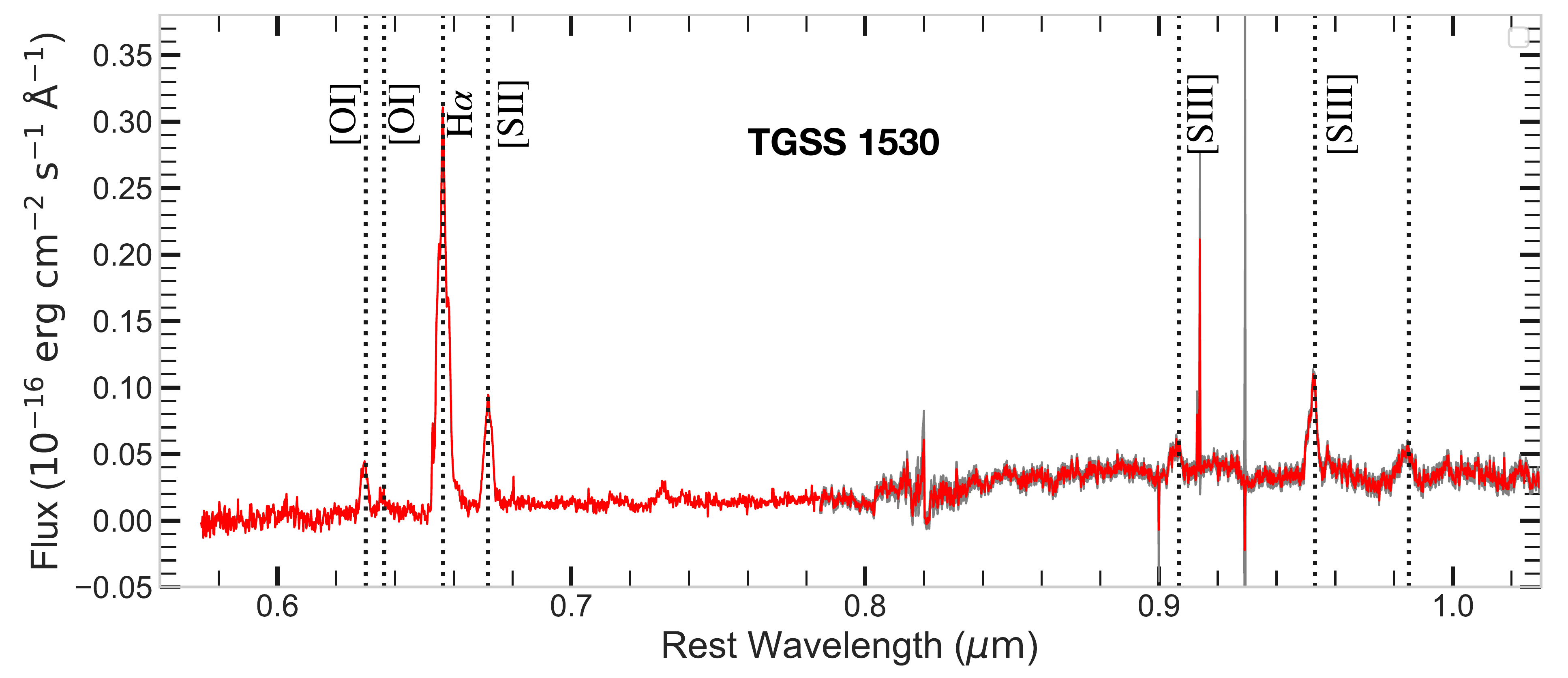}
    \caption{ The JWST/NIRSpec IFU spectra extracted from around the detected radio core in the northern part of TGSS1530. The spectrum shown are integrated within a box aperture of 2.5$''$ each side, centered at the location of the radio core: (232.708$^{\circ}$, 10.825$^{\circ}$). The gray-shaded region shows 1-$\sigma$ errors in the measured fluxes. The rest-frame optical nebular emission lines visible in the spectral window are marked by black dashed lines.   } 
    \label{fig:integrated_spectra}
\end{figure*}

The data reduction steps, outlined in the previous section, produce wavelength and flux calibrated, cleaned and exposure-combined science-ready  data cube for each source. The extracted spectra have spatial pixel (spaxel) dimensions of 0.1$''$ $\times$ 0.1$''$. Figure \ref{fig:integrated_spectra}  shows an example integrated spectrum from TGSS1530 with the G395H grating. The spectrum was obtained by summing over all spaxels within a box-shaped aperture of 2.5'' per side, centered at the location of the radio detection. 
Multiple emission lines are detected in all sources, including [OIII] $\lambda \lambda$4959, 5007, H$\alpha$, H$\beta$, [NII]  $\lambda \lambda$6548, 6584, and [SII]  $\lambda \lambda$6717, 6731, except for TGSS150 where only H$\alpha$, [NII], [SII] and [SIII] fall within the observed wavelength window.

The systemic redshift  for TNJ1338 was reported to be $z = 4.104 \pm 0.001$ in \cite{roy24}, which was derived from the   [OIII] $\lambda$5007 emission line at the host galaxy center. This is in agreement with the value derived using He II emission from the MUSE rest UV spectra \citep{swinbank15}. For TGSS1530, we derive a redshift of $z = 4.002 \pm 0.0007$ from the H$\alpha$ emission line at the continuum center, which is assumed to be the host galaxy center. 
Systemic redshifts for the remaining 4 HzRG sources $-$ 4C+19.71, TNJ0205+2242, 4C+03.24, and TNJ0121+1320$-$ are adopted from \cite{wang25}, based on HeII $\lambda 1640$ and [CI](1–0) emission at the radio core. The values are  $z = 3.5892$ (4C+19.71), $z = 3.5060$ (TNJ0205+2242), $z = 3.5657$ (4C+03.24), and $z = 3.5190$ (TNJ0121+1320). We use these redshifts to shift the observed wavelength grids in the spectral cube to the rest frame before further analysis. All emission lines in each galaxy are constrained to have the systemic redshifts mentioned above.

We collapse the three-dimensional spectral cubes into two-dimensional ``slices'' along the wavelength axis in $\sim$ few \AA \  intervals centered around strong emission lines such as [OIII] $\lambda$5007 or H$\alpha$. We construct these narrow band slices from the blueshifted to redshifted sides of the emission line. The images (later shown in Figure \ref{fig:tgss_slice} and subsequent figures) trace the spatial distribution of emission line flux across velocity space and allow visual identification of extended blue- and redshifted gas, indicative of gas outflows. 

Our primary goal is to characterize the emission line profiles and outflow characteristics in detail. Hence, to isolate the emission lines, the first step is to subtract the underlying continuum level for each spaxel. We follow the method by \cite{roy24}. For [OIII] $\lambda$5007, we fit a second-order polynomial within the continuum window $-$ chosen to be 4900$-$ 4925 \AA \ and 5075$-$ 5100 \AA\ in the rest frame, excluding the emission line itself. 
For TGSS1530, where [OIII] is not covered, we use H$\alpha$ and perform similar continuum subtraction within the windows 6300$-$ 6320 \AA\ and 6395$-$ 6415 \AA. We use \texttt{SCIPY}'s optimization routine to derive the best fit polynomial by the least square minimization technique. The fitted continua are subtracted spaxel-by-spaxel to produce continuum-subtracted emission-line cubes. This method provides a robust  subtraction of the continuum level for each line \citep[as demonstrated in][]{roy24}.

From these emission line cubes, we generate moment maps: moment-0 (flux), moment-1 (velocity), and moment-2 (velocity dispersion). We use [OIII$\lambda$5007 for all the other sources, and H$\alpha$ emission for TGSS1530. All our analyses are restricted to spaxels where the S/N of the emission line $\geq$ 2. We quantify the gas turbulence/disturbance using the non-parametric line width $\rm W_{80}$, defined as the width containing 80\% of the line flux. This is calculated directly from the data, by measuring the cumulative distribution function (CDF(x) = P(X $\leq$ x) of the emission line flux. The velocities $v_{10}$ and $v_{90}$ corresponding to the 10th and 90th percentiles of the flux are derived from the CDF, and $\rm W_{80} = v_{90} - v_{10}$. This width measurement does not assume a specific line profile, and is independent of any model fitting constraints. However, $\rm W_{80}$ is very sensitive to the presence of broad wings.

The final step is to calculate the outflow properties and map their impact on the host galaxy ISM. We compute the outflow properties using dust-corrected [OIII]$\lambda$ 5007 emission line (H$\alpha$ for TGSS1530), following the formalism in \citep{cano-diaz12, veilleux20, roy24}.  We estimate the local electron densities of the gas by calculating [SII]$\lambda$6717/$\lambda$6731 line ratio for every spaxel with [SII] detection. To isolate the outflow component, we consider only emission from high-velocity gas, defined as having $|v| > 500 \ \rm km \ s^{-1}$, which exceeds typical rotational velocities for the stellar masses of our sources and is thus attributed to outflows. Using the line luminosities, electron densities and outflow velocity maps, we calculate the spatially resolved gas mass (M), mass outflow rate ($\rm \dot{M}$), outflow velocity (v), momentum flux ($\rm \dot{p} = \dot{M} v$), and the kinetic power ($\rm \dot{E} =  \frac{1}{2} \dot{M} v^2$) of the outflowing gas. These quantities allow us to locally evaluate the feedback efficiency of the jets and assess whether the energy imparted to the ISM is sufficient to unbind or reconfigure the gas on galactic scales.

A key goal of this study is to examine how gas kinematics and outflow energetics vary spatially in relation to the orientation of the radio jets.  Hence, we measure radial gradients of line widths ($\rm W_{80}$), ionized gas mass outflow rates ($\rm \dot{M}$), and outflow kinetic power ($\rm \dot{E}$) along the projected radio axis as a function of radial distance from the center. We first determine the projected position angle of each galaxy from the [OIII] emission-line moment map. We define the major axis as the axis along which the second-order spatial moments of the [OIII] flux distribution are maximized. The corresponding position angle is measured counterclockwise from the vertical (y-axis).
All relevant maps—including [OIII] flux, line width ($\rm W_{80}$), velocity centroid ($\rm v_{50}$), outflow velocity $\rm v_{out} = \sqrt{V_{50}^2 + W_{80}^2} $, and electron density are rotated to align the galaxy's major axis along the vertical y-spaxel direction. 
Along this axis, we stack a series of rectangular boxes, each 1 kpc in height (along the major-axis) and bounded laterally by the extent of [OIII] emission.
Within each 1 kpc box, we compute the total [OIII] flux, the mean electron density, the average line width $\rm W_{80}$ values and an effective outflow velocity, defined as:

\begin{equation} \label{eq:vout}
    v_{\text{out}} = \sqrt{\langle v_{50} \rangle^2 + \langle W_{50} \rangle^2}
\end{equation}

Here, $\langle v_{50} \rangle$ and $\langle W_{50} \rangle$ are the mean velocity centroid and line width respectively within the parcel. 

These spatially binned measurements provide radially resolved profiles of the gas mass and outflow velocities of the ionized gas, enabling us to measure how the mass, momentum, and kinetic energy outflow rates vary as a function of distance from the galaxy center along the radio jet axis.

Given the spatial resolution of our data, it is also important to consider the accuracy of cross-matching between the radio and NIRSpec coordinate frames when interpreting spatial correlations.
Astrometric alignment between JWST/NIRSpec IFU observations and radio (VLA or VLBI) data is subject to inherent uncertainties from both instruments. 
The astrometric uncertainty  in JWST/NIRSpec datacube can be anywhere between $0.1''-0.3''$, if target acquisition is not performed. In contrast, radio interferometric observations, especially from VLA and VLBI, achieve much higher positional precision, often better than $0.02''$. As a result, when comparing spatial features across these datasets, offsets of up to $0.2''-0.3''$ can naturally arise. For our redshift range ($3.5 < z < 4$), this indicates a spatial scale of $\sim 1-2 $ kpc. Therefore, small positional offsets less than $\sim 2$ kpc between radio structures (e.g., cores or lobes) and optical/NIR emission-line regions are likely attributable to astrometric uncertainties and are not interpreted as physically significant.

\begin{table*}[ht]
\resizebox{1.05\textwidth}{!}{
\begin{tabular}{lccccccccc}
\hline\hline
\textbf{Source} & RA, DEC (Radio) & Redshift & $\log_{10} M_{\star}$ & $\log_{10} L_{1.4\mathrm{GHz}}$ & Size$_{1.4}$ & Radio morphology & $\log_{10} L_{\mathrm{AGN}}$ & $\log_{10} L_{\mathrm{jet}}$ & $\log_{10} L_{\mathrm{shock}}$ \\
 & [hh:mm:ss, dd:mm:ss] &  & [$M_{\odot}$] & [erg s$^{-1}$] & [kpc] &  & [erg s$^{-1}$] & [erg s$^{-1}$] & [erg s$^{-1}$] \\
\hline
TGSS & 15:30:49.890, +10:49:30.02 & 4.002 & $<$10.25 & 43.20 & 3 & Double-lobed & 46.82 & 45.53 & 46.02 \\
TNJ0205 & 02:05:10.676, +22:42:50.57 & 3.506 & 10.82 & 43.97 & 22 & Unresolved & 47.31 & 46.05 & 46.18 \\
4C03 & 12:45:38.377, +03:23:21.14 & 3.566 & $<$11.27 & 44.93 & 30 & Double-lobed & 47.65 & 46.65 & 46.81 \\
4C19 & 21:44:07.512, +19:29:14.58 & 3.589 & $<$11.13 & 44.77 & 60 & Double-lobed & 46.30 & 46.55 & 46.16 \\
TNJ0121 & 01:21:42.725, +13:20:58.26 & 3.519 & 11.02 & 43.95 & 10 & Double-lobed & 47.43 & 46.02 & 46.81 \\
TNJ1338 & 13:38:26.10, -19:42:31.1 & 4.104 & 10.97 & 44.44 & 36 & Double-lobed & 48.33 & 46.35 & 46.57 \\
\hline\hline
\end{tabular}
}
\caption{Source properties of the six HzRGs analyzed in this work. (1) Source name, (2) RA, DEC of the radio center (3) redshift, (4) stellar mass, (5) Radio 1.4 GHz luminosity, (6) radio size at 1.4 GHz luminosity, (7) radio morphology (8) AGN bolometric luminosity from [OIII]5007, (9) jet mechanical luminosity, and (10) luminosity associated with shock heated gas. For determining (2) -- for the TGSS source, no radio core is detected; we adopt the midpoint between the two compact lobes as its radio center position. TNJ1338 lacks a detectable core in the 4.8 GHz image but reveals a faint core component at 8.4 GHz in higher-frequency VLA observations, from which the core coordinates are measured. The redshift (Column 3) of TGSS is derived from H$\alpha$ emission at the peak of the continuum, while redshifts for the remaining sources are adopted from \cite{wang25, roy24}. Reported radio end-to-end sizes, morphological classifications, and 1.4 GHz luminosities ($L_{1.4\mathrm{GHz}}$) are compiled from various archival radio programs \citep{pentericci00, debreuck99, gabanyi18, van97, nesvadba07, nesvadba17, wang24}. $\log_{10} L_{\mathrm{jet}}$ is measured from Eq.~\ref{eq:ljet}, described in \S\ref{subsec:ljet}. Calculation of $\log_{10} L_{\mathrm{shock}}$ is described in \S\ref{subsubsec:ionization}.}
\label{tab:1}
\end{table*}

\section{Results} \label{sec:results}

In this section, we present the spatially resolved ionized gas kinematics, their spatial correlation with the radio lobes and jet morphology and the corresponding outflow rates and kinetic power in a sample of six high redshift radio galaxies at z$\sim$ 3.5 to 4.0. Our analyses generally follow the  methodology outlined in \S\ref{subsec:data_analyses}, and \cite{roy24}. Given the diversity of radio morphology and outflow characteristics across the sample, we adopt a case-by-case basis.  We provide an in-depth analysis of TGSS1530+1049, as this is the first detailed study of the source. For the remaining five galaxies already presented in \citet{wang24}, \citet{wang25}, and \citet{roy24}, we highlight key results on jet–gas coupling and energetics of the outflow.

\subsection{TGSS1530+1049} \label{subsec:tgss}
\subsubsection{Emission Line morphology}

\begin{figure*}
    \centering
    \includegraphics[width=\textwidth]{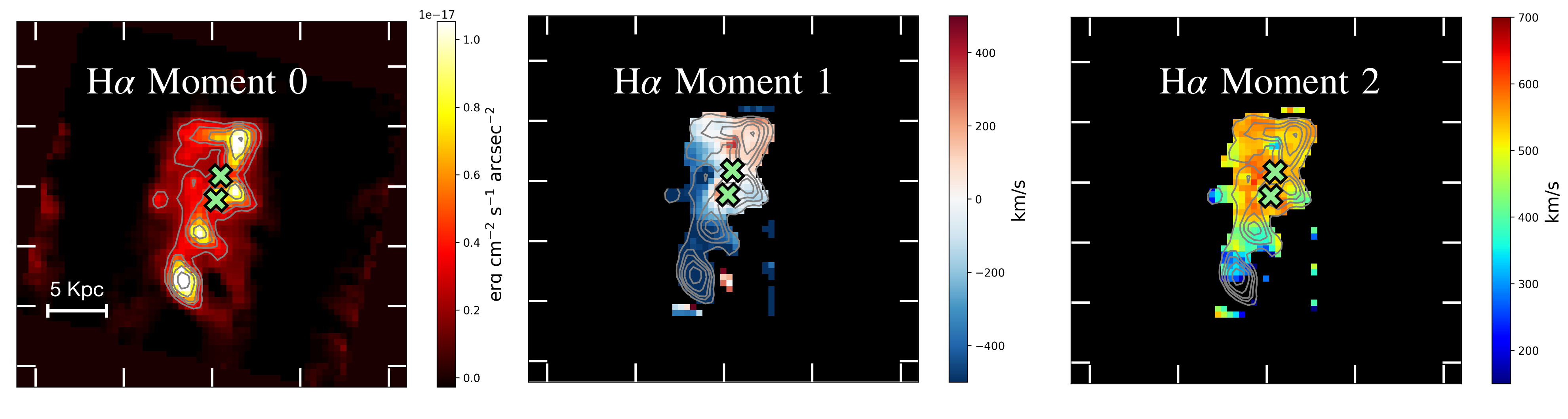}
    \caption{ Moment 0 [left panel], moment 1 [middle panel] and moment 2 [right panel] maps of the H$\alpha$ emission line of TGSS1530. The overlaid gray contours represent the $3\sigma, 5\sigma, 7\sigma, 10\sigma $ and $15\sigma$ of [OIII] moment 0, and the two green stars mark the position compact double radio source. The moment 1 (velocity centroid) map exhibits a sharp velocity gradient ($\sim 800 \ \rm km \ s^{-1}$) and transitions from blueshift to redshift around the radio location. The moment 2 values  are very high in the spatial locations coincident with the radio center, which indicates broad line widths ($\sigma > 800 \ \rm km \ s^{-1}$), turbulent gas motions and jet driven outflows.   } 
    \label{fig:tgss_moments}
\end{figure*}

TGSS1530 (z $\sim$ 4.0) is a radio galaxy with small-scale jets and one of the most luminous sources identified in the TGSS ADR1 survey \citep{intema06, intema17}. It exhibits a steep radio spectral index ($\alpha = -1.3$), was unresolved in VLA 1.4 GHz imaging at $2.5''$ resolution, but resolved with higher resolution VLBA imaging \citep{gabanyi18} into a double source with a separation of 0.4 arcsec ($\sim 2.8$ kpc) with a roughly N/S orientation.
These characteristics are consistent with compact steep spectrum (CSS) sources, that represent young radio-loud AGNs in an early evolutionary phase with small-scale jets that are still confined within the dense ISM of the host galaxy \citep{odea98, odea21}. 
 
 Among the rest-frame optical emission lines detected in the JWST/NIRSpec G395H grating, H$\alpha$ is the most prominent emission line. Hence, it is an effective tracer of the ionized gas distribution in the ISM and potential outflow activity. Figure.~\ref{fig:tgss_moments}  presents the moment-0 map of the continuum subtracted H$\alpha$ emission,  integrated over a velocity range of approximately $\pm 1500 \rm \ km\ s^{-1}$ (corresponding to $\pm \rm 25 $\AA\ in the observed frame).  The ionized gas exhibits a complex, filamentary morphology with an ``8''-shaped structure oriented roughly north-south. The gas is aligned with the axis of the compact radio source, but extends over a larger region ($\sim$ 13 kpc). The radio source is indicated by a pair of stars.
 

\begin{figure*}
    \centering
    \includegraphics[width=\textwidth]{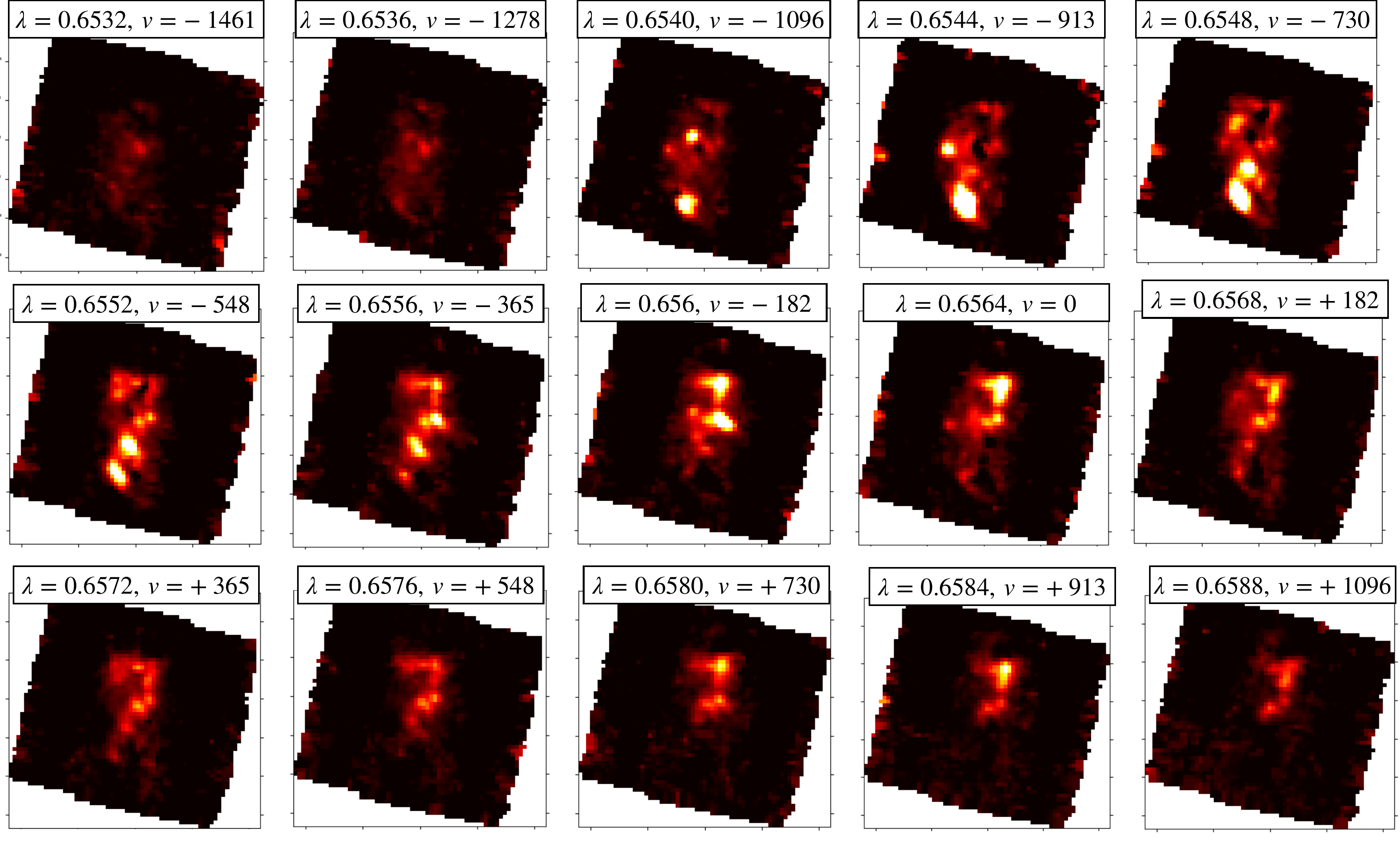}
    \caption{ Narrowband wavelength slices of the H$\alpha$ flux distribution in TGSS1530, constructed from the NIRSpec datacube. Each panel shows the spatially resolved morphology at successive spectral slices, starting from the blueshifted side of the H$\alpha$ line ($\lambda = 0.6532 \ \mu$m, v  $= -1461 \rm \ km \ s^{-1}$) to the redshifted side ($\lambda = 0.6588 \ \mu$m, v$= +1096 \rm \ km \ s^{-1}$), with a $\Delta \lambda = 4$ \AA \ interval (corresponding to $\Delta v \approx 182 \ \rm km \ s^{-1}$). These maps reveal how the ionized gas morphology evolves across the different kinematically distinct gas components extracted from the line profile. The blueshifted slices ($\lambda \lesssim 0.6552 \ \mu$m) show extended emission predominantly bright in the south of the radio nucleus, indicating high-velocity outflowing gas along that direction. At systemic and mildly redshifted velocities ($\lambda \sim 0.6564 - 0.6580 \ \mu$m), the emission adopts an `8'-shaped configuration, with ionized gas distributed symmetrically about the nucleus. In the most redshifted slices ($\lambda \gtrsim 0.6580 \ \mu$m), the emission brightens in the northern region, spatially coincident with the location of the radio source.  }
    \label{fig:tgss_slice}
\end{figure*}

Figure \ref{fig:tgss_slice} shows a sequence of narrow-band wavelength slices centered on the H$\alpha$ emission line, constructed from the blue to the red side of the line in 3$-$6 \AA\ intervals. It is clear that the emission is dominated by two distinct H$\alpha$-bright regions: the southern side of the galaxy shows strong H$\alpha$ emission in the blue-shifted side of the line. Interestingly, the southern structure shows negligible to no continuum associated with it, indicating a purely line emitting region dominated by ionized gas with no stellar component. The emission however, becomes increasingly faint as one moves towards the redshifted part of the H$\alpha$ emission line.
On the other hand, the northern part of the galaxy becomes H$\alpha$ bright primarily in the wavelength slice that correspond to the red-shifted part of the line relative to the systemic velocity. The peak emission is roughly coincident with the spatial location of the compact double radio source (marked as stars in the moment map). These emission structures strongly resemble
those seen in other powerful high redshift radio galaxies  \citep{nesvadba08,mukherjee16, roy24}, where expanding cocoons entrain and displace large amount of ionized gas via outflows. In the next section, we will demonstrate that these extended H$\alpha$-bright regions exhibit broad emission line profiles, high velocity dispersions, and extreme velocity offsets that are signposts  of AGN-driven outflows \citep{harrison14}.

\subsubsection{Gas kinematics and their alignment with radio emission}

\begin{figure*}
    \centering
    \includegraphics[width=\textwidth]{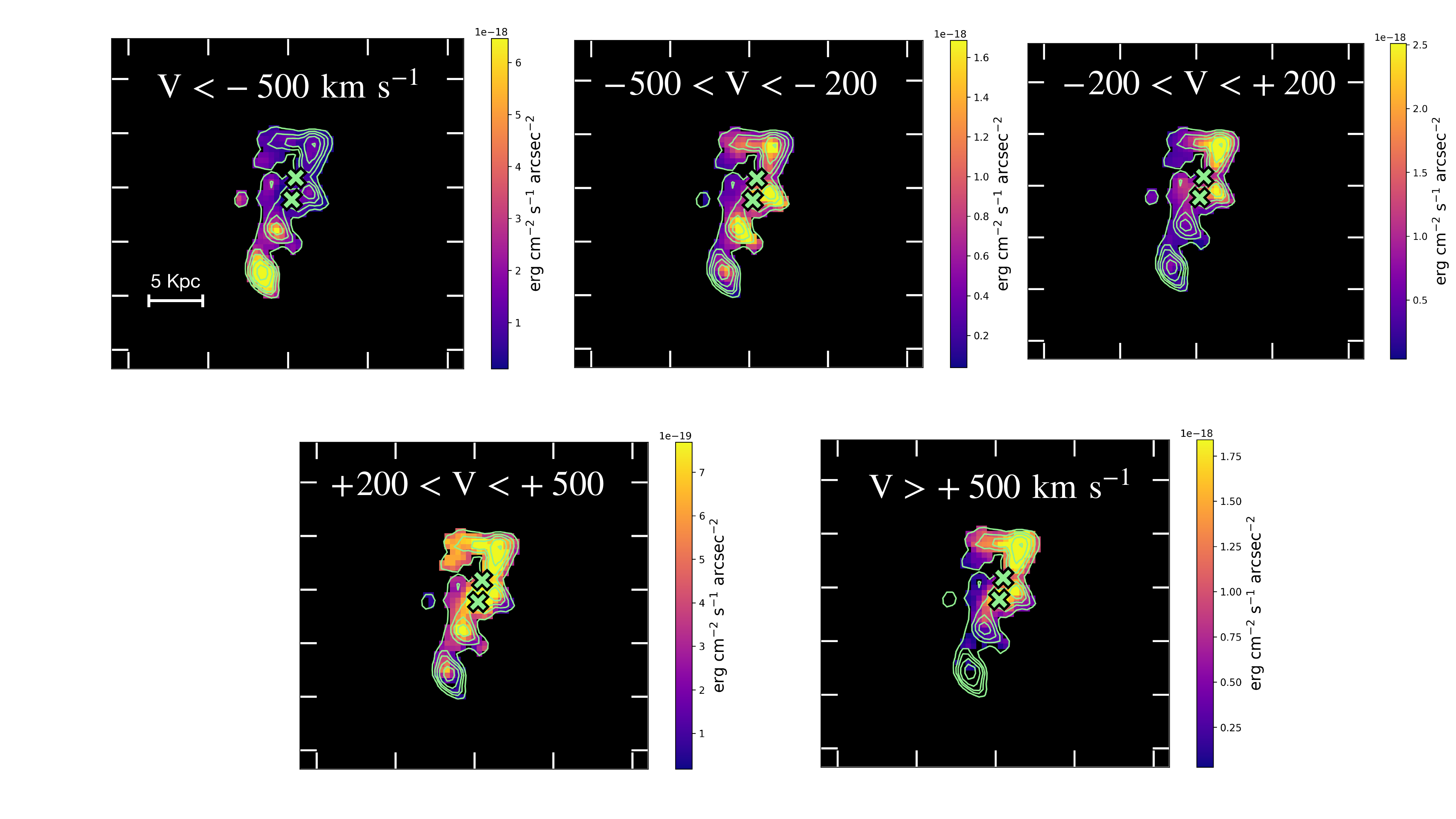}
    \caption{Velocity channel maps of the H$\alpha$ line emission in TGSS1530. The panels show the spatial distribution of the ionized gas at different velocity intervals relative to the systemic velocity ($v\_{\rm sys}$).
The top-left panel shows the high-velocity blueshifted gas ($v < -500 \ \rm km \ s^{-1}$), corresponding to the fastest approaching material along the line of sight, likely tracing the near side of an expanding outflow cone. The next four panels (top-middle to bottom-left) display the intermediate velocity channels within $\pm 500 \ \rm km \ s^{-1}$, which contain the majority of the extended ionized gas emission. The bottom-right panel shows the high-velocity redshifted gas ($v > +500 \ \rm km \ s^{-1}$), tracing the receding side of the outflow structure.
The green crosses indicates the location of the compact double radio source derived from European VLBI observations. The overall morphology and velocity distribution are consistent with a biconical outflow geometry driven by an expanding cocoon of shocked gas inflated by the radio jet. 
} 
    \label{fig:tgss_velchannels}
\end{figure*}

In this section, we highlight the kinematic structure of the peculiar filamentary ionized  gas morphology described above. 
We analyze both velocity-resolved channel maps and moment maps derived from the continuum-subtracted H$\alpha$ emission line. The velocity channel maps shown in Figure \ref{fig:tgss_velchannels} are constructed by integrating emission within five velocity bins: $V < -500$, $-500 < V < -200$, $-200 < V < 200$, $200 < V < 500$, and $V > 500$ km s$^{-1}$, relative to the systemic redshift. These maps trace how the spatial morphology of the ionized gas changes with velocity.

The H$\alpha$ emission component with blueshifted velocities ($V < -500$ km s$^{-1}$) is predominantly concentrated in the southern region of the galaxy, which is devoid of detectable continuum with S/N $>$ 5. Indeed, the
majority of the rapidly moving gas emission lying in the blue ``wing'' of the H$\alpha$ line is located in the
nebular-line-dominated region.  Near systemic velocities ($-200 < V < 200$ km s$^{-1}$), the emission becomes more centrally concentrated, overlapping with the radio lobes (marked by stars), while redshifted channels ($V > 500$ km s$^{-1}$) show emission peaking in the northern extension of the  source. This velocity asymmetry is consistent with a bipolar outflow structure, oriented approximately north–south along the radio lobes, with the southern outflow component approaching and the northern component receding.

The moment-1 (velocity centroid) map (Figure.~\ref{fig:tgss_moments}) confirms this pattern, revealing a clear velocity gradient of $\sim 800$ km s$^{-1}$ across the nebula. Blueshifted gas (up to $-400$ km s$^{-1}$) is concentrated toward the south, while redshifted gas (up to $+400$ km s$^{-1}$) dominates in the northern half. This gradient aligns closely with the morphological axis of the ionized nebula and is consistent with a biconical outflow scenario mentioned above \citep[e.g.,][]{nesvadba08, jarvis19, roy24}.
The moment-2 (velocity dispersion) map (Figure.~\ref{fig:tgss_moments}) shows broad line widths throughout the central and northern regions of the galaxy. The moment-2 value peaks in the regions coinciding with the radio core, and exceeds 800 km s$^{-1}$. Such high gas dispersions near the radio emission are characteristic of highly disturbed gas. In contrast, the southern blueshifted region shows moderately lower dispersions ($\sim 300-500$ km s$^{-1}$), which may reflect a more coherent or collimated outflow component.

\begin{figure*}
    \centering
    \includegraphics[width=\textwidth]{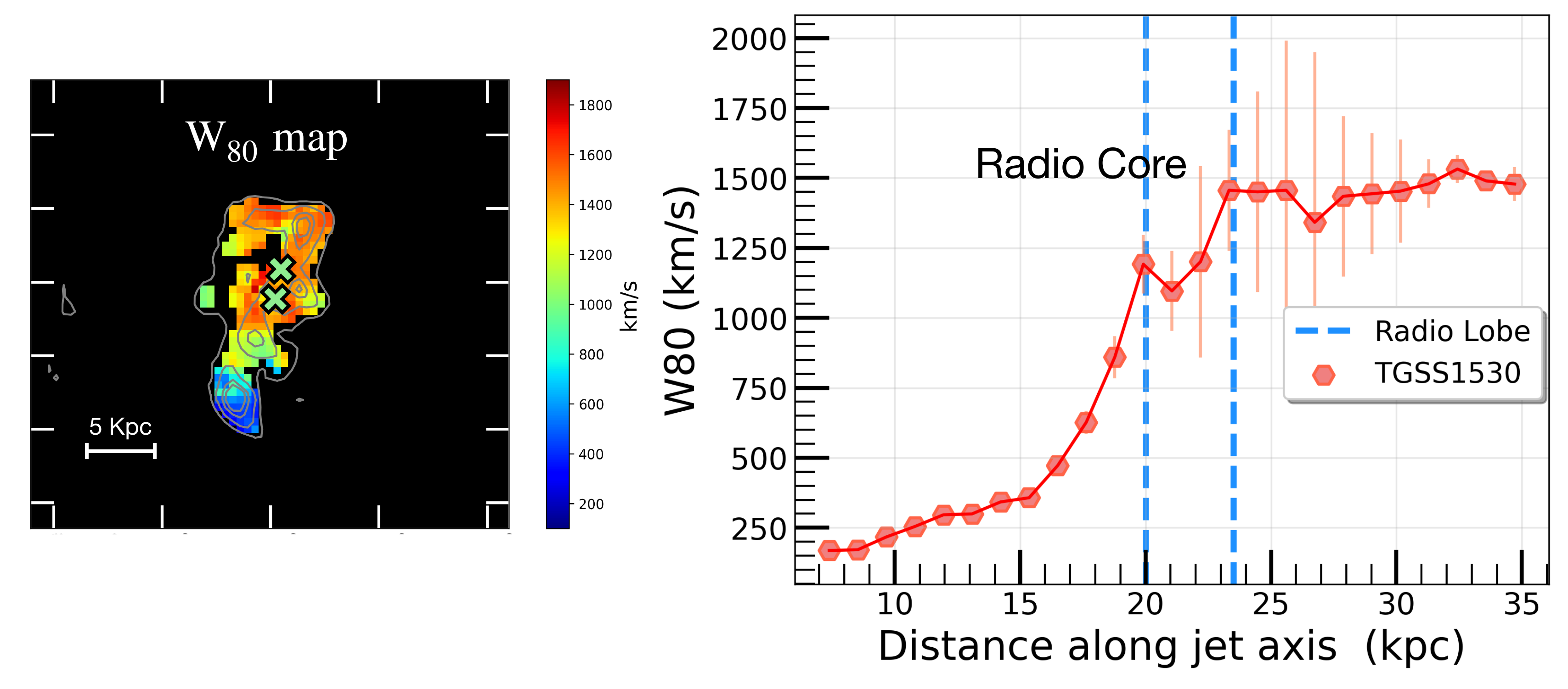}
    \caption{ Left panel: Map of $W_{80} = v_{90} - v_{10}$ from H$\alpha$, a standard gas kinematic tracer, which is the line width containing 80\% of the emission line flux. Here $v_{90}$ and $v_{10}$ are velocities at 90th and 10th percentile of the overall emission line profile in each spaxel. Broad line widths are prevalent in the emission line dominated nebular region coincident with the radio center. Contours represent the H$\alpha$ flux map. Right panel: The radial profile of $W_{80}$ with distance measured from the origin set at the southernmost tip of the galaxy. The location of the two central radio coordinates are shown by the dashed vertical lines. The profile clearly demonstrates the pattern noticed in the 2D map:  a steady increase in the line widths as we approach the northern part of the galaxy towards the radio source. } 
    \label{fig:tgss_w80}
\end{figure*}

To further quantify the turbulent motions, we construct non-parametric $\rm W_{80}$ map, defined as the width of the emission line containing 80\% of the integrated flux (Figure.~\ref{fig:tgss_w80}). The $\rm W_{80}$ map shows values in the range of $800-2000$ km s$^{-1}$ across much of the ionized nebula. Very similar to the moment-2 values, $\rm W_{80}$ map also exhibits the most extreme line widths near the radio core with values $>$ 2000 $\rm km \ s^{-1}$, about 4 times higher than the southern side of the galaxy. These large variations in the line width along the ionized nebula and the spatial coincidence of the broad, disturbed, turbulent gas  with the radio source implies that this extreme gas kinematics is likely a signature of jet driven outflow. The kinematics cannot be produced by gravitational motions alone. 
Figure.~\ref{fig:tgss_w80} (right panel) shows the variation of $\rm W_{80}$ as a function of radial distance  (see \S\ref{subsec:data_analyses} for further details on the construction of the radial profiles).

We construct rectangular apertures aligned with the galaxy’s major axis, with fixed heights of 1 kpc and widths defined by the lateral extent of the [OIII] emission. The average $\rm W_{80}$ from the spaxels within each 1 kpc-wide box is computed and plotted as a function of distance along the major axis, with the origin set at the southernmost tip of the galaxy.  
The locations of the two radio lobes are indicated by dashed lines. Here also a clear trend emerges: $\rm W_{80}$ gradually increases as one approaches from the southern side of the galaxy towards the northern side close to the radio source, and eventually the $\rm W_{80}$ value flattens with a maximum average value $\sim$ 1400 km/s. The strong spatial correlation apparent in the radial profile provide direct evidence that the radio structure and possibly the small-scale embedded jet hiding within the gaseous envelope, is dynamically influencing the kinematics of the ionized gas. The channel maps, velocity centroids, dispersion, and $\rm W_{80}$ diagnostics reveal a consistent picture that TGSS1530 exhibit a dynamic, multi-component outflow extending over $\sim 20$ kpc. TGSS1530 is possibly undergoing early-stage of jet-mode feedback, where compact radio jets are beginning to break out of the host galaxy and drive fast, turbulent outflows into the surrounding medium, consistent with predictions from simulations of jet-ISM coupling \citep[e.g.,][]{wagner12, mukherjee18}. 

\subsubsection{Outflow rate and kinetic power} \label{subsec:tgss_energy}

\begin{figure*}
    \centering
    \includegraphics[width=\textwidth]{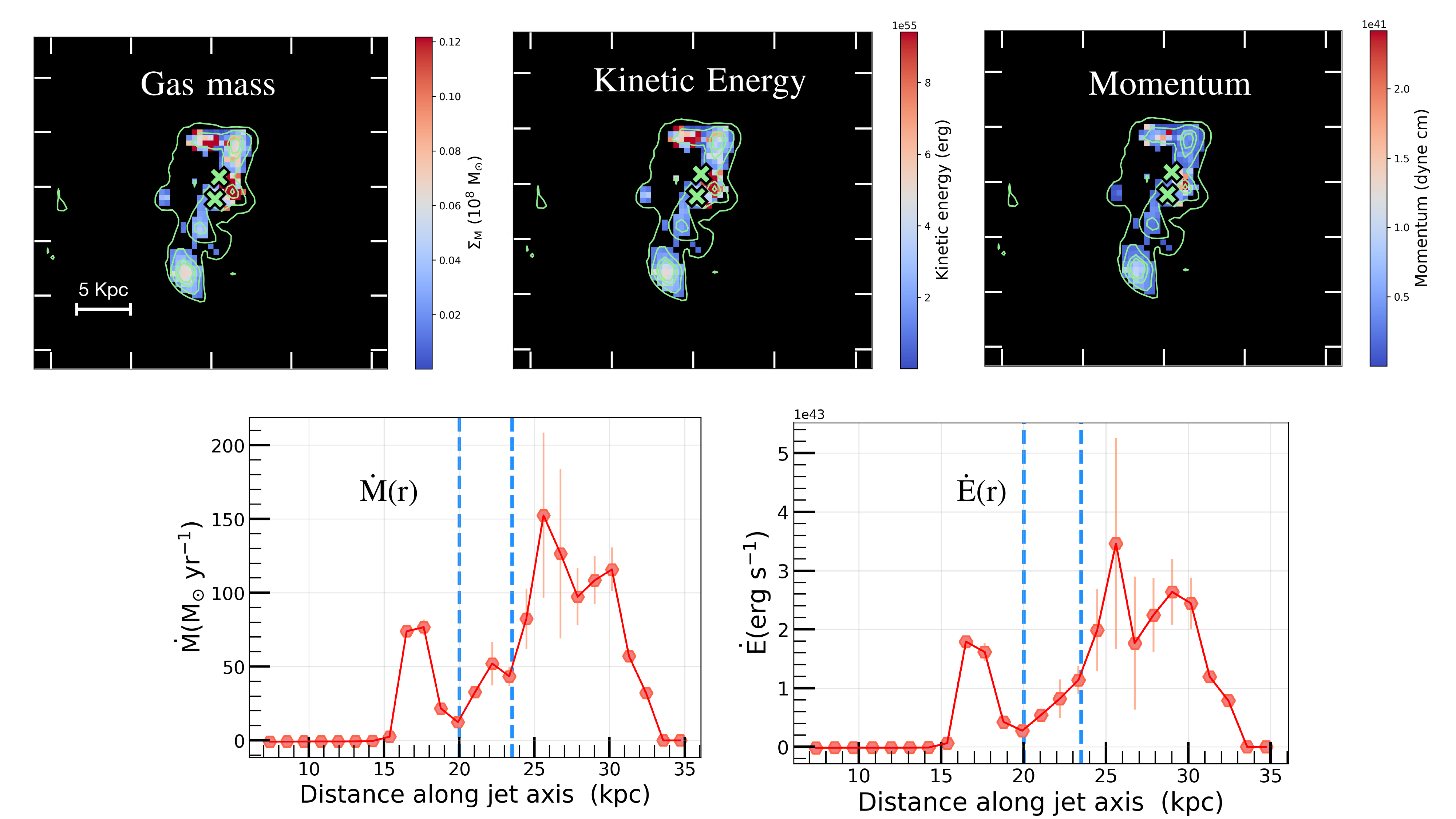}
    \caption{ Spatially resolved maps of ionized gas mass, mass outflow rate, momentum and outflow power in TGSS1530, along with radial trends of mass outflow rate and outflow kinetic power shown along the jet axis.
    Upper left: spatially resolved map of ionized gas mass surface density, in units of $10^8 \ \rm M_{\odot}$ per spaxel, derived from extinction-corrected H$\alpha$ luminosity and electron density estimates from the [SII] doublet (see Section \ref{subsec:tgss_energy} for details).  The total ionized gas mass integrated over the whole galaxy is  $\rm 1.0 \ \times 10^9 \ M_{\odot}$. Overlaid contours correspond to H$\alpha$ moment-0 flux, and the green crosses mark the location of the double radio lobes. Upper middle: Spatial distribution of outflow kinetic energy in units of erg. The kinetic energy peaks around the regions coincident with the radio lobes. Upper right: Spatial distribution of the momentum  in units of dyne-cm. Bottom left panel: The radial profile of the mass outflow rate ($\dot{M}_{\rm out}$ in units of $\rm M_{\odot} \ yr^{-1}$). The origin for the radial distances are set to be the southernmost tip of the galaxy. Thus the left and right vertical dashed lines in the profiles indicate the locations of the southern and northern radio lobes respectively. $\dot{M}{\rm out} (r)$ exhibit a sharp enhancement 
    around $\sim 25$ kpc, which is near the approximate location of the northern radio lobe. This  indicates more mass is being displaced close to the northern lobe, which indicates that the radio jets are pushing into the surrounding dense gas and driving turbulence, resulting in enhanced line widths, greater gas motions and also larger amount of ejected mass. Bottom right panel: The radial profile of kinetic power $\rm \dot{E} (r)$ in units of  erg $\rm s^{-1}$.  The kinetic power show very similar  trend as $\dot{M}_{\rm out}$, peaking near the radio lobe. The total kinetic power integrated over the ionized outflow is $\dot{KE}_{\rm outflow} \sim 1.65 \times 10^{43} \ \rm erg \ s^{-1}$. The measurements of total gas mass, mass outflow rate, kinetic energy, power and momentum are provided in Table 2.    } 
    \label{fig:tgss_energy}
\end{figure*}

In the previous section, we present clear detection of outflowing gas probed in H$\alpha$ emission via broad line width ($\rm W_{80} > 1500 \ km \ s^{-1}$) and high velocity gradient across the nebular gas morphology. In this section, we will study the spatial distribution of the ionized gas mass distribution, mass outflow rate, kinetic power and momentum flux of the outflowing gas in the warm ionized ($\rm T \sim 10^4 \ K$) phase. We accurately measure electron density $n_e$ at every spatial location within the extended nebular region, by fitting single-component Gaussian profile to each of the [SII] $\lambda \lambda$6717, 6731 doublet lines  and computing their flux ratio. We used the parameterization of \cite{sanders16}, and the
assumption of $\rm T \sim 10^4 \  K$ to convert to electron density. We find that the resulting electron density ranges from 400$-$950 $\rm cm^{-3}$. Since H$\beta$ is not covered within the observed wavelength range, we assume a mean extinction $A_v = 0.8$ following $z = 4.1$ Radio galaxy \cite{roy24}, and dust-correct the H$\alpha$ flux. 
The ionized gas mass distribution for each of the 1 kpc-wide bins can be calculated using H$\alpha$ emission associated with the outflow assuming case B recombination, following \cite{cresci23}:

\begin{equation} \label{mion_ha}
    \rm M_{out} = 3.2 \times 10^5 \ \frac{L_{out, H\alpha}}{10^{40} \ erg \ s^{-1}}\frac{100 \ cm^{-3}}{n_{e, out}} \ M_{\odot}
\end{equation}

where $\rm L_{out, H\alpha}$ is the extinction corrected H$\alpha$ luminosity of the outflow component and $n_e$ is the measured electron density. We only include flux from H$\alpha$ corresponding to the velocity channel $\rm |v| > 500 \ km \ s^{-1}$ to include the outflowing component. Figure.~\ref{fig:tgss_energy} shows the distribution of ionized gas mass associated with the outflowing component. The ionized gas mass surface density ranges from $\rm (0.4 - 1.2) \times 10^7\ M_{\odot} $ per spatial pixel, with a total outflowing ionized gas mass of $\rm 1 \ \times 10^9 \ M_{\odot}$. 

Once the outflow mass is derived, the spatially resolved outflow rate for each 1-kpc-wide radial bin along the radio axis is:
\begin{equation} 
    \rm \dot{M}_{out} = \frac{M_{out}v_{out}}{\Delta R_{out}}
\end{equation}

Here $\Delta R_{out}$ is equal to 1 kpc. 
We take the velocity in each pixel $\rm v_{out} = \sqrt{W^{2}_{50} + \Delta v^2}$, where W50 is the line width corresponding to 50\% of the emission-line flux, and $\Delta$v corresponds to the velocity
offset from the systemic velocity. 
The bottom-left panel of Figure \ref{fig:tgss_energy} shows the radial profile of $\dot{M}$, with the positions of the radio lobes marked by dashed lines. This reveals a clear spatial correspondence between elevated outflow rates and the radio structure.
The total mass outflow rate computed in each 1kpc-rectangular box shows a 3$x$ enhancement around the radio core compared to the mean outflow rate in the rest of the galaxy. 
The radial profile of outflow kinetic power (Figure~\ref{fig:tgss_energy}, bottom right) mirrors the profile of the mass outflow rate, and also exhibit a pronounced peak near the radio lobes.  This spatial alignment strongly supports a direct coupling between the radio jet and the ionized gas.
 This is consistent with the scenario observed in the $z=4.1$ radio galaxy TNJ1338 $-$ another source in our sample $-$ previously analyzed in \cite{roy24}. For TGSS1530, we estimate an average mass outflow rate $\rm \dot{M}_{out} \sim 76 \  M_{\odot} \ yr^{-1}$.  The resolved maps of kinetic energy of the outflows and the momentum flux are also shown in Figure \ref{fig:tgss_energy}. The total kinetic power of the outflow is $\rm 1.65 \times 10^{43} \ erg \ s^{-1}$, which is indicative of powerful feedback energy capable of affecting the galaxy on the largest scale. 

\subsection{4C+03.24} \label{subsec:4c03}

\begin{figure*}
    \centering
    \includegraphics[width=\textwidth]{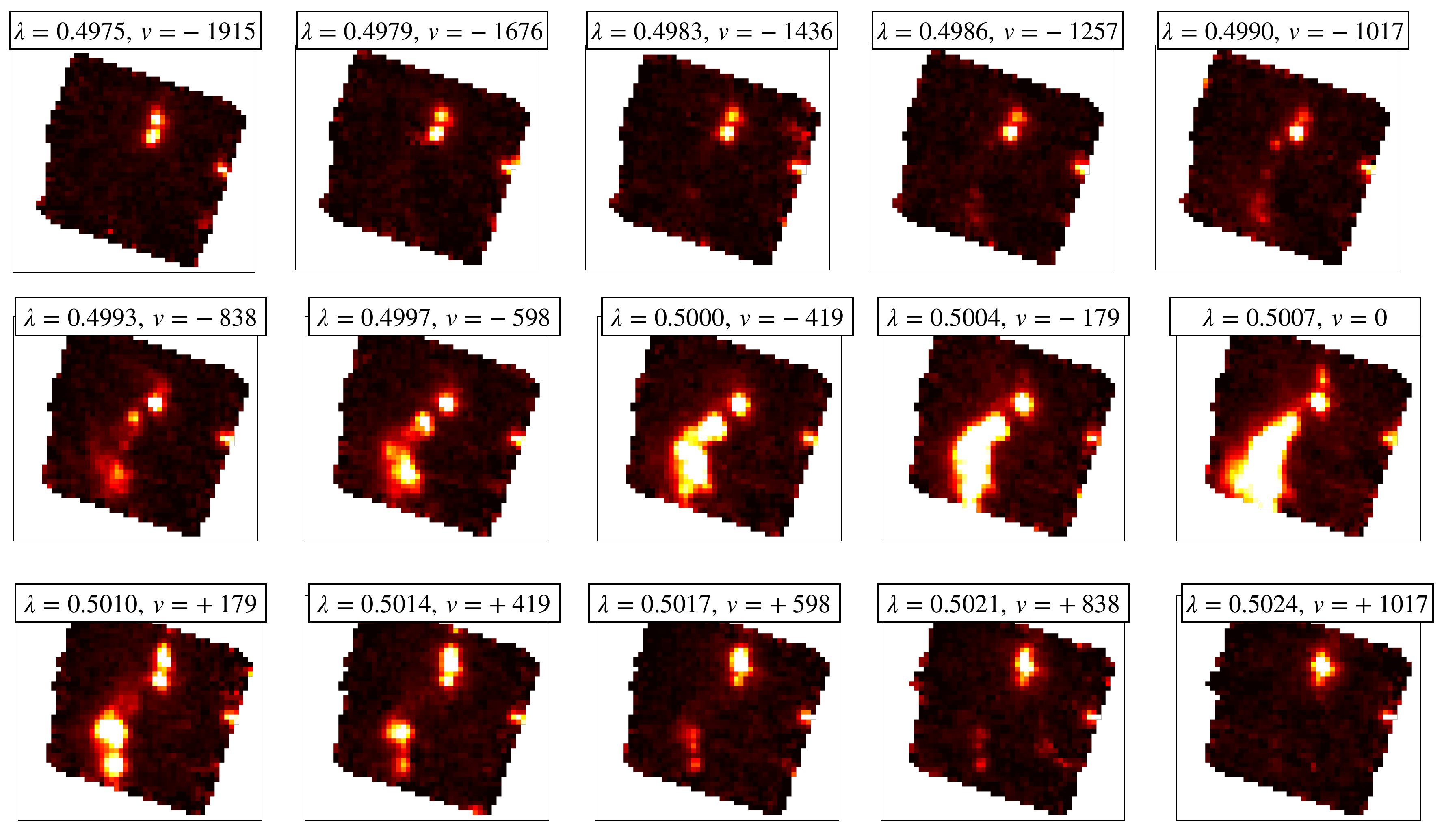} 
    \caption{Narrowband channel maps of [OIII]$\lambda5007$ emission in 4C03+24. Each panel shows the spatial distribution of [OIII] flux along successive spectral slices from the blueshifted side of the line ($ v \approx -1900 \ \rm km \ s^{-1}$) to the redshifted side ($v \approx +1010 \ \rm km \ s^{-1}$) in  $\sim 3-4$ \AA\ intervals.
The blueshifted slices reveal a compact [OIII] morphology, primarily coincident with the radio core. As the velocity approaches the systemic redshift of the galaxy, the ionized gas becomes significantly more extended and luminous, connecting the central radio source with the southern radio lobe. In the redshifted slices, the nebular emission becomes more compact again, with enhanced flux localized near the northern and southern extremities of the galaxy. } 
    \label{fig:4c03_slice}
\end{figure*}

4C03 has an extended radio morphology, as revealed by VLA 15 GHz radio observation. It has three peaks in radio emission, the center one of which is determined as the core, The other two radio emission blobs resemble extended jets of a lobed radio galaxy. Although the central radio emission is referred to as the core, it is spatially extended (end to end size of the central component $\sim 0.8''$). Thus, it might physically be a part of one of the extended radio lobes seen on either side. All the major emission lines are detected using G235H grating of JWST/NIRSpec IFU. We use [OIII]5007 primarily for our outflow analyses since it is the brightest emission line emitted by the outflow within the rest-frame visible range
covered by the NIRSpec data cube. It is also well separated in
wavelength from neighboring emission lines, so decomposition of blended lines is not required.

Figure.~\ref{fig:4c03_slice} shows the sequence of narrow band wavelength slices around [OIII]$\lambda$5007 emission line. These slices span from $\sim$30\AA \ blue ward of the line wavelength to 20\AA\ to the red side of the line in the rest frame. These slices correspond to a velocity range of roughly -1800 km/s to +1200 km/s, and offer a velocity-resolved view of the ionized gas distribution across the galaxy. 
The images  reveal a bright, extended, and morphologically complex nebular emission from the warm ionized component along the south-north direction. In the blue channels which capture the blue-shifted components of the emission line ($-1500 < v < -500 $ km/s), the emission is compact and concentrated near the northern portion of the system, coincident with the central radio core. As the slices move towards the systemic velocity, the nebular [OIII] emission becomes increasingly bright ($ \times 5$) and extended ($\sim 30$ kpc) and spans the full length of the field, reaching projected distances of $\sim$ 20 kpc. The emission morphology, thus, transitions from compact to contiguous and elongated morphology and starts to bridge the radio core at the northern part of the galaxy with the southern radio lobe along the putative jet axis.  
At redder wavelengths ($\lambda >$0.5007 $\mu$m), corresponding to redshifted velocities, the emission breaks into more spatially distinct clumps on both the north and south side of the galaxy. The patchy ionized gas possibly traces decelerating or shocked gas at the jet edges.

\begin{figure*}
    \centering
    \includegraphics[width=\textwidth]{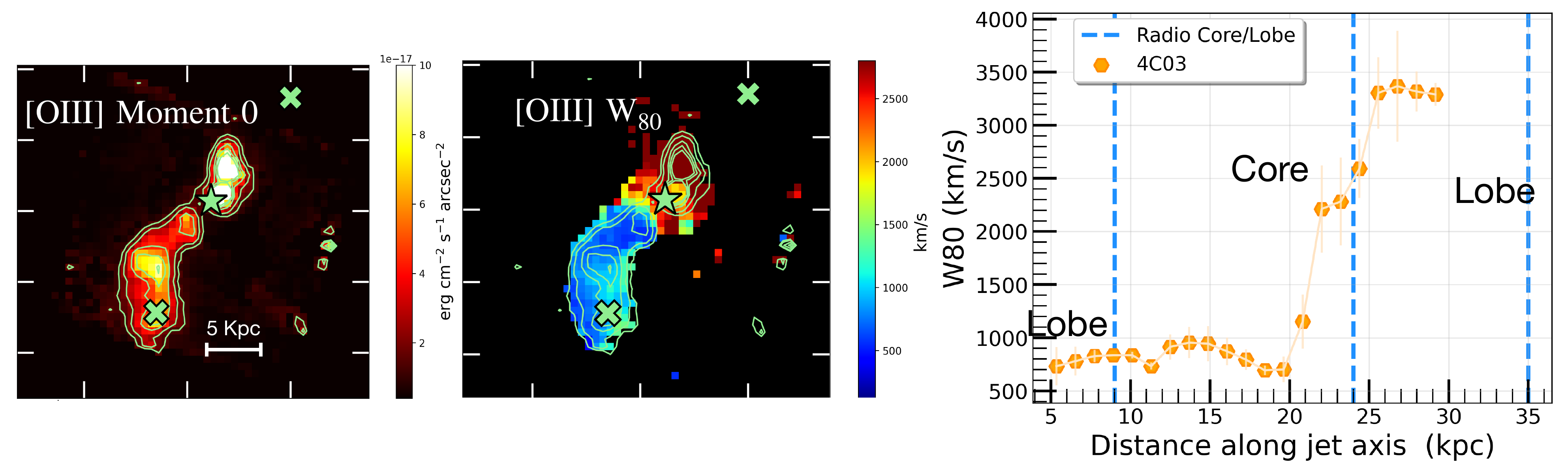}
    \caption{ Spatially resolved maps of ionized gas in 4C03+24 traced by [OIII]5007 emission line. The star and the two crosses show the putative radio core and lobes. The left panel shows the [OIII] moment-0 (integrated flux) map, revealing the extended ionized gas morphology aligned broadly with the radio axis. The middle panels shows [OIII] $W_{80}$ map, a non parameteric tracer of line width (similar to Figure \ref{fig:tgss_w80}).
     $W_{80}$ increases drastically from the southern side of the galaxy to the northern end, peaking  near the radio center. There is a moderate enhancement in line width along the southern lobe as well. 
     The right panel displays $W_{80}$ as a function of projected distance along the radio axis, with the origin defined at the southern edge of the galaxy, near the southern radio lobe. A modest increase in $W_{80}$ is observed near the southern lobe, followed by a sharp rise toward the radio core: consistent with the 2D map. These trends suggest that the ionized gas becomes progressively more kinematically disturbed along the jet axis near the radio hotspots. We hypothesize that an expanding cocoon driven by the shocked jet fluid accelerates the ambient gas outwards to large velocities.
} 
    \label{fig:4c03_moments}
\end{figure*}

Figure \ref{fig:4c03_moments} shows the moment-0 (line integrated flux) map for [OIII] $\lambda$5007. 
 The gas distribution follows the direction connecting the radio core to the lobes, suggesting a strong spatial and physical coupling between the ionized outflow and the underlying radio jet. The morphology is broadly consistent with that of the high-redshift radio galaxy TNJ1338 at $z=4.1$, also observed with JWST/NIRSpec IFU by \cite{roy24}. Thus our conclusion is that this filamentary, clumpy structure of the ionized gas arises from jet propagating through a non-uniform ambient medium  shock-heats the gas, rather than illumination from double AGNs or companions as proposed by \cite{wang25}.

The $\rm W_{80}$ linewidth map (Figure.~\ref{fig:4c03_moments}, middle panel) provides kinematic information to complete this picture. We find extremely high gas dispersion and broad lines, with $\rm W_{80}$ exceeding $2500 \rm km \ s^{-1}$, coincident with the [OIII]-bright region near the radio core in the northern part of the galaxy. A distinct jump in line width is observed when moving from the southern regions of the galaxy toward the north, particularly near the radio core (marked by the star symbol) and extending toward the radio lobe. Given the astrometric uncertainties discussed earlier, the apparent offset between the core and the region of maximum line broadening is likely within the positional error margins, suggesting they are indeed spatially coincident.  The [OIII] flux also increases by a factor of $\sim 3-5$ there. Regions close to the southern radio hot spot (or ``lobe''; marked by cross) also exhibit broad [OIII] line widths with $\rm W_{80} > 1400 \ \rm km \ s^{-1}$. However, $\rm W_{80}$ averages much lower $: 600-750  \ \rm km \ s^{-1}$ in the rest  of the galaxy.  Thus, regions with the brightest [OIII]-emitting regions, and closest to any of the radio core/lobes are most turbulent in gas motions, showing elevated gas velocity dispersions. This is a definite signature of a fast moving outflow, and are incompatible with rotational motion associated with the host galaxy. The spatial coincidence of increased line widths with the radio hot spots likely suggest that the large scale outflows are triggered by the radio jet.

\begin{figure*}
    \centering
    \includegraphics[width=\textwidth]{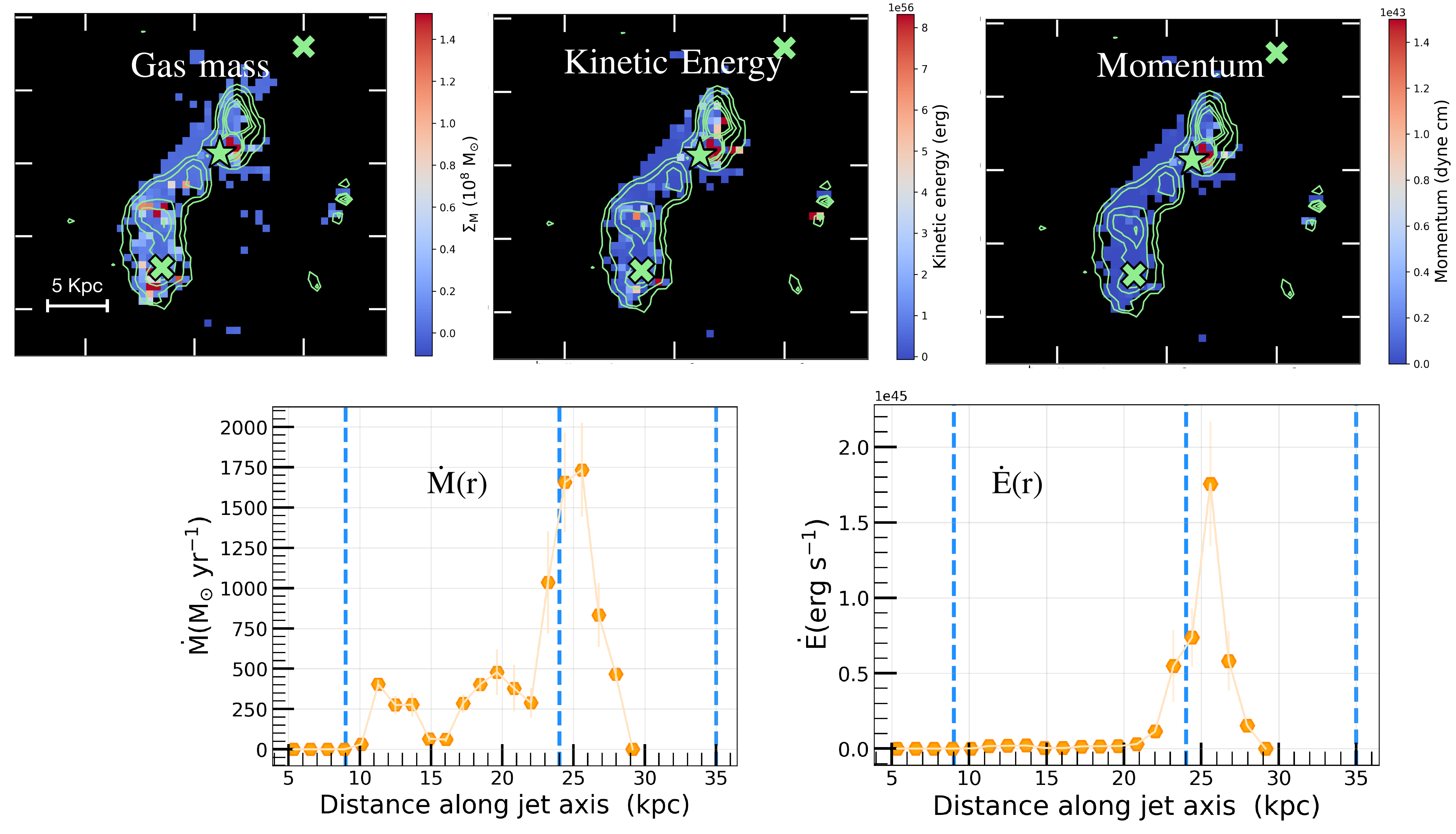}
    \caption{ Top row (left to right): Spatially resolved maps of ionized gas mass, kinetic energy, and momentum flux in 4C03+24. The color scale, units, and contour definitions follow those used in Figure~\ref{fig:tgss_energy}. Bottom row: Radial profiles of mass outflow rate and kinetic power along the jet axis, following the same convention as Figure~\ref{fig:tgss_energy}. The radio core and lobes are marked as a star and crosses in the maps, and as vertical dashed lines in the radial profiles. The origin of the radial axis is defined at the southernmost extent of the galaxy, such that the dashed lines from left to right correspond to the southern lobe, core, and northern lobe, respectively. The radial profiles of both the mass outflow rate and kinetic power exhibit a clear enhancement—by a factor of 2–3—in regions coincident with the central radio core. The total gas mass, kinetic energy, outflow rates and momentum measurements are given in Table 2.} 
    \label{fig:4c03_energy}
\end{figure*}

Figure.~\ref{fig:4c03_energy} (right panel) shows the variation of $\rm W_{80}$ as a function of  distance along the radio jet axis which demonstrates these features in a clearer, and more definite way (similar to Figure.~\ref{fig:tgss_w80} shown before). The location of the three radio peaks are marked as dashed lines.  A clear trend emerges: $\rm W_{80}$ shows a modest increase as one approach from the southern side of the galaxy. It shows a small jump near the southern radio emission, then shows a sharp jump in line width (by a factor of 2) around the radio emission near the center, and the line width increases further and eventually the $\rm W_{80}$ value flattens near the northern part of the galaxy, achieving the highest value $\sim$ 3500 km/s. Such a tight spatial correlation between line width and radio structure can only arise when the radio jet drives large scale gas outflows and modulate the kinematics. Similar behavior has been previously observed in another high redshift ($z=4.1$) Radio galaxy TNJ1338 \citep{roy24}. The kinematics have been interpreted as radio jet shock heating the surrounding jet fluid, and driving laterally expanding bubble/cocoons of shocked material, which may accelerate the warm ionized gas outwards in radial direction.

Since the [OIII]$\lambda$5007 emission line is covered within our observed wavelength range, we adopt an [OIII]-based prescription to estimate the ionized gas mass, rather than using H$\alpha$ diagnostics given in Eq.~\ref{mion_ha}. Specifically, we use the expression from \cite{cano-diaz12, veilleux23}: 

\begin{equation}
    \rm  M_{ionized} = 5.3\times 10^8 \frac{C_e L_{44}([OIII]\lambda 5007)}{n_{e,2}10^{[O/H]}} M_{\odot}
\end{equation}

Here, $\rm L_{44}([OIII]\lambda 5007)$ is the dust-corrected luminosity of the [OIII] $\lambda$5007 line, normalized to $\rm 10^{44} \ erg\ s^{-1}$, $\rm n_{e,2}$ is the electron density normalized to $\rm 10^2 cm^{-3}$, $C_e$ is the electron density clumping factor which can be assumed to be of order unity, and $\rm 10^{[O/H]}$ is the oxygen-to-hydrogen abundance ratio relative to solar abundance. We assume solar abundance, so [O/H] = 0, and thus $\rm 10^{[O/H]}$ is 1. 
We apply this [OIII]-based method uniformly in all the rest of the radio galaxies in our sample
to be consistent with our previous work \citep{roy24}.

The resolved maps of gas mass, kinetic energy of the outflows and the momentum flux are shown in the top row of Figure.~\ref{fig:4c03_energy}. The radial profiles of the mass outflow rate and kinetic power of the ionized gas outflows are shown in the bottom panel, similar to Figure \ref{fig:tgss_energy}. 
The regions exhibiting the highest gas turbulence ($\rm W_{80} > 2500 \ km \ s^{-1}$ in Figure \ref{fig:4c03_moments}) concentrated around the central radio core (marked with a star) correspond to the highest surface densities of both outflowing mass and kinetic energy injection. This direct spatial coincidence between the ionized gas mass displacement in the radio jet direction and the energy injection rate to the ambient medium is also apparent in the radial profile of mass outflow rate and the kinetic power (Figure \ref{fig:4c03_energy} bottom row), which shows a sharp rise in $\rm \dot{M}$ and $\rm \dot{E}$ near the radio center.
This signifies that majority of the outflowing gas may still be concentrated near the central radio blob, and then expand towards the larger area. The average mass outflow rate computed from the radial profile $ = \rm 654 \ M_{\odot} \ yr^{-1}$, and the total kinetic power $= \rm 9.1 \times 10^{44} \ erg \ s^{-1}$.

\subsection{4C+19.71} \label{subsec:4c19}

\begin{figure*}
    \centering
    \includegraphics[width=\textwidth]{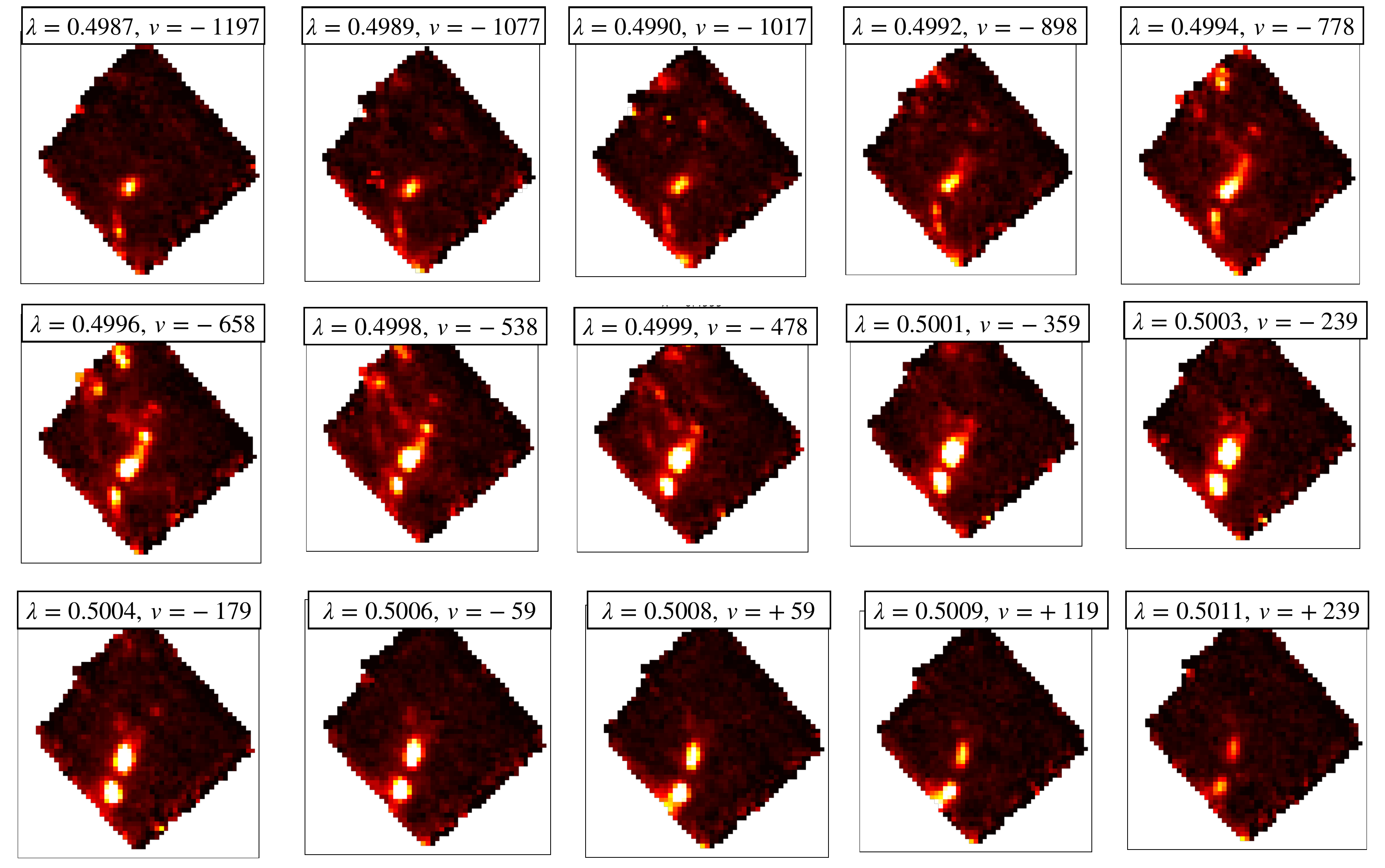}
    \caption{Narrowband channel maps of [OIII]$\lambda5007$ emission in 4C+19.71, constructed along wavelength slices from the blue to red side of the emission line, similar to Figure~\ref{fig:4c03_slice}. The slices span a total velocity range of $\sim 1430 \ \rm km \ s^{-1}$. The ionized gas morphology is initially compact and concentrated near the radio core, but becomes increasingly extended in subsequent velocity slices, revealing a faint, bent bridge of emission connecting the central region to the northern radio lobe across $\sim$ 30 kpc. The northern lobe is located near the edge of the NIRSpec IFU field of view. The kinematically varying nebular structure and the evolving [OIII] emission morphology across velocity channels suggest disturbed, complex, multi-component ionized gas that extends preferentially along the radio axis.  } 
    \label{fig:4c19_slice}
\end{figure*}

4C+19.71 is a luminous radio galaxy at $z = 3.5892$ with double radio lobes on either side of the continuum peak extending $\sim 60$ kpc across, as seen in VLA 4.8 GHz imaging. Both lobes lie outside the JWST/NIRSpec IFU field of view obtained from JWST GO-1970 program \citep{wang24, wang25}. Although the radio core was initially undetected in the radio image, recent reprocessing of VLA C-band observations by \cite{wang25} revealed that the central radio core component is faint, but present, and is coincident with the ALMA continuum emission.

Figure.~\ref{fig:4c19_slice} shows the wavelength-sliced narrow band [OIII]$\lambda$5007 maps, that reveal extended, multiple morphologically complex emission. In the bluer velocity channels (e.g., $\lambda \lesssim 0.4999\ \mu\rm m$), a diffuse extended tail of emission emerges to the north. At systemic velocity and redshifted slices ($\lambda \gtrsim 0.5001\ \mu\rm m$) the emission coalesces into two bright knots extending $1.5''$ ($\sim$11.2 kpc). These knots align with the projected jet axis and may exhibit a morphologically symmetrical structure around the radio core. This suggest a physical connection between the ionization of the gas and the presence of the AGN.

\begin{figure*}
    \centering
    \includegraphics[width=\textwidth]{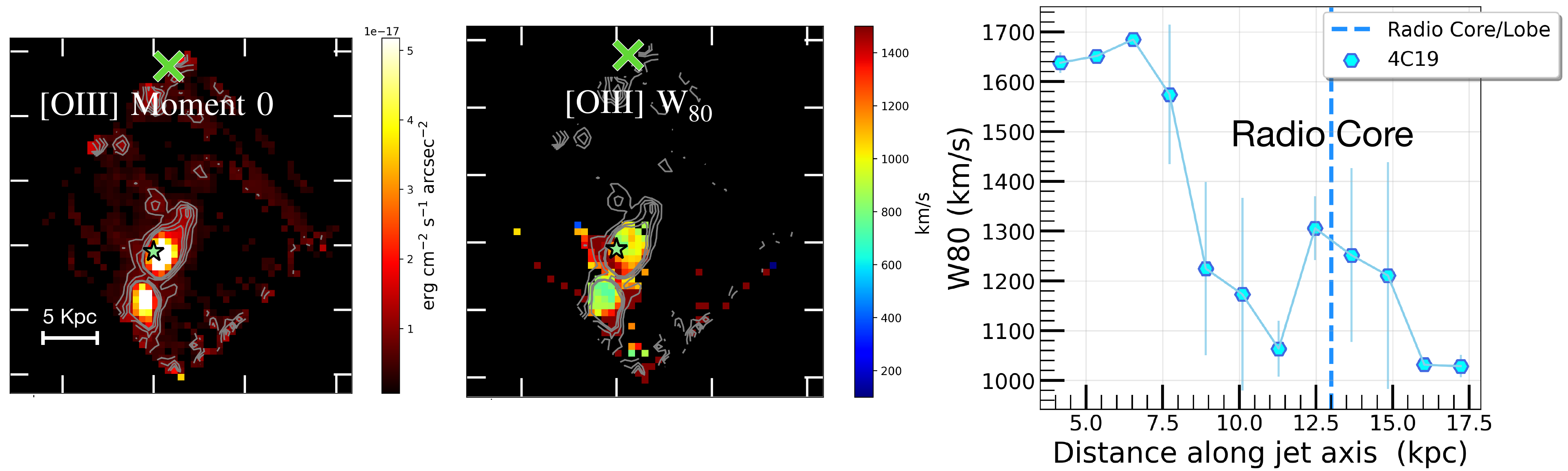}
    \caption{ Spatially resolved maps of [OIII]5007 moment 0 flux map (left panel), $\rm W_{80}$ line width map (middle panel) and radial profile of $\rm W_{80}$ as a function of distance along the putative jet axis (right panel) in 4C+19.71. The symbols, contours, and annotations follow the same conventions as Figure~\ref{fig:4c03_moments}. The star symbol marks the position of the radio core, and the green cross indicates the northern radio lobe. From the 2D $\rm W_{80}$ map, it is clear that the broadest lines with $\rm W_{80} > 1500 \ \rm km \ s^{-1}$ are concentrated near the radio core. When we compute average line widths in 1kpc-wide rectangular boxes and compute the radial profile setting the origin as the southern most edge of the galaxy, we find that $\rm W_{80}$ generally declines with increasing distance. It should be noted that the measurements at small radial distance are affected by limited spaxel sampling, and line widths could not be reliably measured at the northern radio lobe location either due to low S/N. A clear enhancement in $\rm W_{80}$ is still observed near the radio core, consistent with localized kinematic disturbance. Although the hypothesis that the strongest kinematic disturbance is concentrated near the AGN still holds for this source, the radial profiles are less indicative  of the scenario. } 
    \label{fig:4c19_moments}
\end{figure*}

The $\rm W_{80}$ map, which quantifies the width of the emission line enclosing 80\% of the total flux, is shown in Figure.~\ref{fig:4c19_moments}. The velocity dispersion remains relatively uniform across the field of view, with a mean of $\rm W_{80, avg} = 996 \pm 6 \ km \ s^{-1}$ for $>97\%$ of the spaxels. The line widths are significantly broader than expected from virial motions or disk rotation alone, indicating that the gas is dynamically disturbed.

While the majority of the map exhibits flat kinematics, there are localized enhancements in $\rm W_{80}$ to $\sim 1500-2000 \ \rm  km\  s^{-1}$ near the central few spaxels situated between the two bright [OIII] knots, spatially coincident with the position of the radio core (marked by a star). A $2x$ increase in $\rm W_{80}$ is also observed toward the lower edge of the map, close to the direction of the southern radio lobe located just outside the IFU field. However, the spaxels are too close to the IFU edge and suffer from low S/N spectra. Figure~\ref{fig:4c19_moments} (right panel) shows the radial profile of average $\rm W_{80}$ along the major axis. The line widths peak near the southern tip of the galaxy, defined as the origin, and exhibit a steady decline with increasing radial distance. A localized enhancement in $\rm W_{80}$ is observed near the radio core; however, the preferential increase in line widths along the radio jet axis and their spatial correlation are less pronounced compared to the other sources in our sample. Since the radio lobes themselves lie beyond the IFU footprint, this dataset cannot fully capture the effect of the lobe termination shocks or the lateral expansion of the cocoon on the outer ISM. 
Nevertheless, the extremely high line widths are most consistent with an extended outflowing gas launched close to the radio core. An alternative scenario of a nearby companion has been proposed by \cite{wang25}, which is a possibility that cannot be ruled out fro the current data.

\begin{figure*}
    \centering
    \includegraphics[width=\textwidth]{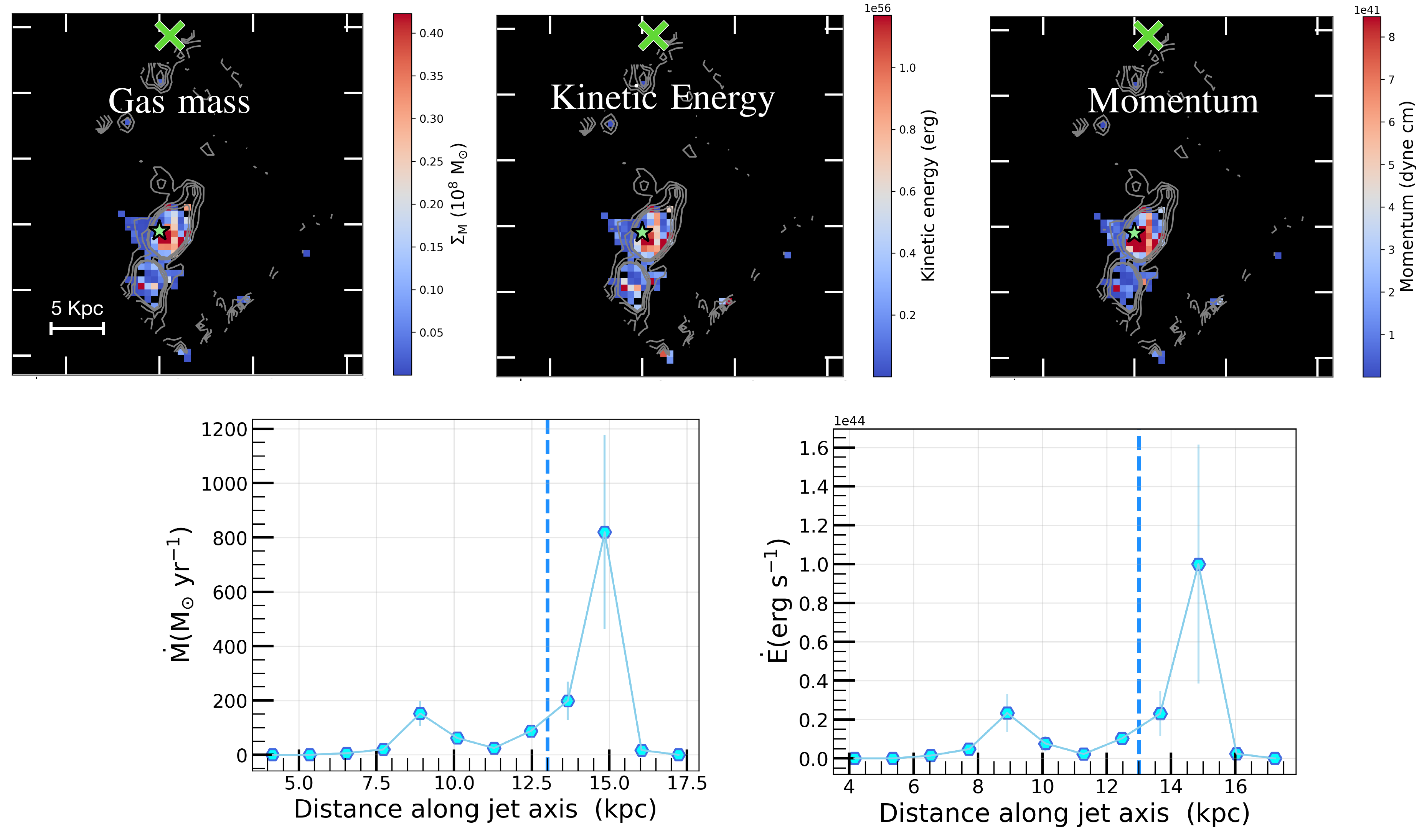}
    \caption{Same as Figure~\ref{fig:tgss_energy}, but for source 4C+19.71. Both mass outflow rates (bottom left panel) and kinetic power (bottom right panel) exhibit a clear enhancement near the radio core. The radial profiles show a sharp peak in both mass outflow rate and kinetic power at $\sim 15$ kpc, spatially coincident with the radio core position (blue dashed line). Although the $\rm W_{80}$ radial trend did not show a strong localized enhancement in the $\rm W_{80}$ radial profile in Figure \ref{fig:4c19_moments}, the outflow rates and energetics maps confirm that substantial mass and energy are being injected near the AGN. } 
    \label{fig:4c19_energy}
\end{figure*}

To quantify the gas energetics, we derive spatially resolved maps of ionized gas mass, kinetic energy and momentum, along with radial profiles of the mass outflow rate and kinetic power (Figure \ref{fig:4c19_energy}). These measurements are based on dust-corrected [OIII] emission, using local electron densities estimated from the [SII] $\lambda$6716/$\lambda$6731 ratio. A velocity threshold of $|v| > 500\ \rm km\ s^{-1}$ is applied to isolate the outflowing component. Both kinetic energy and momentum are enhanced near the radio center. 
Figure.~\ref{fig:4c19_energy} 
bottom row shows the radial profile of mass outflow rate $\rm \dot{M}$ and kinetic power $\rm\dot{E}$, both of which reveal a sharp (factor of $\sim 4-5$) rise at location approximately coincident with the radio core. This indicates that most of the mass and energy transfer is occurring near the central engine which drives expanding cocoon and ejects gas outwards. The integrated total outflow rate is $\dot{M}_{\rm out} \approx 314 \ M_\odot\ \rm yr^{-1}$, which is very high and consistent with large scale ejection of gas via outflows. 
The corresponding total kinetic power  $\dot{KE}_{\rm outflow} \approx 4.0 \times 10^{43}\ \rm erg\ s^{-1}$. This corresponds to a kinetic efficiency of $\epsilon_{\rm kin} \sim 0.15\%$, where the AGN bolometric luminosity  is calculated to be $L_{\rm bol} = 2.6 \times 10^{46}\ \rm erg\ s^{-1}$. This is higher than that observed in other HzRGs like TNJ1338 which show large scale outflows along the radio jet driven by expanding  cocoons \citep{roy24}. 
The total integrated momentum flux value is $\dot{p}_{\rm out} \sim 1.2 \times 10^{36}\ \rm dyne$. 

Altogether, the morphology, kinematics, and resolved energetics of the [OIII]-emitting gas in 4C+19.71 provide compelling evidence of a coherent ionized outflow driven from the central radio core. This source exhibits less extreme gas kinematics compared to the other Radio AGNs (e.g., TGSS1530, 4C03), and no apparent spatial correlation between enhanced gas velocity dispersions and the radio jet direction; however the estimated energy, momentum and mass outflow rate are sufficiently high to perturb the galaxy-scale ISM.

\subsection{TNJ0205+2242} \label{subsec:tnj0205}

TNJ0205+2242 is a powerful radio galaxy at z = 3.506 exhibiting a clear double-lobed radio morphology and a central radio core detected in high resolution VLA imaging. The JWST/NIRSpec IFU observations, reported in \cite{wang25}, spatially resolve the [OIII]$\lambda$5007 emission and detect several [OIII]-emitting blobs and complex nebular structures aligned with the radio axis. Both the radio lobes lie within the field of view of the IFU observations, and thus enable a direct probe of the impact of jet driven cocoons on the ambient gas. 

\begin{figure*}
    \centering
    \includegraphics[width=\textwidth]{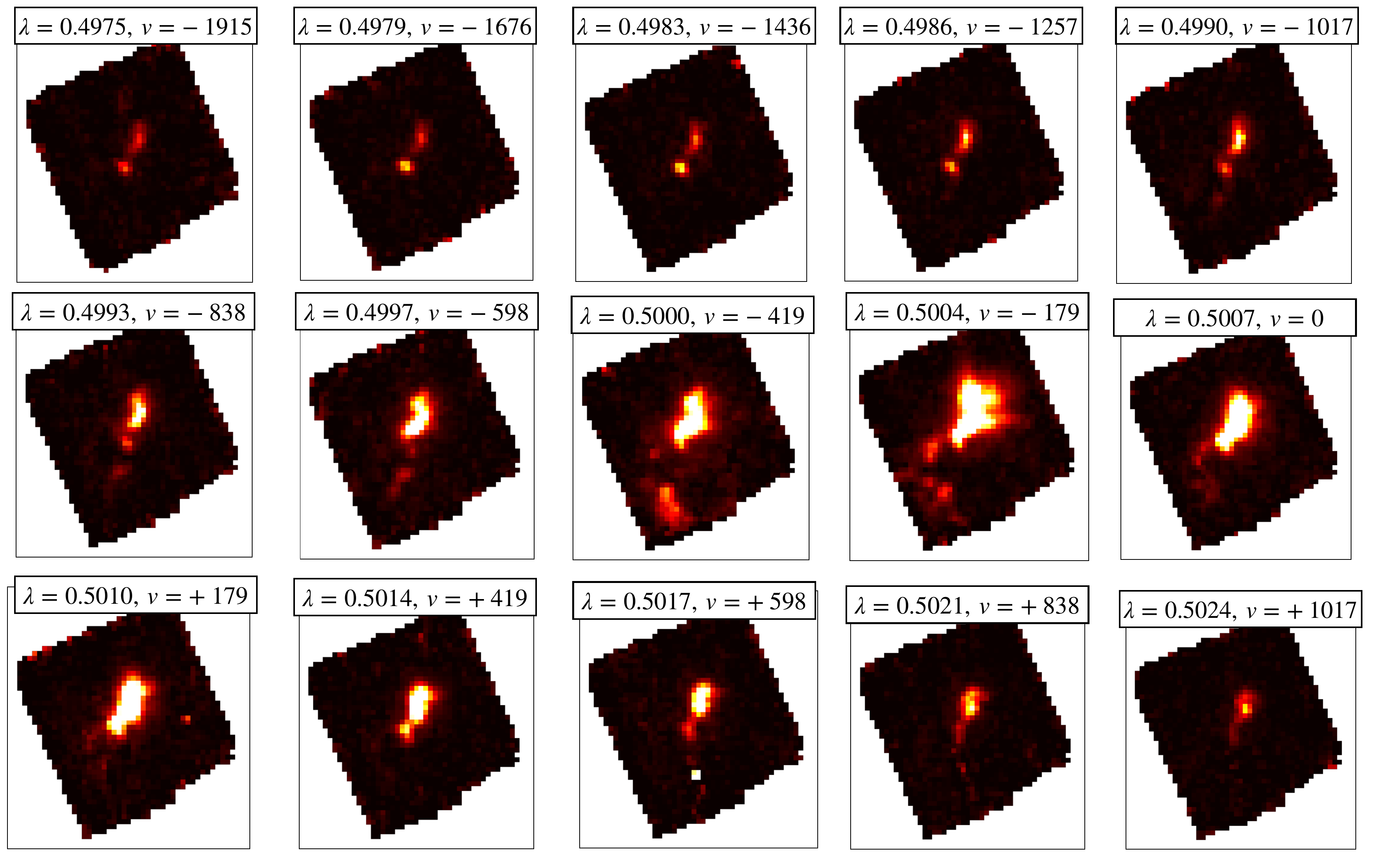}
    \caption{ Narrowband channel maps of [OIII]$\lambda$5007 emission in TNJ0205+2242. The panels show flux slices along the wavelength axis from $\lambda$ = 0.4975 $\mu$m to $\lambda$ = 0.5025 $\mu$m in $3-4$\ \AA\ intervals, siimilar to Figure \ \ref{fig:4c03_slice}. At blueshifted velocities, the line emission is confined to two locations: one compact component near the radio core and an extended emission region aligned with the northern radio lobe. As velocities approach systemic and redshifted values, these regions merge into a continuous ionized bridge extending toward the southern lobe, spanning $\sim$ 1.5–2.5$''$ (12–20 kpc). The flux increases by over an order of magnitude near systemic velocities, particularly around the denser northern lobe. The line emission remains faint toward the southern lobe indicating lower gas density and limited jet-ISM interaction.} 
    \label{fig:tnj0205_slice}
\end{figure*}

Narrowband [OIII] slices spanning $\lambda = 0.4981 - 0.5018 \ \mu m$ in Figure.~\ref{fig:tnj0205_slice} show that the line emission  is initially concentrated in  two distinct components in the blueshifted region: a compact nucleus coincident with the radio core and an extended feature aligned with the northern radio lobe. As the slices approach systemic redshift and then further redward in wavelength space, these distinct [OIII] blobs merge, forming a continuous ionized gas bridge and  extending all the way up to the location of the southern lobe spanning projected distances of $\sim 1.5-2.5''$ (12-20 kpc). The flux also increases 10 fold near the systemic redshift compared to the slice corresponding to $ v \sim -1000 \ \rm km \ s^{-1}$, particularly in regions co-spatial with the northern radio lobe where the gas density is higher. The line emission is much fainter in the southern lobe region, where the radio lobe is also located farther away from the center indicating that the gas density is possibly lower $-$ so the jet interacts with less gas and results in much less shock heated ionized gas in that region.
This filamentary structure and spatially extended emission along the jet axis suggest interaction between the radio plasma and the ambient medium.

\begin{figure*}
    \centering
    \includegraphics[width=\textwidth]{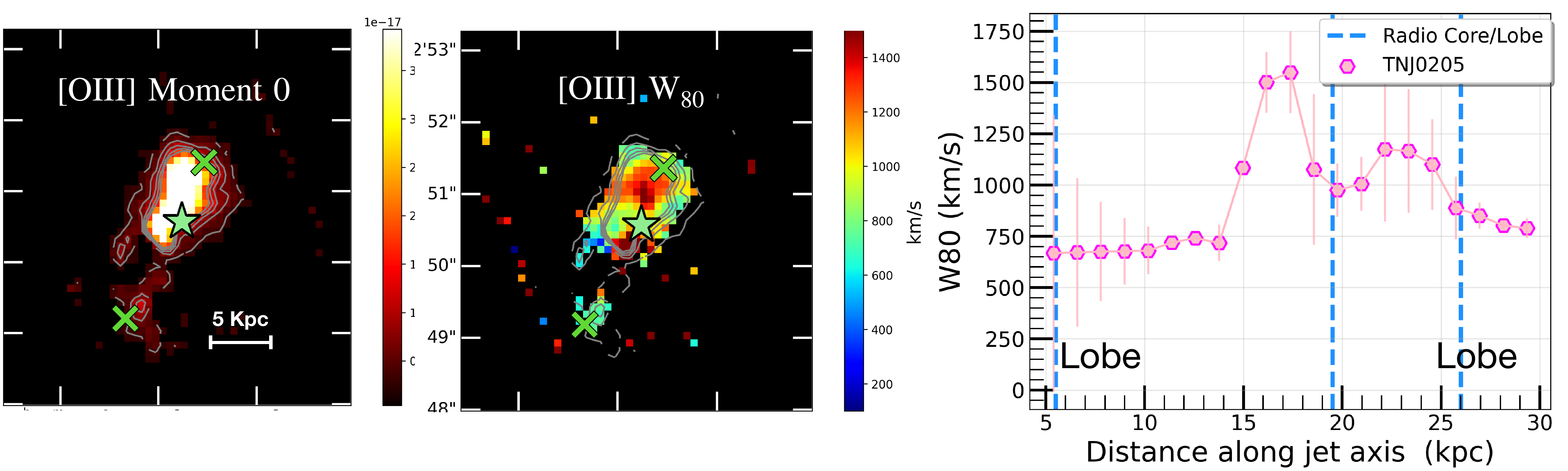}
    \vspace{-30pt}
    \caption{Left: Integrated [OIII] moment-0 (flux) map of TNJ0205 showing extended ionized gas emission aligned along the radio axis. The green star marks the radio core and green crosses denote the radio lobes. Middle: Map of non-parametric line width ($W_{80}$), which traces kinematic broadening. The line widths peak between the radio core and the northern lobe, with $W_{80}$ exceeding $\sim 1400 \ \rm km \ s^{-1}$. Right: Radial profile of $W_{80}$ along the jet axis, with zero distance defined at the southern end of the galaxy. A distinct increase in $W_{80}$ is observed around the core and the northern lobe, indicating stronger kinematic disturbance in that region, while the southern lobe shows weaker coupling to the gas. The asymmetric enhancement of $W_{80}$ between the north and southern lobe suggests directional interaction between the radio jet and the ambient ISM, due to the prevalence of dense gas. } 
    \label{fig:tnj0205_moments}
\end{figure*}

Figure~\ref{fig:tnj0205_moments} presents the spatially resolved [OIII]$\lambda5007$ moment-0 (flux) map and corresponding $\rm W_{80}$ line width map, tracing the ionized gas distribution and kinematic structure in TNJ0205+2242.
The $\rm W_{80}$ velocity width map  shows a marked enhancement in line width ($\rm W_{80} >  1500 \ km \ s^{-1}$)
 near the radio core and northern lobe. In contrast, the southern region shows significantly lower velocity dispersions and fainter emission. This asymmetry in both flux and kinematics confirms a denser or more turbulent medium near the northern lobe, resulting in a one-directional interaction between the radio jet and the ambient ISM.
 The $\rm W_{80}$ vs radial distance along the radio jet axis  shows $\rm W_{80, avg} = 700 \ km \ s^{-1} $ in the major part of the galaxy,  with the line width increasing to $\rm W_{80, avg} > 1000 \ km \ s^{-1} $ near the core and the northern lobe —consistent with jet driven shocked gas expanding large scale cocoon predominantly on  one side and entraining ionized gas causing outflows.

\begin{figure*}
    \centering
    \includegraphics[width=\textwidth]{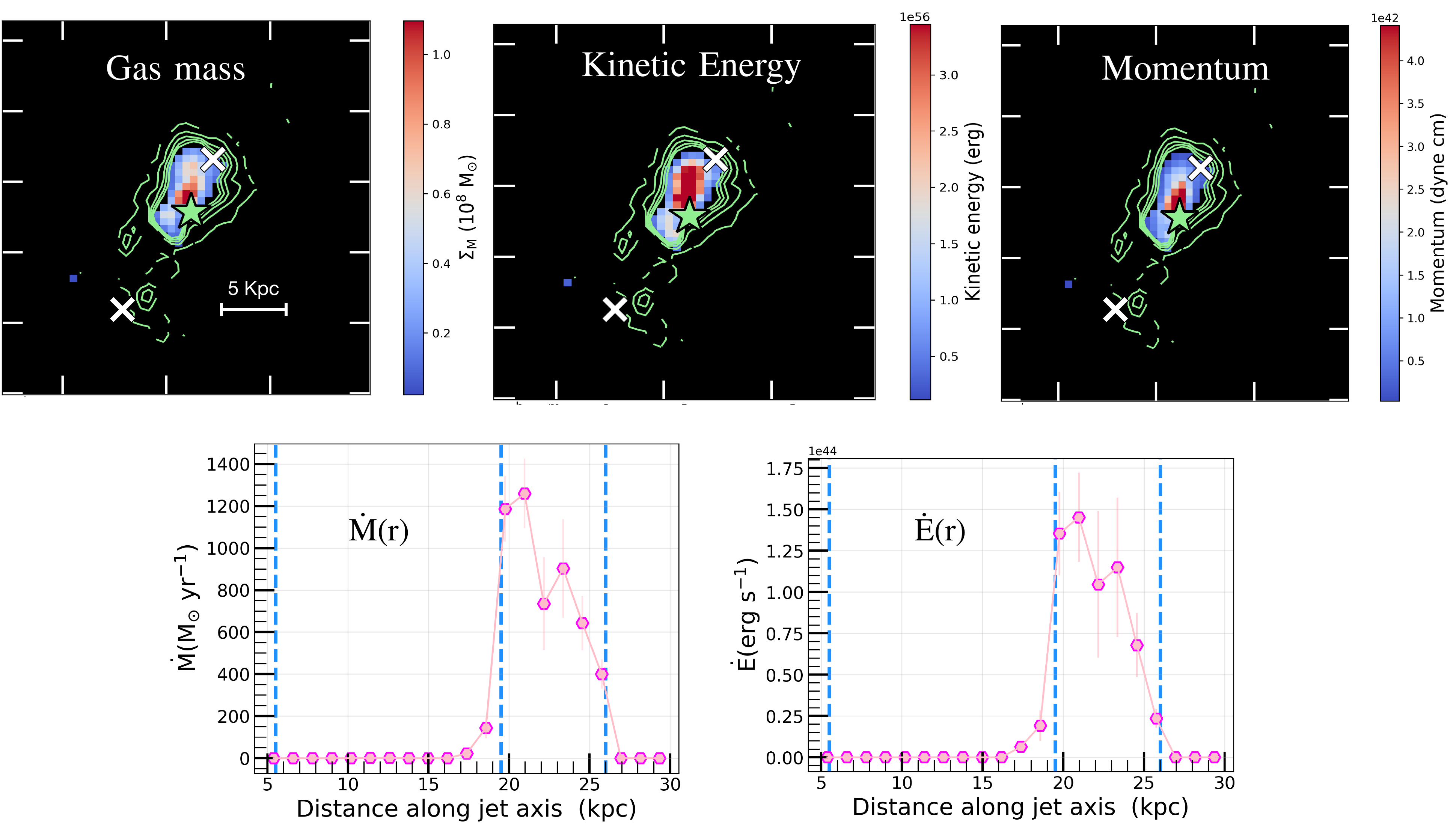}
    \caption{
    Spatially resolved maps of ionized gas mass surface density, kinetic energy and momentum in TNJ0205+2242, similar to Figure \ref{fig:tgss_energy} and Figure \ref{fig:4c03_energy}. The green star and crosses denote the positions of the radio core and the two lobes, respectively. All quantities exhibit clear enhancements along the axis connecting the radio core to the northern lobe. The kinetic power and outflow rates increase by $6-7$ times around the radio core-lobe location. All measurements are given in Table 2. 
} 
    \label{fig:tnj0205_energy}
\end{figure*}

We derive a total ionized mass outflow rate of $288 \rm \ M_{\odot} \ yr^{-1}$, with the radial profile showing the most intense outflows concentrated around the core and northern lobe where the gas density is also higher. The kinetic energy of the outflow also increases by $4x$ from $\sim 1 \times 10^{56} \ \rm erg$ south of the radio nucleus to $\sim 4\times 10^{56} \ \rm erg $ at the core and northern lobe location. The total kinetic power of the outflow is $\rm 3.5 \times 10^{43} \ erg/s$, and the total momentum flux reaches $\rm 1.1\times 10^{36} \ \rm dyne$. These values are consistent with AGN jet-driven feedback and are comparable in power to sources like TNJ1338 \citep{roy24}.

\subsection{TNJ0121+1320} \label{subsec:tnj0121}

\begin{figure*}
    \centering
    \includegraphics[width=\textwidth]{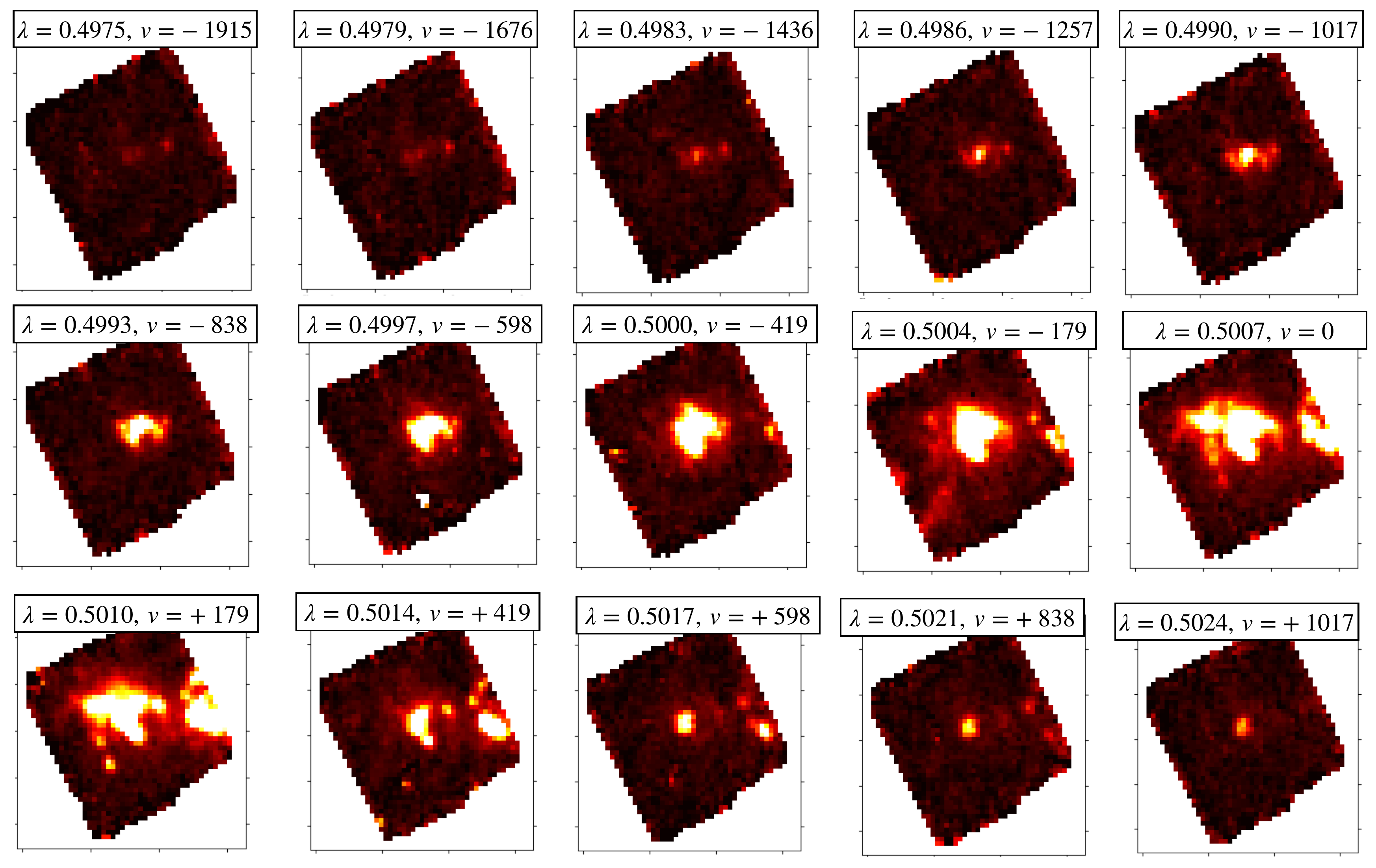}
    \caption{  Velocity-resolved [OIII]$\lambda5007$ channel maps in TNJ0121+1320 illustrate the kinematic evolution of the ionized gas across the emission line profile. In the most blueshifted slices ($\lambda < 0.5000 \ \mu$m), the emission is concentrated in two compact knots separated by $\sim 1'' (\sim 7.5 \ \rm kpc)$. These features likely trace high-velocity, blueshifted gas near the AGN . As the slices approach the systemic wavelength, [OIII] emission exhibits an extended, asymmetric  structure, bridging the two knots and revealing they are part of a coherent ionized structure coincident with the radio AGN location. When the slices go progressively redder, a third emission component $\sim 2''$ east of the radio AGN core becomes prominent at these velocities. This feature which was completely absent in the blue wings may reflect lower-velocity gas or a distinct companion system as suggested by \cite{wang25}.
 } 
    \label{fig:tnj0121_slice}
\end{figure*}

TNJ0121 exhibits a compact radio morphology without a clearly resolved radio core. The radio emission is unresolved even in the highest resolution ($\sim$ 0.3$''$) 8.4GHz VLA image and lacks any prominent extended lobe structures. However, the radio morphology shows a slight elongation along the east-west direction.  Following \cite{wang25}, we adopt the radio core position to be the peak of the western elongation, which also closely aligns with the central peak of the [OIII]$\lambda$5007 emission.

Figure.~\ref{fig:tnj0121_slice} wavelength-sliced [OIII]$\lambda$5007 channel maps of TNJ0121 reveal a distinct kinematic evolution of the ionized gas morphology across velocity space. In the bluest slices ($\lambda < 0.5000 \mu$m), the emission is dominated by two compact emission blobs located within $\sim 1 ''$ of each other, both confined to the region encompassed by the western elongation of the unresolved radio structure. These features likely trace high-velocity blueshifted gas near the AGN and its immediate radio environment. As the wavelength approaches the systemic redshift and extends into the redshifted velocity channels ($\lambda > 0.5007 \ \mu$m), these knots are bridged together by extended emission, indicating they are likely part of a common ionized structure rather than two isolated companions. Then,
a third emission component $\sim 2''$ to the east becomes  prominent. This spatially offset emission feature which was absent in the high-velocity blue wings, may correspond to lower-velocity ionized gas, local obscuration effects or a dynamically distinct structure. The possibility of a nearby companion, as suggested by \cite{wang25} cannot be ruled out. 

\begin{figure*}
    \centering
    \includegraphics[width=\textwidth]{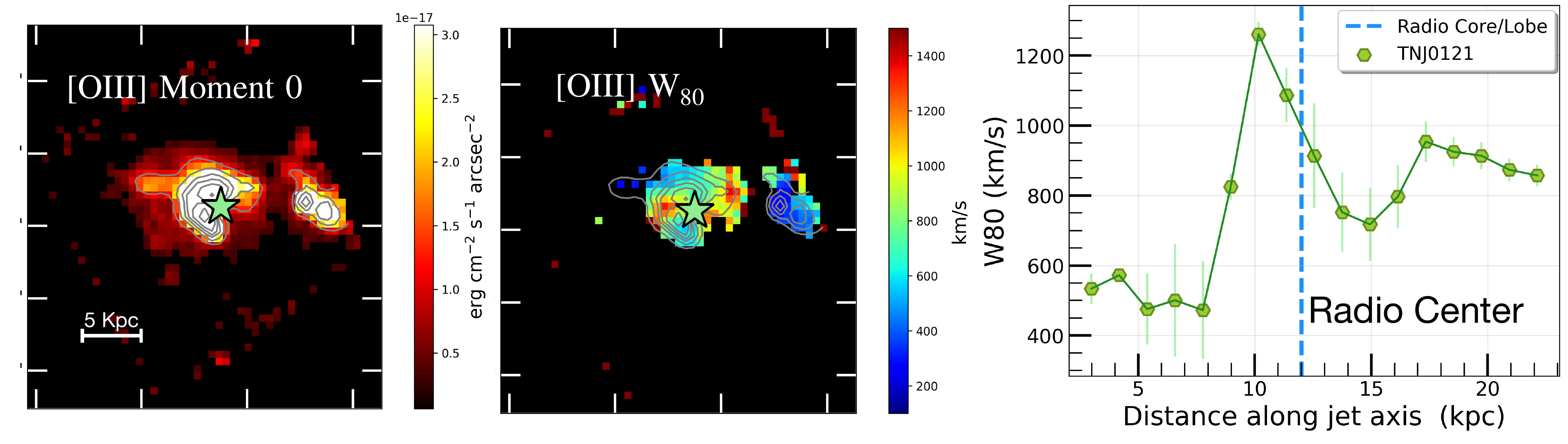}
    \caption{ Spatially resolved [OIII]$\lambda$5007 emission in TNJ0121. Left: [OIII] moment-0 map the clumpy and asymmetric ionized gas distribution, with the brightest emission concentrated near the radio core (green star) and a secondary, spatially offset clump lying $\sim 2''$ to the east. Middle: [OIII] $\rm W_{80}$ map  tracing the velocity width of the [OIII] line, which peaks in the vicinity of the radio center and shows localized broadening ($>1200 \rm \ km \ s^{-1}$)  indicative of AGN-driven kinematic disturbance. Right: Radial $\rm W_{80}$  radial profile along the axis connecting the two emission regions, where the origin is taken as the western edge of the galaxy. Similar to the 2D map, the average $\rm W_{80}$ radial profile also shows relatively modest line width ($\sim 500-800 \ \rm km \ s^{-1}$) across most regions but exhibit a spike to $>1200 \ \rm km \ s^{-1} $ near the radio core. The compact radio morphology in TNJ0121 might host a small scale jet underneath, that is pushing the gas and disrupting the kinematics.
} 
    \label{fig:tnj0121_moments}
\end{figure*}

The velocity dispersion map, traced by $\rm W_{80}$, is shown in Figure~\ref{fig:tnj0121_moments}. The $\rm W_{80}$ distribution is relatively flat across the system, with values predominantly in the range of 600–800 $\rm km \ s^{-1}$, which is significantly lower than the broader profiles observed in the more extended radio galaxies in our sample. A modest enhancement in line width ($\sim 1100$–$1200 \ \rm km \ s^{-1}$) is detected near the presumed radio core, representing a $\sim$1.5$\times$ increase relative to the surrounding gas.

The radial profile of $\rm W_{80}$, measured along the axis connecting the western and eastern [OIII] components, exhibits a similar trend. It remains nearly constant at $\sim 600 \ \rm km \ s^{-1}$ but show an enhanced  peak near the core. The morphology and kinematics are broadly consistent with an AGN-driven outflow, possibly linked to a compact radio jet. The presence of a jet cannot be confirmed in this radio galaxy from the existing radio data. Due to the unresolved radio morphology and lack of an extended jet axis, the $\rm W_{80}$ distribution in TNJ0121 provides limited directional information compared to the larger double-lobed sources, where jet-driven gas propagation is more clearly traced. Interestingly, this source show the highest mass outflow rate ($\dot{M} \sim 950 \rm \ M_{\odot} \ yr^{-1}$), signifying large scale displacement of gas without the presence of large scale radio jets.

\begin{figure*}
    \centering
    \includegraphics[width=\textwidth]{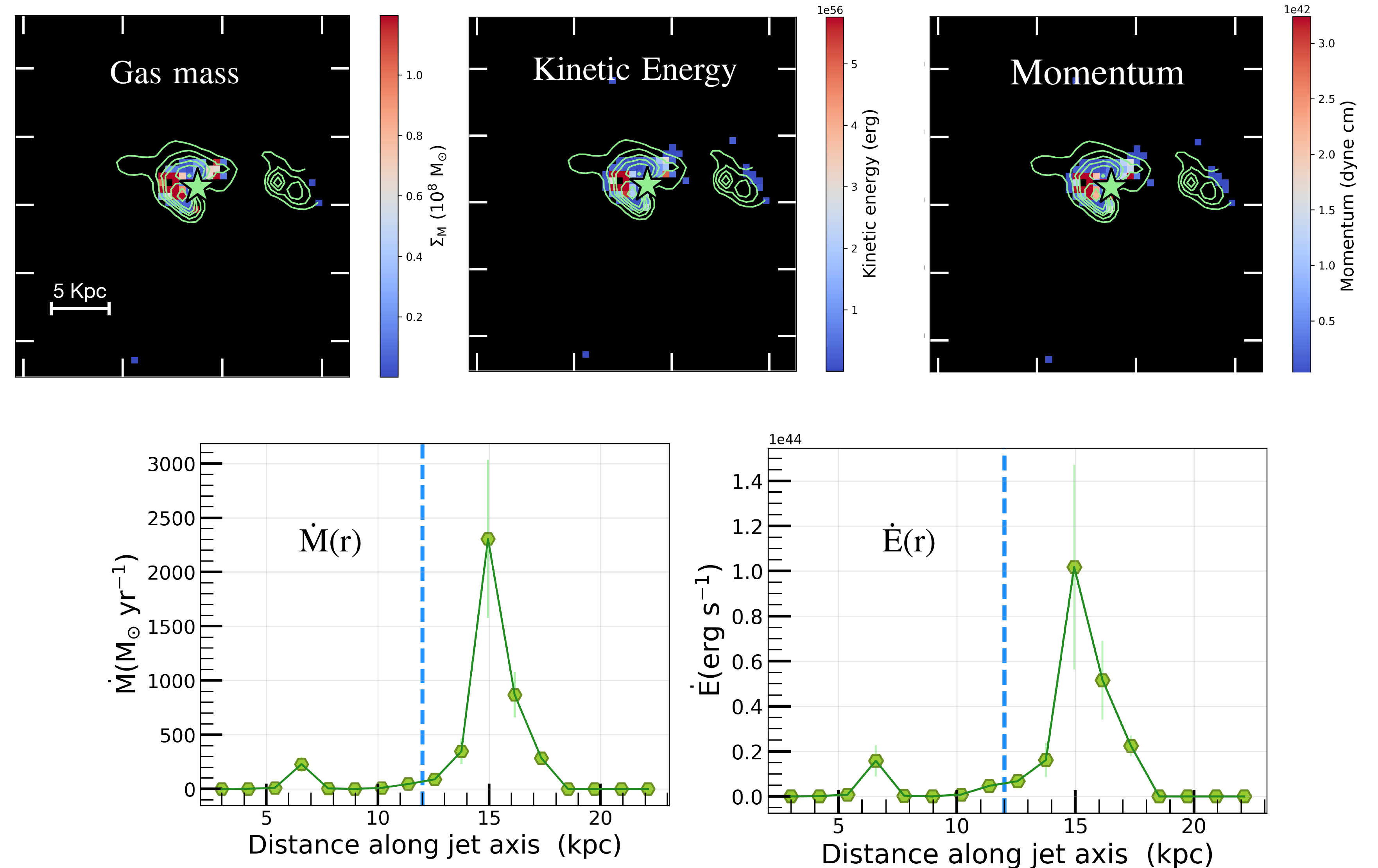}
    \caption{Same format as Figure \ref{fig:tgss_energy}, now shown for TNJ0121+1320. The source exhibits a compact radio morphology, marked by the green star. The radial profiles reveal sharp enhancements in both the mass outflow rate and kinetic power at $\sim$15 kpc, closely corresponding to the radio position, though possibly slightly offset. Due to the unresolved radio morphology and lack of an extended jet axis, the jet-gas interaction is unclear. } 
    \label{fig:tnj0121_energy}
\end{figure*}

\subsection{TNJ1338-1942} \label{subsec:tnj1338}  

\begin{figure*}
    \centering
    \includegraphics[width=\textwidth]{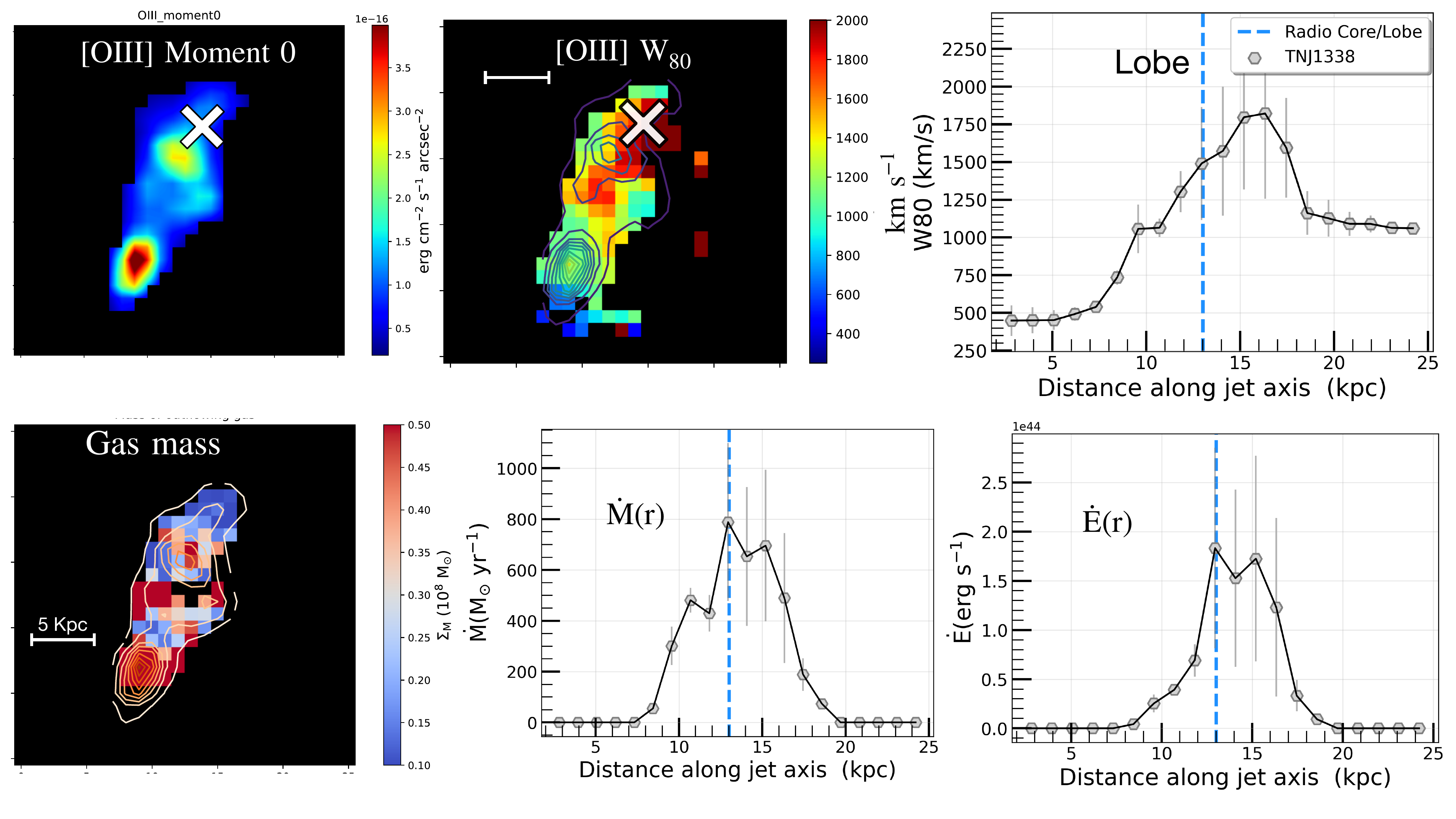}
    \caption{Spatially resolved [OIII]$\lambda5007$ emission-line maps for TNJ1338. Left: [OIII] moment-0 (integrated flux) map with black contours tracing the [OIII] flux and the white cross marking the position of the northern radio lobe. A significant concentration of ionized gas is observed offset from the center, near the lobe. Middle: [OIII] $\rm W_{80}$ map tracing the line width, revealing localized broadening exceeding $2000 \ \rm km \ s^{-1}$ in the vicinity of the northern radio lobe. As discussed in \cite{roy24}, these broad profiles indicate strong outflows and enhanced turbulence driven by jet-ambient medium interactions near the bow shock. Right: Radial profile of $\rm W_{80}$ along the jet axis, showing a $\sim 3\times$ increase in line widths near the radio lobe, consistent with large-scale, lobe-associated outflows.
} 
    \label{fig:tnj1338_moments}
\end{figure*}

We do not include a detailed discussion of TNJ1338 here, as the source has already been thoroughly analyzed in \cite{roy24}.
For TNJ1338, the radio morphology consists of a prominent northern lobe and a fainter southern lobe which is located outside the JWST/NIRSpec IFU field of view. While the central radio core is not directly detected and thus not shown, the northern lobe is fully enclosed within the IFU footprint and provides critical spatial context for interpreting the gas kinematics (Figure.~\ref{fig:tnj1338_moments}). As previously reported by \cite{roy24}, the ionized gas in TNJ1338 exhibits large-scale outflows that are preferentially aligned along the direction of the northern radio lobe, rather than being centered on the core. This trend is clearly evident in the $\rm W_{80}$ velocity dispersion map, which shows line widths exceeding 1500–2000 km/s along the northern axis where the radio lobe is located. The profile of $\rm W_{80}$ as a function of spaxel distance (Figure.~\ref{fig:tnj1338_moments}) further confirms this alignment: the line width increases gradually and monotonically toward the northern lobe, with no enhancement near the undetected central core. This steady rise in $\rm W_{80}$ with distance—peaking at values $>$2000 km/s—is a strong signature of a jet-driven outflow, with energy being deposited at large scales via the advancing radio jet front. 

\begin{table*}[ht]
\centering
\setlength{\tabcolsep}{5pt}
\renewcommand{\arraystretch}{1.2} 
\resizebox{\textwidth}{!}{
\begin{tabular}{l c c c c c c c c c}
\hline\hline
\textbf{Source} & $L_{\mathrm{[OIII]}}$ & $M_{\rm gas}$ & $\dot{M}_{\rm out}$ & $E_{\mathrm{th}}$ & $KE_{\rm outflow}$ & $\dot{KE}_{\rm outflow}$ & $\dot{p}_{\rm outflow}$ & $v_{\rm out, avg}$ & Size$_{\rm outflow}$ \\
 & [erg s$^{-1}$] & [$M_\odot$] & [$M_\odot$ yr$^{-1}$] & [erg] & [erg] & [erg s$^{-1}$] & [dyne] & [km s$^{-1}$] & [kpc] \\
\hline
TGSS       & $8.5 \times 10^{44\ \star}$ & $1.0 \times 10^9$  & 76  & $4.4 \times 10^{58}$ & $4.8 \times 10^{56}$ & $1.65 \times 10^{43}$ & $4.0 \times 10^{35}$  & 832 & 21.0 \\
TNJ0205    & $3.5 \times 10^{45}$        & $2.6 \times 10^9$  & 288 & $3.9 \times 10^{58}$ & $1.3 \times 10^{57}$ & $3.49 \times 10^{43}$ & $1.12 \times 10^{36}$ & 474 & 14.9 \\
4C03       & $1.5 \times 10^{46}$        & $8.4 \times 10^9$  & 654 & $3.1 \times 10^{59}$ & $1.0 \times 10^{58}$ & $9.1 \times 10^{44}$  & $8.8 \times 10^{36}$  & 687 & 21.5 \\
4C19       & $3.4 \times 10^{45}$        & $1.6 \times 10^9$  & 314 & $7.25 \times 10^{58}$& $1.5 \times 10^{57}$ & $4.0 \times 10^{43}$  & $1.2 \times 10^{36}$  & 696 & 14.9 \\
TNJ0121    & $1.5 \times 10^{46}$        & $8.0 \times 10^9$  & 951 & $1.2 \times 10^{59}$ & $2.5 \times 10^{57}$ & $4.1 \times 10^{43}$  & $2.0 \times 10^{36}$  & 442 & 7.5 \\
TNJ1338    & $6.0 \times 10^{45}$        & $5.0 \times 10^9$  & 503 & $8.0 \times 10^{58}$ & $2.7 \times 10^{58}$ & $1.0 \times 10^{44}$  & $2.2 \times 10^{36}$  & 527 & 21.0 \\
\hline\hline
\end{tabular}
}
\caption{Measured derived properties of the six HzRGs analyzed in this work. (1) Source name, (2) dust-corrected [OIII]$\lambda$5007 luminosity. For TGSS, where [OIII] is not covered, we use H$\alpha$ luminosity. The subsequent columns report (3) ionized gas mass ($M_{\rm gas}$), (4) mass outflow rate ($\dot{M}_{\rm out}$), (5) total thermal energy ($E_{\mathrm{th}}$) in the shock-heated hot ($\sim 10^7 \rm \ K$) gas -- $E_{\rm th} = 3/2 P \Delta V$, (6) total kinetic energy of the ionized outflow ($KE_{\rm outflow}$), (7) outflow kinetic power ($\dot{KE}_{\rm outflow}$), (8) momentum rate ($\dot{p}_{\rm outflow}$), (9) average outflow velocity ($v_{\rm out, avg} = \sqrt{W_{50}^2 + \Delta v^2}$), and (10) spatial size of the ionized outflow. See main text for further details.}
\label{tab:2}
\end{table*}

\section{Discussion} \label{sec:discussion} 

\subsection{Enhanced ionized gas kinematics in HzRGs} \label{subsec:w80}

\begin{figure}
    \centering
    \includegraphics[width=0.48\textwidth]{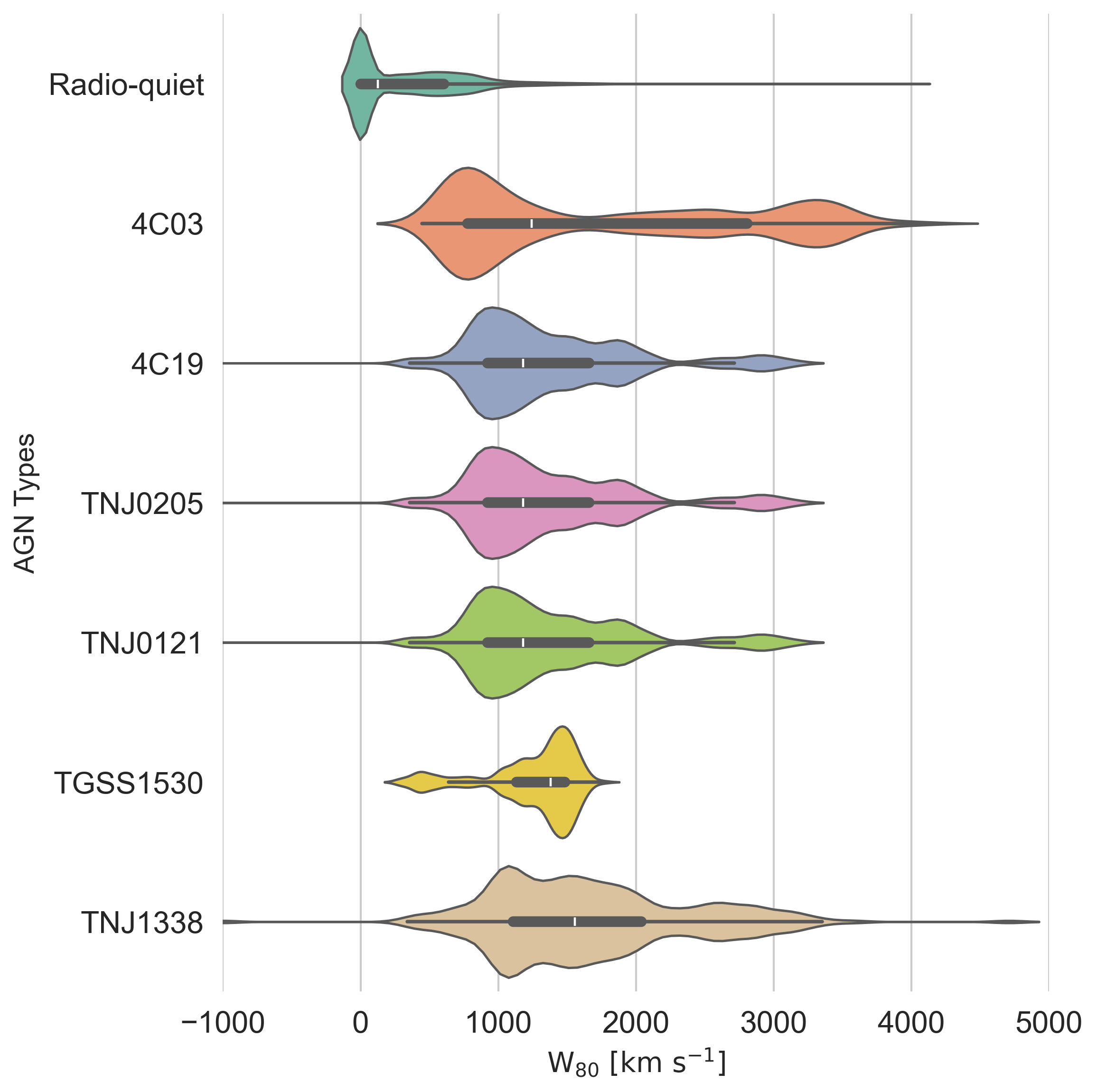}
    \caption{ Violin plots showing the distribution of $\rm W_{80}$ line widths across all spaxels in the combined sample of radio-quiet quasars observed with JWST/NIRSpec IFU \citep[][: top row]{veilleux23, vayner23, marshall23, cresci23}. This is compared to the spaxel-level $\rm W_{80}$ distributions in each of the six high-redshift radio galaxies presented in this work ($\rm 2^{nd} - 7^{th}$ row). $\rm W_{80}$ is a non-parametric measure of the line width and serves as a proxy for the strength of AGN-driven outflows and the degree of kinematic disturbance. The radio-quiet quasars exhibit typical $\rm W_{80}$ values of $\sim 600 \ \rm km \ s^{-1}$, significantly lower than the radio galaxy sample, where median $\rm W_{80}$ values range from $\sim 950$ to $2100 \ \rm km \ s^{-1}$, with extended high-velocity tails exceeding $2500 \ \rm km \ s^{-1}$. This highlights the stronger gas turbulence and outflow activity associated with radio jets. } 
    \label{fig:w80_violinplot}
\end{figure}

We investigate the spatial distribution of ionized gas turbulence across our sample of six high-redshift radio galaxies by analyzing the line width parameter $\rm W_{80}$, defined as the width encompassing 80\% of the line flux. $\rm W_{80}$ is a robust, non-parametric tracer of gas kinematics, which is sensitive to broad wings and multiple components. It is  commonly used to identify AGN-driven outflows and turbulence. 
We find that all high-redshift radio galaxies in our JWST/NIRSpec sample (4C03, 4C19, TGSS1530, TNJ0205, TNJ0121, and TNJ1338) exhibit broad [OIII]$\lambda$5007 emission line profiles across their spatial extents. The velocity widths ($\rm W_{80}$) we measure range from $\sim$700 to $>$3000 km/s which is well above what is typically observed in radio-quiet AGNs. The distributions of $\rm W_{80}$ across spaxels in each galaxy are shown in Figure ~\ref{fig:w80_violinplot}. The median $\rm W_{80}$ values span $\sim 950-2100 \ \rm km \ s^{-1}$, with the extended tails of the distributions reaching $>2500 \ \rm km \ s^{-1}$ in almost all sources. By contrast, the comparison sample of radio-quiet Type-I quasars observed with JWST/NIRSpec IFU \citep{marshall23, veilleux23, cresci23, perna23} (first row in Figure ~\ref{fig:w80_violinplot}) shows a sharply peaked distribution of $\rm W_{80}$ values centered at $\rm  \sim 600 \ km \ s^{-1} $, and the individual spaxel values rarely exceeds 1000 km/s. This stark contrast highlights that radio AGNs host enhanced gas turbulence compared to radio quiet systems, possibly due to widespread mechanical energy deposition by radio jets.
These differences are consistent with prior results from JWST/NIRSpec studies of high-z radio-quiet AGNs  \citep[e.g., ][]{perna23, marshall23, ubler23} and with stacked IFU analyses of z$\sim$2 quasars \citep{schreiber19}, reinforcing the idea that radio galaxies at high redshift exhibit more extreme gas turbulence than their radio-quiet counterparts.

This enhancement in line widths is also more pronounced than that seen in most low-redshift radio AGNs. For example, the MURALES survey \citep{balmaverde19} and MAGNUM project \citep{venturi21} show that while low-z radio galaxies also host multiphase outflows, their  $\rm W_{80}$ rarely exceed $\rm 1000 \ km \ s^{-1}$, and are typically concentrated within the central few kiloparsecs. Our high-z radio sources, on the other hand, exhibit disturbed, large scale outflowing kinematics over extended regions (often $> 10 \ \rm kpc$) with $\rm W_{80}$ exceeding $\rm 1500 \ km \ s^{-1}$ particularly near the lobes. This suggests that the presence of denser, gas-rich environments and more powerful jets at earlier epochs lead to feedback processes that are more spatially distributed and energetic.
This interpretation is supported by simulations of AGN jets in clumpy, high-z ISM environments \citep[e.g., ][]{mukherjee16, mukherjee18, bicknell18}, where expanding cocoon structures can propagate turbulence and entrain gas over tens of kiloparsecs.

A key insight from our analysis is the directionality of the line broadening. 
We observe a pronounced spatial alignment between high-$\rm W_{80}$ regions and the radio jet axes in nearly all sources. As discussed in previous sections, every galaxy in our sample shows either a slow-monotonic or rapid-asymmetric increase in $\rm W_{80}$ towards the direction of the radio lobes. In 4C03, $\rm W_{80}$ rises from $\sim$800 km/s near one end of the galaxy to $>$3000 km/s near one of the lobes. Similar trends are observed in TGSS1530, 4C19, and TNJ1338. TNJ0205 also shows a sharp rise in $\rm W_{80}$ toward one radio lobe, although the radio morphology is more complex. These enhancements strongly support a scenario in which the advancing radio jets and lobes drive expanding cocoons that shock the surrounding ISM, consistent with earlier interpretations \citep{nesvadba08, mukherjee18, bicknell18, collet16}.

Even in TNJ0121 where the radio morphology is compact, a mild $\rm W_{80}$ enhancement ($\sim$950 km/s) is seen toward the western elongation where the peak of radio and [OIII] emission overlap. However, due to the unresolved radio structure, the alignment with the jet axis is ambiguous and the $\rm W_{80}$ vs. distance profile remains relatively flat. In contrast, TNJ1338, whose large-scale outflow structure has already been characterized in \cite{roy24}, shows a 
steady increase of $\rm W_{80}$ along the northern radio lobe, which is the only lobe within the IFU footprint. 
These spatially resolved line width patterns provide strong evidence that the ionized gas turbulence is not isotropic, but is instead directionally aligned with the radio jet propagation. This argues that radio-mode AGN feedback, via jet–ISM interactions, plays a dominant role in disturbing and expelling gas from galaxies at high redshift.

These spatially extended, jet-aligned high-velocity ionized outflows have significant implications for the regulation of star formation and the fate of the circumgalactic medium (CGM) in massive high-redshift galaxies. If even a modest fraction of the gas with  $\rm W_{80} > 1500 - 2000 \ km \ s^{-1}$ is unbound, this implies mass-outflow rates on the order of hundreds of solar masses per year, and mass loading factors of order unity. Recent JWST studies of luminous AGNs \citep[e.g., ][]{marshall23, perna23, ubler23}
 have begun to reveal similar  outflow signatures, but the presence of highly collimated, directionally coherent $\rm W_{80}$ structures in our radio galaxy sample distinguishes the jet-driven mode of feedback from radiation-driven winds. Moreover, our findings are consistent with predictions from recent high-resolution 3D hydrodynamical simulations of radio jet–ISM interactions in clumpy, gas-rich galaxies \citep{nesvadba08, mukherjee18, bicknell18, meenakshi22}. These simulations show that relativistic jets drill through an inhomogenous ISM and inflate expanding cocoons. They generate strong bow shocks, and entrain dense clouds along the jet axis, producing asymmetric, high-velocity outflows with spatial structures strikingly similar to those observed in our sample. In particular, the simulations suggest that gas removal is most efficient when jets couple to multi-phase, high-density conditions, which is likely very common in $z>2$. Our results thus provide observational support for a mode of mechanical AGN feedback capable of expelling large gas reservoirs, and  indicate that radio galaxies at $z>3$ are significantly more kinematically disturbed than their radio-quiet or low-redshift analogs.


\subsection{Kinetic Energy and momentum deposition by ionized outflows in HzRGs} \label{subsec:energy}

\begin{figure*}
    \centering
    \includegraphics[width=\textwidth]{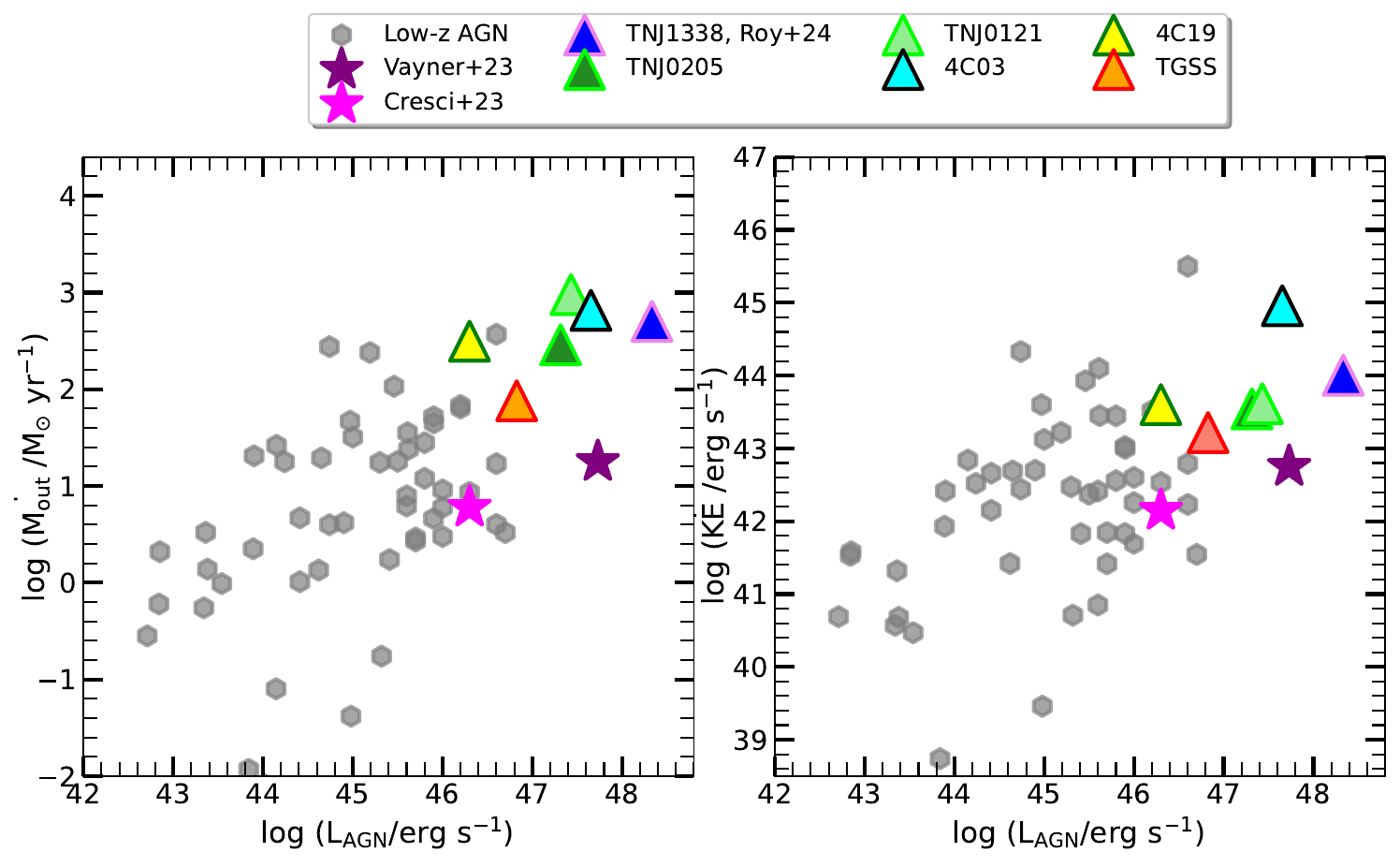}
    \caption{ Left panel: globally integrated mass outflow rate vs. AGN bolometric luminosity reported in the literature using a variety of IFU/long-slit spectroscopy.
Gray symbols indicate low-redshift AGNs from the combined literature \citep{nesvadba06,  brusa16, revalski18, oliveira21, revalski21, kakkad22, ruschel-dutra21, ulivi24, greene12}. The colored stars indicate JWST/IFU observations of radio-quiet quasars at high redshifts which have resolved outflow measurements and used electron density measurements from [SII]6717, 31 lines to estimate mass outflow rates \citep{cresci23, vayner23}. While a few additional high-redshift Type I quasars have been reported to host extended ionized gas outflows based on NIRSpec IFU observations \citep{marshall23, perna23}, these studies rely on assumed electron densities rather than direct measurements. As a result, we exclude those sources from our comparative analysis. The high-z radio galaxies analyzed in this work are shown with colored triangles. The HzRGs show the highest mass outflow rates but span a large range in 
the AGN bolometric luminosity observed. 
Right panel: kinetic outflow
power vs. AGN bolometric luminosity for literature compiled objects (gray symbols and colored stars) compared to HzRGs (triangle). The kinetic outflow power of the HzRGs are $\sim 0.5-2$\% of 
 the bolometric AGN luminosity and are some of the  strongest values ever measured at similar redshifts in AGN host galaxies. } 
    \label{fig:energy}
\end{figure*}

We determine the effectiveness of radio jets in redistributing gas mass, energy and momentum within and beyond the host galaxy along the radio jet axis by calculating spatially resolved mass outflow rate, kinetic power and momentum flux from the outflows in the ionized gas phase. As presented in previous sections, the spatial maps of these quantities highlight localized regions of enhanced outflow activity, particularly near radio cores and along jet axes.  In Figure.~\ref{fig:energy} (left panel), we show the  total mass outflow rate ($\dot{M}_{\rm out}$) integrated over all spaxels as a function of AGN bolometric luminosity ($L_{\rm bol}$) for our sample of six $z > 3.5$ radio galaxies  (triangle). These are compared with high redshift Type I quasars observed with  JWST/NIRSpec IFU  \citep[stars:][]{ vayner23, veilleux23, cresci23}, 
and a compilation of low-redshift ($z<0.1$) AGNs with similar spatially resolved outflow measurements from the literature  \citep[gray symbols: ][]{nesvadba06, brusa16, revalski18, shimizu19, oliveira21, revalski21, kakkad22, ruschel-dutra21, ulivi24, greene12}. Note that for the high-redshift Type I quasars, we include only those sources that exhibit extended outflow signatures in existing IFU spectra and have directly measured electron densities,  in order to be consistent with our own analyses of HzRGs. Therefore, we exclude the JWST/IFU observed quasar outflow results from \citet{marshall23, perna23} in Figure \ref{fig:energy}, as these studies lack direct electron density measurements necessary for robust outflow rate estimates.
   Our comparison sample of low-z AGNs is selected using the same criteria. 
 We intentionally exclude sources where outflows are detected via absorption features (e.g., BAL quasars), or traced in other gas phases (e.g., molecular or neutral gas), or inferred from globally integrated spectra rather than spatially resolved IFU observations \citep[e.g.,][]{baron19, xu20}, to ensure an apples-to-apples comparison of spatially resolved ionized outflow properties as discussed above.


The high redshift radio galaxies in our sample show significantly elevated outflow energetics compared to their radio-quiet counterparts. The mass outflow rates span a wide range from $\sim 40$ to $350 \ \rm M_\odot \ \rm yr^{-1}$, with the most extreme values observed in TNJ1338 and 4C19. Correspond kinetic powers range between $\dot{KE}_{\rm outflow} \sim 10^{42.5} - 10^{44} \ \rm erg \ s^{-1}$. This range corresponds to kinetic efficiencies of $\epsilon_{\rm kin} = \dot{KE}_{\rm outflow} / L_{\rm bol} \sim 0.15\% - 8\%$. These values fall within or exceed the theoretically predicted thresholds for efficient AGN feedback. 
Various feedback models and simulations predict that kinetic powers $\gtrsim 5-7\%$ of $L_{\rm bol}$ are needed for feedback to drive galaxy-scale blastwaves sufficiently powerful to entrain gas and to inject adequate energy into the ISM \citep[e.g.,][]{dimatteo05, hopkins06}. On the other hand, a `two-stage' model was proposed by \cite{hopkins10} that stated that initial feedback from the central quasar needs to only initiate a moderate wind in the low-density hot gas, which reduces the required energy budget for feedback by an order of magnitude. In this model, a $\sim 0.5\%$ coupling efficiency is sufficient. Our observations reveal that multiple radio galaxies in the sample meet or exceed these thresholds. These results strengthen the case that radio jets in massive, gas-rich galaxies at $z > 3.5$ are capable of launching galaxy-scale, high-efficiency outflows, which are far more powerful than typical radiatively-driven outflows in quasars. Crucially, the measured outflows are not only high in velocity but are spatially extended, morphologically coherent, and preferentially aligned along the radio jet axis. This spatial coupling implies a direct causal connection between jet-driven energy injection and the disturbed ionized gas kinematics observed across tens of kiloparsecs.



\subsection{A new $ L_{jet}- L_{radio}$ scaling relation: 
Estimating Jet power from radio power}  \label{subsec:ljet}

Studies of nearby radio galaxies in clusters for decades have demonstrated that jets can shock  heat the surrounding gas and create cavities in the hot X-ray gas \citep{mcnamara07, mcnamara12, fabian12}. This direct signature of AGN feedback has revealed that the energy from the jets 
 can supply enough energy to regulate star formation and suppress cooling of the hot halos of galaxies and clusters \citep{birzan04, dunn06, rafferty06, cavagnolo10, heckman14}. 
 To assess how AGN feedback couples to the thermodynamic state of the host environment, empirical scaling relations have emerged between radio luminosity ($L_{\rm 1.4GHz}$) and mechanical jet power ($L_{\rm jet}$), where the latter is inferred from the enthalpy of X-ray cavities inflated by radio jets \citep{birzan04, cavagnolo10, heckman14}. However, each of these studies have restricted to a sample spanning a narrow radio luminosity range while calibrating this relationship. 

\begin{figure}
    \centering
    \includegraphics[width=0.49\textwidth]{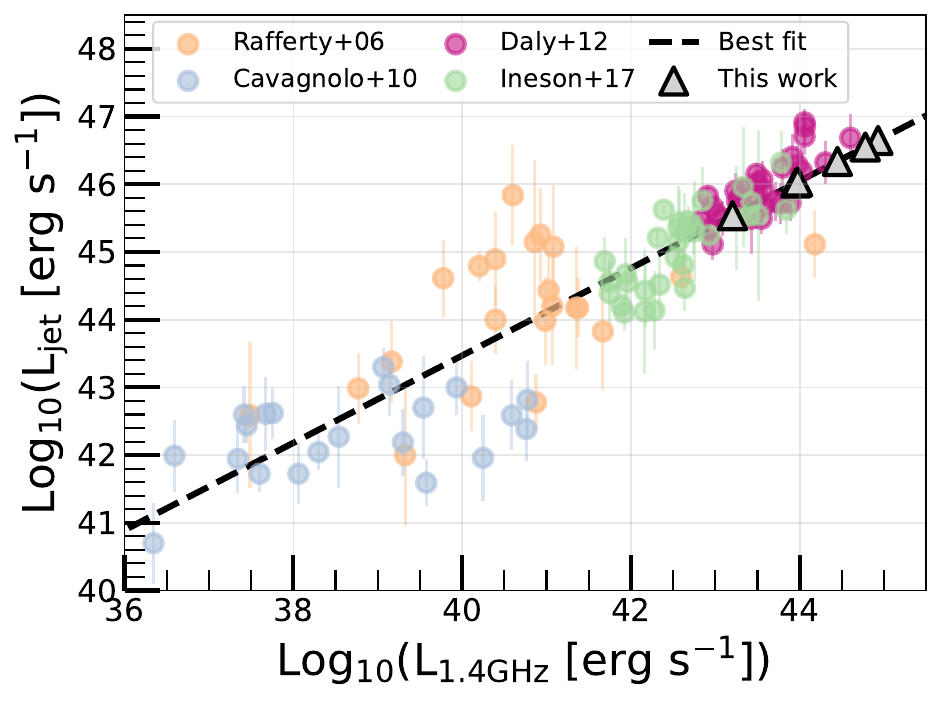}
    \caption{ Jet power vs. radio power calibration in radio galaxies. Blue, yellow, green and red circles show the observed measurements presented in \cite{rafferty06, cavagnolo10, daly12, ineson17}. Their combined sample spans $\sim 9 $ orders of magnitude in radio power. The dashed line represents our best-fit linear relation derived from a robust least-squares regression (see Equation~\ref{eq:ljet}). The estimated jet powers for the high-redshift radio galaxies (HzRGs) analyzed in this work and computed using our calibration, are shown as triangles.  } 
    \label{fig:ljet}
\end{figure}

Here, we present a new best fit $\rm L_{jet}$ $vs.$  $\rm L_{1.4GHz}$ relationship, where the radio luminosity of the radio galaxies: $\rm L_{1.4GHz}$, measured at 1.4 GHz frequency, span $\sim 10^{36} - 10^{45} \ \rm erg \ s^{-1}$. We use the combined sample of radio galaxies presented in \cite{rafferty06, birzan04, cavagnolo10, daly12, ineson17}. We take into account the reported $\rm L_{jet}$ errors and employ a linear least square fitting algorithm. The resultant best-fit relation, as shown by the dashed line in Figure \ref{fig:ljet}, is:

\begin{equation} \label{eq:ljet}
    \rm log \  L_{jet} = 0.64(\pm0.03)\ log \ L_{1.4GHz} + 17.68(\pm1.23)
\end{equation}

where $\rm L_{jet}$ and $\rm L_{1.4GHz}$ are in units of $\rm erg \ s^{-1}$. We utilize this relation to estimate the jet power for the high redshift radio galaxies in our sample (triangle symbols in Figure \ref{fig:ljet}).  Our HzRGs have $\rm L_{1.4GHz} \sim 10^{43.2} - 10^{44.9} \ erg \ s^{-1}$, which indicate corresponding $\rm L_{jet}$ to be $\rm \sim 10^{45.5} - 10^{46.7} \ erg \ s^{-1}$.

\subsection{Jet-ISM interaction} \label{subsec:jetism} 

\begin{figure*}
    \centering
    \includegraphics[width=\textwidth]{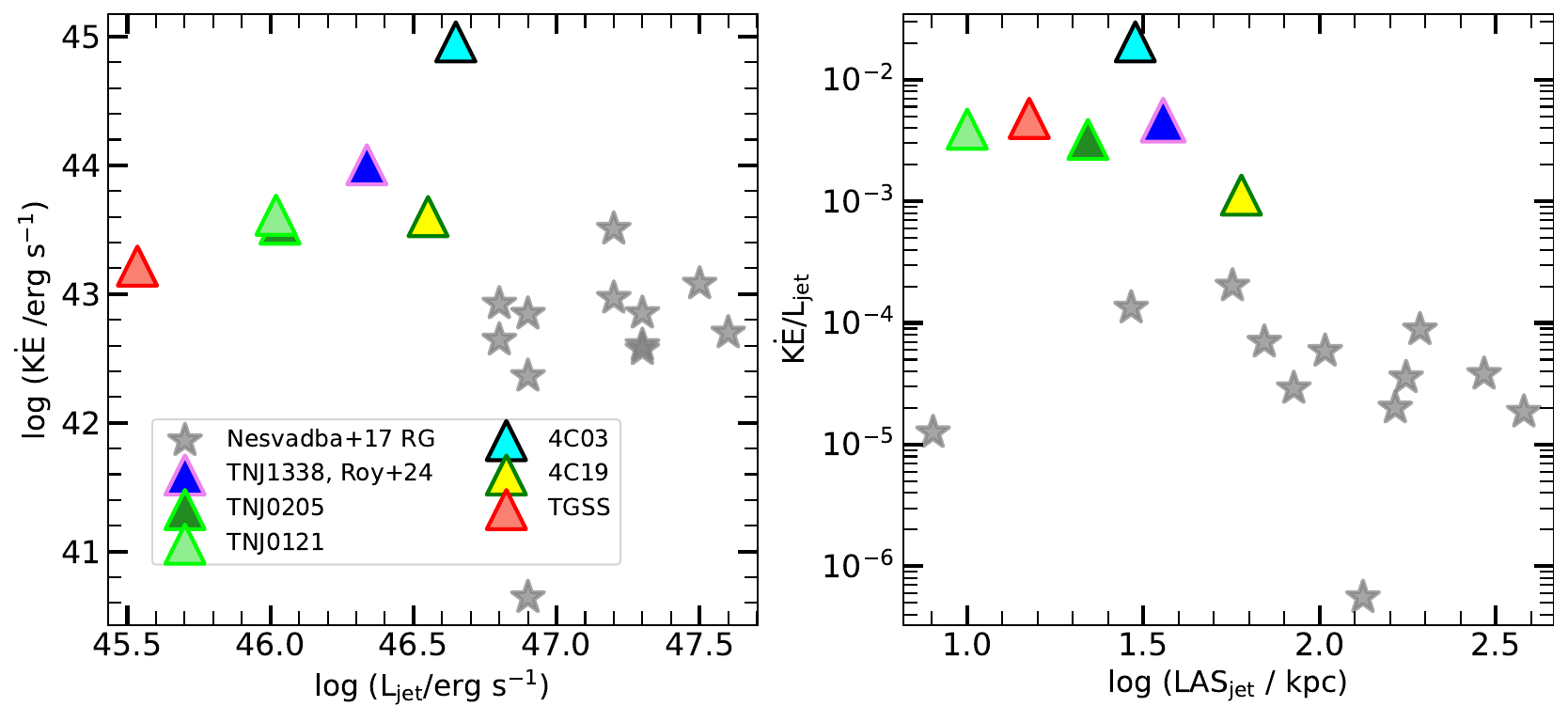}
    \caption{Left panel: Kinetic power of the ionized outflows versus radio jet mechanical power for the $z \sim 2.5$ radio galaxy sample from \cite{nesvadba17} (gray stars) and the $z > 3.5$ radio galaxies presented in this work (triangles) We have computed the jet and outflow powers in the the Nesvadba sample using methodologies consistent with ours. Right panel: Ratio of outflow kinetic power to jet mechanical power as a function of radio source size. The high-redshift radio galaxies in this work exhibit the most energetic outflows, with $\dot{E}_{\rm outflow}$ exceeding the median of the \cite{nesvadba17} sample by nearly an order of magnitude.  A clear negative trend emerges in the right panel, where smaller radio sources show higher coupling efficiency between the jet and the ionized gas. This suggest that the
transfer of kinetic energy from the jet to the gas is facilitated by the higher gas densities expected at smaller radii. } 
    \label{fig:jet}
\end{figure*}

In this section, we will quantify the nature of the jet/ISM interaction in our sources. 
We compare the energetics of the outflow in our six HzRGs to those of the other high-z radio galaxies investigated by \cite{nesvadba17}[N17] using ground-based near-infrared IFU data. To do so, we have computed the jet and outflow powers in the the Nesvadba sample using methodologies consistent with ours.
The left panel of Figure.~\ref{fig:jet} shows the relation between the kinetic power of the outflow and the jet mechanical power. It is clear that there is no strong correlation between the two quantities.  The kinetic energy injection rates for our six sources span a wide range of $\rm log \ \dot{KE}_{outflow}  \sim 42.5 - 44.2  \ erg \ s^{-1}$ , while the jet power ranges between $\rm log \ L_{jet}  \sim 45.5 - 46.7 \ erg \ s^{-1}$. Note that, we recalculated the jet powers of both our sample and the N17 sample following Eq \ref{eq:ljet}, and we do not use the reported jet powers published in N17 to be consistent across the sample. Except TNJ0121, all the remaining five HzRGs studied here show elevated $\rm \dot{KE}_{outflow}$ values compared to the median of N17 sample.

The right panel of Figure.~\ref{fig:jet} plots the $\rm \dot{KE}_{outflow}/L_{jet} $ 
compared with the projected largest angular size (LAS) of the radio source. A clear anti-correlation is observed: compact sources exhibit higher $\rm \dot{KE}_{outflow}/L_{jet} $  ratios compared to more extended sources in the N17 sample. Although this trend is affected by small number statistics, the emerging trend, if holds true, supports the interpretation that compact radio sources may be more efficient at transferring energy into the ISM, likely due to interactions occurring in denser environments closer to the galaxy nucleus. As discussed in \cite{roy24}, this enhanced efficiency could be due to stronger jet-ISM coupling during the early stages of radio source evolution, before jets break out of the dense interstellar medium into the more tenuous circumgalactic medium.

Quantitatively, most of our sources have $\rm \dot{KE}_{outflow}/L_{jet} \sim 10^{-3} - 10^{-2}$, i.e. $0.1-1\%$. This is consistent with expectations from high-resolution hydrodynamic simulations of radio jets interacting with multi-phase media \citep{mukherjee18, meenakshi22}. These studies found that while jets can drive shocks and turbulence, the efficiency of transferring jet power into kinetic energy of the large-scale, coherent outflows is typically less than 1\%. Most of the jet energy is dissipated as heat or contributes to turbulent motions within the ISM. This suggest that AGN jet feedback, while energetically significant, couples inefficiently to the bulk motion of the ISM, especially in the early stages of jet-ISM interaction. 
These simulations also show that radio jet cocoons can accelerate and displace large amounts of gas, particularly in gas-rich systems, which is consistent with our previous results.

\subsection{Where Does the Jet Energy Go?}

As discussed above (and see Figure 23), only a small fraction of the estimated jet kinetic energy ($\sim 0.3\%$) is being carried by the emission-line gas in the observed outflows. The obvious question then is where does the bulk of the jet energy go. Some will reside in the internal energy of the relativistic/magnetized jet-fluid that inflates the cocoon (e.g. Begelman \& Cioffi 1989). However, for feedback to occur some of the jet energy must be transferred to the ISM/CGM. In this section we consider two possibilities: that it resides in a hot gas phase created by the jet, or that is it is used to shock-ionize the emission-line gas.

\subsubsection{Thermal Energy in Hot Gas}

The calculations of kinetic power and mass outflow rates derived from JWST/NIRSpec IFU spectra offer crucial insights into the energetics of the warm ionized phase ($T \sim 10^4 \ \rm K$) of the outflows. But they represent only a subset of the total energy budget imparted by the AGN jets. Radio jets are known to interact with multiple phases of gas in the surrounding ISM and CGM, and deposit significant energy in other forms or phases not probed by our observations, including turbulent- and shock-heated hot gas. 

To address this broader energy budget, we estimate the total thermal energy content ($E_{\rm th}$) in the hot gas surrounding the radio jets. Taking the energy density in the gas $U_{th}$ to be 3/2 $P$, this is calculated as:

\begin{equation}
    E_{\rm th} = 3/2 P \, \Delta V
\end{equation}

where $P$ is the thermal pressure of the hot gas and $\Delta V$ is the volume it occupies.

To calculate the energy, we approximate the volume using a rectangular box geometry, where the length is set by the end-to-end size of the nebular emission region (``a'': listed in Table 2), and the breadth and width are assumed to match the thickness of the [OIII] or H$\alpha$ emitting region (``b''). The computed volume is thus calculated to be $\Delta V = \rm a \times b^2$.  The thermal pressure $P$ is estimated using electron densities derived from the [SII] $\lambda\lambda6716,6731$ doublet, assuming a photoionized gas temperature of $T_e \sim 10^4\,{\rm K}$: $P = 2 n_e k T$ and $U_{th} = 3 n_e kT$. This corresponds to clouds confined by (in pressure balance with) the hot gas. Following arguments used to evaluate the energy content of cavities excavated by jets in X-ray gas (e.g. McNamara \& Nulssen 2007), the total energy delivered by the jet is the sum of this thermal energy plus the work done to inflate the cavity ($P\Delta V$).

Across the six $z > 3.5$ radio galaxies in our sample, we find total estimates of this energy to vary almost an order of magnitude, between 
$ \sim 6.5 \times 10^{58} - 5.2 \times 10^{59} \,{\rm erg}$. Thus, $E_{th}$ is typically about 2 orders of magnitude larger than the corresponding kinetic energies of the warm ionized gas phase in the outflowing gas ($\rm KE_{outflow} \sim 4.8 \times 10^{56} - 2.7 \times 10^{58}\,{\rm erg}$; see \S\ref{subsec:energy} and Table \ref{tab:2}). 

This disparity between $E_{th}$ and $KE_{\rm outflow}$ is expected and also physically meaningful. It supports the picture emerging from hydrodynamical simulations \citep{mukherjee18, meenakshi22}, where radio jets inflate over-pressured cocoons that create turbulence and drive shocks into the ISM or CGM, leading to significant thermalization of the deposited energy. The simulations consistently show that while a small fraction (typically $\lesssim 1\%$) of the jet’s mechanical energy is transferred into bulk motion of ionized gas, a much larger portion is dissipated as heat, particularly in the hot gas phases that remain undetected in optical emission lines. Our thermal energy estimates suggest that while warm ionized gas traces the visible imprint of AGN feedback, it constitutes only the tip of the iceberg. The majority of the jet’s energy more likely resides in hot, X-ray-emitting plasma which has to be considered in order to fully account for the total AGN feedback energy budget at high redshift.

There is observational evidence that this hot gas may exist, based on $Chandra$ observations of three of our sample galaxies \citep{smail12, smail13}. In all three cases, diffuse X-ray emission is detected that is aligned with the radio source and has a size similar to the emission-line gas. \cite{smail12}  report 12 other high-z ($z =$ 1.8 to 3.8) radio galaxies with similar spatially-resolved X-ray emission. While these authors interpret this as inverse Compton emission, the possibility remains that it is hot ($\sim 10^7 $K) gas that has been shock-heated by the radio source. We will investigate this possibility in a future paper.



\subsubsection{How is the Gas Ionized?} \label{subsubsec:ionization}

In further considering the energy budget for the jets, here we evaluate the possibility that much of this kinetic energy is used to shock-ionize the observed emission-line gas \citep{allen08}.  Given the very high velocities we observe, these shocks would be in the regime in which the emission-line luminosity would be produced both by gas cooling behind the shock, and by gas located upstream of the shock that is photoionized by the radiation produced by the hot gas behind the shock front. To reproduce the emission-line ratios seen in the radio galaxies \citep{saxena24, wang25} shock velocities in the range $v_{shock}$ = 600 -- 1000 km/s are required \citep{allen08}. Over this range, the models show that the ratio of the total energy flux in the shock to the [OIII] (H$\alpha$) emission-line luminosity is $\sim$ 3 to 5 (10 to 17). 

We take an illustrative value of $v_{shock}$ = 800 km/s, and then use the extinction-corrected emission-line luminosities of the outflows (Table 2) to estimate the energy flux in shocks needed to produce this line emission. The results are given in Table 1. As can be seen there, the implied shock energy fluxes are similar to the jet kinetic powers (within the uncertainties in the jet kinetic powers - see Figure~\ref{fig:ljet}). This implies that the jets could in principle produce the emission-lines via shocks. However, this would require that the jet energy is converted into shocks with high efficiency. Given the observed densities of $\sim 10^3$ cm$^{-3}$, the shock models also allow us to compute the total surface area of the shock fronts required to produce the observed emission-line luminosities. We find that the areas are in the range from 20 to 80 kpc$^2$, very consistent with the projected surface areas of the outflows. 

A plausible alternative scenario is that the [OIII] and H$\alpha$-emitting outflowing gas in the ISM or CGM are photoionized by the obscured central quasar, as its ionizing radiation escapes along the outflow axis.  We adopt a model in which these nebulae are in photoionization equilibrium with the diffuse AGN radiation field \citep{crawford88, baum89, heckman91}. In this framework, the electron number density at a distance $r$ from an ionizing source of photon luminosity $Q$ and ionization parameter $U$ is given by:
$$ 
n_e = Q/4\pi r^2U c 
$$

We have used the emission-line diagnostic (BPT) diagrams for these HzRG in \cite{saxena24} and \cite{wang25}, together with CLOUDY models \citep{ji20}, to estimate the values of U. Taking the median ionization parameter $U \sim 10^{-2}$, a median density $\rm n_e \sim 10^3 \ \rm  cm^{-3}$ and median r (outflow size) of $\sim $ 10 kpc, we derive a required ionizing photon rate of $Q = 3.6 \times 10^{57} \ s^{-1} $. Assuming a typical quasar spectral energy distribution \citep{elvis94}, the relation between AGN bolometric luminosity (LAGN) and the ionizing luminosity (Q) is $\rm L_{AGN}/Q=9.5 \times 10^{-11} \  erg$, which translates to $\rm L_{AGN} = 3.4 \times 10^{47} \ erg \ s^{-1}$.

For comparison, the observed AGN bolometric luminosities in our sample span $\sim 10^{46.1} - 10^{48.2}\ \rm erg\ s^{-1}$, with a median of $\sim 2 \times 10^{47} \ \rm erg\ s^{-1}$. This indicates that while photoionization by the central AGN is energetically sufficient to power the extended nebulae in some cases (within the uncertainties), additional ionization mechanisms, such as jet-driven shocks discussed above, may play a significant role. We can also use the values of $n_e$, $U$ and the emission-line luminosities to compute the total surface areas of the photoionized gas. We find that these range from about 200 to 1250 kpc$^2$. These are about an order-of-magnitude larger than the projected areas of the outflows, requiring multiple clouds along a given line-of-sight.

In summary, we conclude that both shock-heating by the jets and photoionization by a central quasar may be (within the uncertainties) energetically-sufficient to power the observed emission-lines in the outflow. They may make different relative contributions in the individual HzRGs. The role of the jets in ionizing the gas is critical in terms of feedback, since this directly affects how much jet energy can be delivered to, and retained by, the ambient gas rather than being radiated away as emission-lines.

\section{Conclusions} \label{sec:conclusions}

We have presented  JWST/NIRSpec IFU observations of six high-redshift ($z > 3.5$) radio galaxies, from two JWST GO programs (1964: PI $-$ Overzier and 1970: PI $-$ Wang). We present sub-kiloparsec resolved maps of ionized gas kinematics, morphology, mass outflow rates, kinetic power, and momentum flux to investigate the feedback signatures driven by powerful radio jets. These observations  provide a direct view of feedback at cosmic noon and beyond.

All six sources reveal disturbed ionized gas kinematics with broad emission-line profiles ($\rm W_{80} \sim 950 - 2500 \ \rm km \ s^{-1}$) in extended regions ($\approx 10-30 \ \rm kpc$) of the galaxy across the whole sample. This is indicative of widespread turbulence and high-velocity gas outflows. Importantly, the regions of enhanced line width are spatially coincident with the radio jet structures, confirming that the jets are the dominant driver of kinematic disturbance, and the jets are driving directional feedback, disturbing the ambient ISM along preferred pathways set by the jet geometry.

The measured ionized gas masses associated with the outflowing components span $\sim 1 - 8 \times 10^9 \ \rm M_{\odot}$, with spatially resolved mass outflow rates of $\sim 80 - 950 \ \rm M_{\odot} \ yr^{-1}$. The total kinetic power of the ionized outflows spans $\sim 10^{43.2} -  10^{45} \ \rm erg \ s^{-1}$, corresponding to kinetic coupling efficiencies of $\sim 0.15\% - 2\%$ relative to the AGN bolometric luminosity.  In several systems, these efficiencies exceed the theoretical $\sim 0.5\%$ threshold required for feedback to substantially alter the host galaxy's gas reservoir. A key result from our analysis is that both the energetics and the strongest gas kinematic signatures are aligned with the radio jet axis. This implies that the energy transfer from the AGN to the ISM is concentrated along the expanding jet-driven cocoon structures.

Despite this, when comparing the outflow kinetic energy to the inferred jet mechanical power using our updated L$_{\rm jet}$ – L$_{\rm radio}$ calibration, we find that only $\sim$0.1 – 1\% of the jet's mechanical energy is transferred into large-scale bulk kinetic energy of the warm ionized gas. This is consistent with expectations from high-resolution jet–ISM interaction simulations \citep[e.g., ][]{mukherjee16, mukherjee20, meenakshi22}, which predict that most of the jet's energy dissipates as heat as turbulent- or shock-heating rather than accelerating the gas coherently. Indeed, we find that the high pressures measured for the warm ionized gas imply that the total thermal energy of such a hot gas phase would be about two orders-of-magnitude greater than the kinetic energy of the warm ionized gas in most cases. There is possible observational evidence for this hot gas from X-ray observations of three of the galaxies in our sample. This highlights the need for continued multi-phase, high-resolution observations of HzRGs to capture the full energy budget, including the hot X-ray and cold molecular gas phases.

We also show that a significant fraction of the jet energy may be used to shock-heat the emission-line gas, but photoionization by radiation from the obscured quasar escaping along the outflow axis can also contribute. The relative importance of these two processes has major implications for the amount of jet feedback that can be used to affect the subsequent evolution of the galaxy.

 In summary, our detailed analyses of the warm ionized gas reinforce the picture that radio jets in massive, gas-rich galaxies at $z > 3.5$ efficiently redistribute gas and carry energy and momentum on global scales. 

\begin{acknowledgments}

\end{acknowledgments}

%



\software{Astropy \citep{astropy13, astropy18, astropy22},  
          Scipy \citep{scipy20}, 
          JWST STScI pipeline \citep{stsci}, 
          dynesty \citep{speagle20}
          }




\bibliography{main}{}

\begin{thebibliography}{}
\expandafter\ifx\csname natexlab\endcsname\relax\def\natexlab#1{#1}\fi
\providecommand{\url}[1]{\href{#1}{#1}}
\providecommand{\dodoi}[1]{doi:~\href{http://doi.org/#1}{\nolinkurl{#1}}}
\providecommand{\doeprint}[1]{\href{http://ascl.net/#1}{\nolinkurl{http://ascl.net/#1}}}
\providecommand{\doarXiv}[1]{\href{https://arxiv.org/abs/#1}{\nolinkurl{https://arxiv.org/abs/#1}}}

\bibitem[{{Allen} {et~al.}(2008){Allen}, {Groves}, {Dopita}, {Sutherland}, \& {Kewley}}]{allen08}
{Allen}, M.~G., {Groves}, B.~A., {Dopita}, M.~A., {Sutherland}, R.~S., \& {Kewley}, L.~J. 2008, \apjs, 178, 20, \dodoi{10.1086/589652}

\bibitem[{{Astropy Collaboration} {et~al.}(2013){Astropy Collaboration}, {Robitaille}, {Tollerud}, {Greenfield}, {Droettboom}, {Bray}, {Aldcroft}, {Davis}, {Ginsburg}, {Price-Whelan}, {Kerzendorf}, {Conley}, {Crighton}, {Barbary}, {Muna}, {Ferguson}, {Grollier}, {Parikh}, {Nair}, {Unther}, {Deil}, {Woillez}, {Conseil}, {Kramer}, {Turner}, {Singer}, {Fox}, {Weaver}, {Zabalza}, {Edwards}, {Azalee Bostroem}, {Burke}, {Casey}, {Crawford}, {Dencheva}, {Ely}, {Jenness}, {Labrie}, {Lim}, {Pierfederici}, {Pontzen}, {Ptak}, {Refsdal}, {Servillat}, \& {Streicher}}]{astropy13}
{Astropy Collaboration}, {Robitaille}, T.~P., {Tollerud}, E.~J., {et~al.} 2013, \aap, 558, A33, \dodoi{10.1051/0004-6361/201322068}

\bibitem[{{Astropy Collaboration} {et~al.}(2018){Astropy Collaboration}, {Price-Whelan}, {Sip{\H{o}}cz}, {G{\"u}nther}, {Lim}, {Crawford}, {Conseil}, {Shupe}, {Craig}, {Dencheva}, {Ginsburg}, {VanderPlas}, {Bradley}, {P{\'e}rez-Su{\'a}rez}, {de Val-Borro}, {Aldcroft}, {Cruz}, {Robitaille}, {Tollerud}, {Ardelean}, {Babej}, {Bach}, {Bachetti}, {Bakanov}, {Bamford}, {Barentsen}, {Barmby}, {Baumbach}, {Berry}, {Biscani}, {Boquien}, {Bostroem}, {Bouma}, {Brammer}, {Bray}, {Breytenbach}, {Buddelmeijer}, {Burke}, {Calderone}, {Cano Rodr{\'\i}guez}, {Cara}, {Cardoso}, {Cheedella}, {Copin}, {Corrales}, {Crichton}, {D'Avella}, {Deil}, {Depagne}, {Dietrich}, {Donath}, {Droettboom}, {Earl}, {Erben}, {Fabbro}, {Ferreira}, {Finethy}, {Fox}, {Garrison}, {Gibbons}, {Goldstein}, {Gommers}, {Greco}, {Greenfield}, {Groener}, {Grollier}, {Hagen}, {Hirst}, {Homeier}, {Horton}, {Hosseinzadeh}, {Hu}, {Hunkeler}, {Ivezi{\'c}}, {Jain}, {Jenness}, {Kanarek}, {Kendrew}, {Kern}, {Kerzendorf}, {Khvalko}, {King}, {Kirkby}, {Kulkarni},
  {Kumar}, {Lee}, {Lenz}, {Littlefair}, {Ma}, {Macleod}, {Mastropietro}, {McCully}, {Montagnac}, {Morris}, {Mueller}, {Mumford}, {Muna}, {Murphy}, {Nelson}, {Nguyen}, {Ninan}, {N{\"o}the}, {Ogaz}, {Oh}, {Parejko}, {Parley}, {Pascual}, {Patil}, {Patil}, {Plunkett}, {Prochaska}, {Rastogi}, {Reddy Janga}, {Sabater}, {Sakurikar}, {Seifert}, {Sherbert}, {Sherwood-Taylor}, {Shih}, {Sick}, {Silbiger}, {Singanamalla}, {Singer}, {Sladen}, {Sooley}, {Sornarajah}, {Streicher}, {Teuben}, {Thomas}, {Tremblay}, {Turner}, {Terr{\'o}n}, {van Kerkwijk}, {de la Vega}, {Watkins}, {Weaver}, {Whitmore}, {Woillez}, {Zabalza}, \& {Astropy Contributors}}]{astropy18}
{Astropy Collaboration}, {Price-Whelan}, A.~M., {Sip{\H{o}}cz}, B.~M., {et~al.} 2018, \aj, 156, 123, \dodoi{10.3847/1538-3881/aabc4f}

\bibitem[{{Astropy Collaboration} {et~al.}(2022){Astropy Collaboration}, {Price-Whelan}, {Lim}, {Earl}, {Starkman}, {Bradley}, {Shupe}, {Patil}, {Corrales}, {Brasseur}, {N{\"o}the}, {Donath}, {Tollerud}, {Morris}, {Ginsburg}, {Vaher}, {Weaver}, {Tocknell}, {Jamieson}, {van Kerkwijk}, {Robitaille}, {Merry}, {Bachetti}, {G{\"u}nther}, {Aldcroft}, {Alvarado-Montes}, {Archibald}, {B{\'o}di}, {Bapat}, {Barentsen}, {Baz{\'a}n}, {Biswas}, {Boquien}, {Burke}, {Cara}, {Cara}, {Conroy}, {Conseil}, {Craig}, {Cross}, {Cruz}, {D'Eugenio}, {Dencheva}, {Devillepoix}, {Dietrich}, {Eigenbrot}, {Erben}, {Ferreira}, {Foreman-Mackey}, {Fox}, {Freij}, {Garg}, {Geda}, {Glattly}, {Gondhalekar}, {Gordon}, {Grant}, {Greenfield}, {Groener}, {Guest}, {Gurovich}, {Handberg}, {Hart}, {Hatfield-Dodds}, {Homeier}, {Hosseinzadeh}, {Jenness}, {Jones}, {Joseph}, {Kalmbach}, {Karamehmetoglu}, {Ka{\l}uszy{\'n}ski}, {Kelley}, {Kern}, {Kerzendorf}, {Koch}, {Kulumani}, {Lee}, {Ly}, {Ma}, {MacBride}, {Maljaars}, {Muna}, {Murphy}, {Norman},
  {O'Steen}, {Oman}, {Pacifici}, {Pascual}, {Pascual-Granado}, {Patil}, {Perren}, {Pickering}, {Rastogi}, {Roulston}, {Ryan}, {Rykoff}, {Sabater}, {Sakurikar}, {Salgado}, {Sanghi}, {Saunders}, {Savchenko}, {Schwardt}, {Seifert-Eckert}, {Shih}, {Jain}, {Shukla}, {Sick}, {Simpson}, {Singanamalla}, {Singer}, {Singhal}, {Sinha}, {Sip{\H{o}}cz}, {Spitler}, {Stansby}, {Streicher}, {{\v{S}}umak}, {Swinbank}, {Taranu}, {Tewary}, {Tremblay}, {de Val-Borro}, {Van Kooten}, {Vasovi{\'c}}, {Verma}, {de Miranda Cardoso}, {Williams}, {Wilson}, {Winkel}, {Wood-Vasey}, {Xue}, {Yoachim}, {Zhang}, {Zonca}, \& {Astropy Project Contributors}}]{astropy22}
{Astropy Collaboration}, {Price-Whelan}, A.~M., {Lim}, P.~L., {et~al.} 2022, \apj, 935, 167, \dodoi{10.3847/1538-4357/ac7c74}

\bibitem[{{Balmaverde} {et~al.}(2019){Balmaverde}, {Capetti}, {Marconi}, {Venturi}, {Chiaberge}, {Baldi}, {Baum}, {Gilli}, {Grandi}, {Meyer}, {Miley}, {O'Dea}, {Sparks}, {Torresi}, \& {Tremblay}}]{balmaverde19}
{Balmaverde}, B., {Capetti}, A., {Marconi}, A., {et~al.} 2019, \aap, 632, A124, \dodoi{10.1051/0004-6361/201935544}

\bibitem[{{Baron} \& {Netzer}(2019)}]{baron19}
{Baron}, D., \& {Netzer}, H. 2019, \mnras, 486, 4290, \dodoi{10.1093/mnras/stz1070}

\bibitem[{{Baum} \& {Heckman}(1989)}]{baum89}
{Baum}, S.~A., \& {Heckman}, T. 1989, \apj, 336, 702, \dodoi{10.1086/167044}

\bibitem[{{Begelman} \& {Cioffi}(1989)}]{begelman89}
{Begelman}, M.~C., \& {Cioffi}, D.~F. 1989, \apjl, 345, L21, \dodoi{10.1086/185542}

\bibitem[{{Bicknell} {et~al.}(2018){Bicknell}, {Mukherjee}, {Wagner}, {Sutherland}, \& {Nesvadba}}]{bicknell18}
{Bicknell}, G.~V., {Mukherjee}, D., {Wagner}, A.~Y., {Sutherland}, R.~S., \& {Nesvadba}, N. P.~H. 2018, \mnras, 475, 3493, \dodoi{10.1093/mnras/sty070}

\bibitem[{{B{\^\i}rzan} {et~al.}(2004){B{\^\i}rzan}, {Rafferty}, {McNamara}, {Wise}, \& {Nulsen}}]{birzan04}
{B{\^\i}rzan}, L., {Rafferty}, D.~A., {McNamara}, B.~R., {Wise}, M.~W., \& {Nulsen}, P.~E.~J. 2004, \apj, 607, 800, \dodoi{10.1086/383519}

\bibitem[{{Brusa} {et~al.}(2016){Brusa}, {Perna}, {Cresci}, {Schramm}, {Delvecchio}, {Lanzuisi}, {Mainieri}, {Mignoli}, {Zamorani}, {Berta}, {Bongiorno}, {Comastri}, {Fiore}, {Kakkad}, {Marconi}, {Rosario}, {Contini}, \& {Lamareille}}]{brusa16}
{Brusa}, M., {Perna}, M., {Cresci}, G., {et~al.} 2016, \aap, 588, A58, \dodoi{10.1051/0004-6361/201527900}

\bibitem[{Bushouse {et~al.}(2023)Bushouse, Eisenhamer, Dencheva, Davies, Greenfield, Morrison, Hodge, Simon, Grumm, Droettboom, Slavich, Sosey, Pauly, Miller, Jedrzejewski, Hack, Davis, Crawford, Law, Gordon, Regan, Cara, MacDonald, Bradley, Shanahan, Jamieson, Teodoro, \& Williams}]{stsci}
Bushouse, H., Eisenhamer, J., Dencheva, N., {et~al.} 2023, JWST Calibration Pipeline, 1.11.2,  Zenodo, \dodoi{10.5281/zenodo.8140011}

\bibitem[{{Cano-D{\'\i}az} {et~al.}(2012){Cano-D{\'\i}az}, {Maiolino}, {Marconi}, {Netzer}, {Shemmer}, \& {Cresci}}]{cano-diaz12}
{Cano-D{\'\i}az}, M., {Maiolino}, R., {Marconi}, A., {et~al.} 2012, \aap, 537, L8, \dodoi{10.1051/0004-6361/201118358}

\bibitem[{{Cavagnolo} {et~al.}(2010){Cavagnolo}, {McNamara}, {Nulsen}, {Carilli}, {Jones}, \& {B{\^\i}rzan}}]{cavagnolo10}
{Cavagnolo}, K.~W., {McNamara}, B.~R., {Nulsen}, P.~E.~J., {et~al.} 2010, \apj, 720, 1066, \dodoi{10.1088/0004-637X/720/2/1066}

\bibitem[{{Cicone} {et~al.}(2015){Cicone}, {Maiolino}, {Gallerani}, {Neri}, {Ferrara}, {Sturm}, {Fiore}, {Piconcelli}, \& {Feruglio}}]{cicone15}
{Cicone}, C., {Maiolino}, R., {Gallerani}, S., {et~al.} 2015, \aap, 574, A14, \dodoi{10.1051/0004-6361/201424980}

\bibitem[{{Ciotti} {et~al.}(2010){Ciotti}, {Ostriker}, \& {Proga}}]{ciotti10}
{Ciotti}, L., {Ostriker}, J.~P., \& {Proga}, D. 2010, \apj, 717, 708, \dodoi{10.1088/0004-637X/717/2/708}

\bibitem[{{Collet} {et~al.}(2016){Collet}, {Nesvadba}, {De Breuck}, {Lehnert}, {Best}, {Bryant}, {Hunstead}, {Dicken}, \& {Johnston}}]{collet16}
{Collet}, C., {Nesvadba}, N.~P.~H., {De Breuck}, C., {et~al.} 2016, \aap, 586, A152, \dodoi{10.1051/0004-6361/201526872}

\bibitem[{{Crawford} {et~al.}(1988){Crawford}, {Fabian}, \& {Johnstone}}]{crawford88}
{Crawford}, C.~S., {Fabian}, A.~C., \& {Johnstone}, R.~M. 1988, \mnras, 235, 183, \dodoi{10.1093/mnras/235.1.183}

\bibitem[{{Crenshaw} {et~al.}(2003){Crenshaw}, {Kraemer}, {Gabel}, {Kaastra}, {Steenbrugge}, {Brinkman}, {Dunn}, {George}, {Liedahl}, {Paerels}, {Turner}, \& {Yaqoob}}]{crenshaw03}
{Crenshaw}, D.~M., {Kraemer}, S.~B., {Gabel}, J.~R., {et~al.} 2003, \apj, 594, 116, \dodoi{10.1086/376792}

\bibitem[{{Cresci} {et~al.}(2023){Cresci}, {Tozzi}, {Perna}, {Brusa}, {Marconcini}, {Marconi}, {Carniani}, {Brienza}, {Giroletti}, {Belfiore}, {Ginolfi}, {Mannucci}, {Ulivi}, {Scholtz}, {Venturi}, {Arribas}, {{\"U}bler}, {D'Eugenio}, {Mingozzi}, {Balmaverde}, {Capetti}, {Parlanti}, \& {Zana}}]{cresci23}
{Cresci}, G., {Tozzi}, G., {Perna}, M., {et~al.} 2023, \aap, 672, A128, \dodoi{10.1051/0004-6361/202346001}

\bibitem[{{Croton} {et~al.}(2006){Croton}, {Springel}, {White}, {De Lucia}, {Frenk}, {Gao}, {Jenkins}, {Kauffmann}, {Navarro}, \& {Yoshida}}]{croton06}
{Croton}, D.~J., {Springel}, V., {White}, S. D.~M., {et~al.} 2006, \mnras, 365, 11, \dodoi{10.1111/j.1365-2966.2005.09675.x}

\bibitem[{{Dall'Agnol de Oliveira} {et~al.}(2021){Dall'Agnol de Oliveira}, {Storchi-Bergmann}, {Kraemer}, {Villar Mart{\'\i}n}, {Schnorr-M{\"u}ller}, {Schmitt}, {Ruschel-Dutra}, {Crenshaw}, \& {Fischer}}]{oliveira21}
{Dall'Agnol de Oliveira}, B., {Storchi-Bergmann}, T., {Kraemer}, S.~B., {et~al.} 2021, \mnras, 504, 3890, \dodoi{10.1093/mnras/stab1067}

\bibitem[{{Daly} {et~al.}(2012){Daly}, {Sprinkle}, {O'Dea}, {Kharb}, \& {Baum}}]{daly12}
{Daly}, R.~A., {Sprinkle}, T.~B., {O'Dea}, C.~P., {Kharb}, P., \& {Baum}, S.~A. 2012, \mnras, 423, 2498, \dodoi{10.1111/j.1365-2966.2012.21060.x}

\bibitem[{{De Breuck} {et~al.}(1999){De Breuck}, {van Breugel}, {Minniti}, {Miley}, {R{\"o}ttgering}, {Stanford}, \& {Carilli}}]{debreuck99}
{De Breuck}, C., {van Breugel}, W., {Minniti}, D., {et~al.} 1999, \aap, 352, L51, \dodoi{10.48550/arXiv.astro-ph/9909178}

\bibitem[{{De Breuck} {et~al.}(2010){De Breuck}, {Seymour}, {Stern}, {Willner}, {Eisenhardt}, {Fazio}, {Galametz}, {Lacy}, {Rettura}, {Rocca-Volmerange}, \& {Vernet}}]{debreuck10}
{De Breuck}, C., {Seymour}, N., {Stern}, D., {et~al.} 2010, \apj, 725, 36, \dodoi{10.1088/0004-637X/725/1/36}

\bibitem[{{Di Matteo} {et~al.}(2005){Di Matteo}, {Springel}, \& {Hernquist}}]{dimatteo05}
{Di Matteo}, T., {Springel}, V., \& {Hernquist}, L. 2005, \nat, 433, 604, \dodoi{10.1038/nature03335}

\bibitem[{{Dunn} \& {Fabian}(2006)}]{dunn06}
{Dunn}, R.~J.~H., \& {Fabian}, A.~C. 2006, \mnras, 373, 959, \dodoi{10.1111/j.1365-2966.2006.11080.x}

\bibitem[{{Dutta} {et~al.}(2023){Dutta}, {Sharma}, {Sarkar}, \& {Stone}}]{dutta24}
{Dutta}, R., {Sharma}, P., {Sarkar}, K.~C., \& {Stone}, J.~M. 2023, arXiv e-prints, arXiv:2401.00446, \dodoi{10.48550/arXiv.2401.00446}

\bibitem[{{Elvis} {et~al.}(1994){Elvis}, {Wilkes}, {McDowell}, {Green}, {Bechtold}, {Willner}, {Oey}, {Polomski}, \& {Cutri}}]{elvis94}
{Elvis}, M., {Wilkes}, B.~J., {McDowell}, J.~C., {et~al.} 1994, \apjs, 95, 1, \dodoi{10.1086/192093}

\bibitem[{{Fabian}(2012)}]{fabian12}
{Fabian}, A.~C. 2012, \araa, 50, 455, \dodoi{10.1146/annurev-astro-081811-125521}

\bibitem[{{Falkendal} {et~al.}(2019){Falkendal}, {De Breuck}, {Lehnert}, {Drouart}, {Vernet}, {Emonts}, {Lee}, {Nesvadba}, {Seymour}, {B{\'e}thermin}, {Kolwa}, {Gullberg}, \& {Wylezalek}}]{falkendal19}
{Falkendal}, T., {De Breuck}, C., {Lehnert}, M.~D., {et~al.} 2019, \aap, 621, A27, \dodoi{10.1051/0004-6361/201732485}

\bibitem[{{Fiore} {et~al.}(2017){Fiore}, {Feruglio}, {Shankar}, {Bischetti}, {Bongiorno}, {Brusa}, {Carniani}, {Cicone}, {Duras}, {Lamastra}, {Mainieri}, {Marconi}, {Menci}, {Maiolino}, {Piconcelli}, {Vietri}, \& {Zappacosta}}]{fiore17}
{Fiore}, F., {Feruglio}, C., {Shankar}, F., {et~al.} 2017, \aap, 601, A143, \dodoi{10.1051/0004-6361/201629478}

\bibitem[{{F{\"o}rster Schreiber} {et~al.}(2019){F{\"o}rster Schreiber}, {{\"U}bler}, {Davies}, {Genzel}, {Wisnioski}, {Belli}, {Shimizu}, {Lutz}, {Fossati}, {Herrera-Camus}, {Mendel}, {Tacconi}, {Wilman}, {Beifiori}, {Brammer}, {Burkert}, {Carollo}, {Davies}, {Eisenhauer}, {Fabricius}, {Lilly}, {Momcheva}, {Naab}, {Nelson}, {Price}, {Renzini}, {Saglia}, {Sternberg}, {van Dokkum}, \& {Wuyts}}]{schreiber19}
{F{\"o}rster Schreiber}, N.~M., {{\"U}bler}, H., {Davies}, R.~L., {et~al.} 2019, \apj, 875, 21, \dodoi{10.3847/1538-4357/ab0ca2}

\bibitem[{{Gab{\'a}nyi} {et~al.}(2018){Gab{\'a}nyi}, {Frey}, {Gurvits}, {Paragi}, \& {Perger}}]{gabanyi18}
{Gab{\'a}nyi}, K.~{\'E}., {Frey}, S., {Gurvits}, L.~I., {Paragi}, Z., \& {Perger}, K. 2018, Research Notes of the American Astronomical Society, 2, 200, \dodoi{10.3847/2515-5172/aaec82}

\bibitem[{{Greene} {et~al.}(2012){Greene}, {Zakamska}, \& {Smith}}]{greene12}
{Greene}, J.~E., {Zakamska}, N.~L., \& {Smith}, P.~S. 2012, \apj, 746, 86, \dodoi{10.1088/0004-637X/746/1/86}

\bibitem[{{Harrison} {et~al.}(2014){Harrison}, {Alexander}, {Mullaney}, \& {Swinbank}}]{harrison14}
{Harrison}, C.~M., {Alexander}, D.~M., {Mullaney}, J.~R., \& {Swinbank}, A.~M. 2014, \mnras, 441, 3306, \dodoi{10.1093/mnras/stu515}

\bibitem[{{Heckman} \& {Best}(2014)}]{heckman14}
{Heckman}, T.~M., \& {Best}, P.~N. 2014, \araa, 52, 589, \dodoi{10.1146/annurev-astro-081913-035722}

\bibitem[{{Heckman} \& {Best}(2023)}]{heckman23}
---. 2023, Galaxies, 11, 21, \dodoi{10.3390/galaxies11010021}

\bibitem[{{Heckman} {et~al.}(1991){Heckman}, {Lehnert}, {Miley}, \& {van Breugel}}]{heckman91}
{Heckman}, T.~M., {Lehnert}, M.~D., {Miley}, G.~K., \& {van Breugel}, W. 1991, \apj, 381, 373, \dodoi{10.1086/170660}

\bibitem[{{Heckman} {et~al.}(2024){Heckman}, {Roy}, {Best}, \& {Kondapally}}]{heckman24}
{Heckman}, T.~M., {Roy}, N., {Best}, P.~N., \& {Kondapally}, R. 2024, \apj, 977, 125, \dodoi{10.3847/1538-4357/ad8f3e}

\bibitem[{{Hopkins} \& {Elvis}(2010)}]{hopkins10}
{Hopkins}, P.~F., \& {Elvis}, M. 2010, \mnras, 401, 7, \dodoi{10.1111/j.1365-2966.2009.15643.x}

\bibitem[{{Hopkins} {et~al.}(2006){Hopkins}, {Hernquist}, {Cox}, {Di Matteo}, {Robertson}, \& {Springel}}]{hopkins06}
{Hopkins}, P.~F., {Hernquist}, L., {Cox}, T.~J., {et~al.} 2006, \apjs, 163, 1, \dodoi{10.1086/499298}

\bibitem[{{Ineson} {et~al.}(2017){Ineson}, {Croston}, {Hardcastle}, \& {Mingo}}]{ineson17}
{Ineson}, J., {Croston}, J.~H., {Hardcastle}, M.~J., \& {Mingo}, B. 2017, \mnras, 467, 1586, \dodoi{10.1093/mnras/stx189}

\bibitem[{{Intema} {et~al.}(2017){Intema}, {Jagannathan}, {Mooley}, \& {Frail}}]{intema17}
{Intema}, H.~T., {Jagannathan}, P., {Mooley}, K.~P., \& {Frail}, D.~A. 2017, \aap, 598, A78, \dodoi{10.1051/0004-6361/201628536}

\bibitem[{{Intema} {et~al.}(2006){Intema}, {Venemans}, {Kurk}, {Ouchi}, {Kodama}, {R{\"o}ttgering}, {Miley}, \& {Overzier}}]{intema06}
{Intema}, H.~T., {Venemans}, B.~P., {Kurk}, J.~D., {et~al.} 2006, \aap, 456, 433, \dodoi{10.1051/0004-6361:20064812}

\bibitem[{{Jarvis} {et~al.}(2019){Jarvis}, {Harrison}, {Thomson}, {Circosta}, {Mainieri}, {Alexander}, {Edge}, {Lansbury}, {Molyneux}, \& {Mullaney}}]{jarvis19}
{Jarvis}, M.~E., {Harrison}, C.~M., {Thomson}, A.~P., {et~al.} 2019, \mnras, 485, 2710, \dodoi{10.1093/mnras/stz556}

\bibitem[{{Jarvis} {et~al.}(2021){Jarvis}, {Harrison}, {Mainieri}, {Alexander}, {Arrigoni Battaia}, {Calistro Rivera}, {Circosta}, {Costa}, {De Breuck}, {Edge}, {Girdhar}, {Kakkad}, {Kharb}, {Lansbury}, {Molyneux}, {Mukherjee}, {Mullaney}, {Farina}, {Silpa}, {Thomson}, \& {Ward}}]{jarvis21}
{Jarvis}, M.~E., {Harrison}, C.~M., {Mainieri}, V., {et~al.} 2021, \mnras, 503, 1780, \dodoi{10.1093/mnras/stab549}

\bibitem[{{Ji} \& {Yan}(2020)}]{ji20}
{Ji}, X., \& {Yan}, R. 2020, \mnras, 499, 5749, \dodoi{10.1093/mnras/staa3259}

\bibitem[{{Kakkad} {et~al.}(2022){Kakkad}, {Sani}, {Rojas}, {Mallmann}, {Veilleux}, {Bauer}, {Ricci}, {Mushotzky}, {Koss}, {Ricci}, {Treister}, {Privon}, {Nguyen}, {B{\"a}r}, {Harrison}, {Oh}, {Powell}, {Riffel}, {Stern}, {Trakhtenbrot}, \& {Urry}}]{kakkad22}
{Kakkad}, D., {Sani}, E., {Rojas}, A.~F., {et~al.} 2022, \mnras, 511, 2105, \dodoi{10.1093/mnras/stac103}

\bibitem[{{Kim} {et~al.}(2023){Kim}, {Woo}, {Luo}, {Chung}, {Baek}, {Le}, \& {Son}}]{kim23}
{Kim}, C., {Woo}, J.-H., {Luo}, R., {et~al.} 2023, \apj, 958, 145, \dodoi{10.3847/1538-4357/acf92b}

\bibitem[{{Kolwa} {et~al.}(2023){Kolwa}, {De Breuck}, {Vernet}, {Wylezalek}, {Wang}, {Popping}, {Man}, {Harrison}, \& {Andreani}}]{kolwa23}
{Kolwa}, S., {De Breuck}, C., {Vernet}, J., {et~al.} 2023, \mnras, 525, 5831, \dodoi{10.1093/mnras/stad2647}

\bibitem[{{Marshall} {et~al.}(2023){Marshall}, {Perna}, {Willott}, {Maiolino}, {Scholtz}, {{\"U}bler}, {Carniani}, {Arribas}, {L{\"u}tzgendorf}, {Bunker}, {Charlot}, {Ferruit}, {Jakobsen}, {Rix}, {Rodr{\'\i}guez Del Pino}, {B{\"o}ker}, {Cameron}, {Cresci}, {Curtis-Lake}, {Jones}, {Kumari}, {P{\'e}rez-Gonz{\'a}lez}, \& {Reed}}]{marshall23}
{Marshall}, M.~A., {Perna}, M., {Willott}, C.~J., {et~al.} 2023, \aap, 678, A191, \dodoi{10.1051/0004-6361/202346113}

\bibitem[{{McCully} {et~al.}(2018){McCully}, {Crawford}, {Kovacs}, {Tollerud}, {Betts}, {Bradley}, {Craig}, {Turner}, {Streicher}, {Sipocz}, {Robitaille}, \& {Deil}}]{mccully18}
{McCully}, C., {Crawford}, S., {Kovacs}, G., {et~al.} 2018, {astropy/astroscrappy: v1.0.5 Zenodo Release}, v1.0.5,  Zenodo, \dodoi{10.5281/zenodo.1482019}

\bibitem[{{McNamara} \& {Nulsen}(2007)}]{mcnamara07}
{McNamara}, B.~R., \& {Nulsen}, P.~E.~J. 2007, \araa, 45, 117, \dodoi{10.1146/annurev.astro.45.051806.110625}

\bibitem[{{McNamara} \& {Nulsen}(2012)}]{mcnamara12}
---. 2012, New Journal of Physics, 14, 055023, \dodoi{10.1088/1367-2630/14/5/055023}

\bibitem[{{Meenakshi} {et~al.}(2022){Meenakshi}, {Mukherjee}, {Wagner}, {Nesvadba}, {Bicknell}, {Morganti}, {Janssen}, {Sutherland}, \& {Mandal}}]{meenakshi22}
{Meenakshi}, M., {Mukherjee}, D., {Wagner}, A.~Y., {et~al.} 2022, \mnras, 516, 766, \dodoi{10.1093/mnras/stac2251}

\bibitem[{{Morganti} {et~al.}(2021){Morganti}, {Oosterloo}, {Tadhunter}, {Bernhard}, \& {Raymond Oonk}}]{morganti21}
{Morganti}, R., {Oosterloo}, T., {Tadhunter}, C., {Bernhard}, E.~P., \& {Raymond Oonk}, J.~B. 2021, \aap, 656, A55, \dodoi{10.1051/0004-6361/202141766}

\bibitem[{{Mukherjee} {et~al.}(2016){Mukherjee}, {Bicknell}, {Sutherland}, \& {Wagner}}]{mukherjee16}
{Mukherjee}, D., {Bicknell}, G.~V., {Sutherland}, R., \& {Wagner}, A. 2016, \mnras, 461, 967, \dodoi{10.1093/mnras/stw1368}

\bibitem[{{Mukherjee} {et~al.}(2018){Mukherjee}, {Bicknell}, {Wagner}, {Sutherland}, \& {Silk}}]{mukherjee18}
{Mukherjee}, D., {Bicknell}, G.~V., {Wagner}, A.~Y., {Sutherland}, R.~S., \& {Silk}, J. 2018, \mnras, 479, 5544, \dodoi{10.1093/mnras/sty1776}

\bibitem[{{Mukherjee} {et~al.}(2020){Mukherjee}, {Bodo}, {Mignone}, {Rossi}, \& {Vaidya}}]{mukherjee20}
{Mukherjee}, D., {Bodo}, G., {Mignone}, A., {Rossi}, P., \& {Vaidya}, B. 2020, \mnras, 499, 681, \dodoi{10.1093/mnras/staa2934}

\bibitem[{{Mullaney} {et~al.}(2013){Mullaney}, {Alexander}, {Fine}, {Goulding}, {Harrison}, \& {Hickox}}]{mullaney13}
{Mullaney}, J.~R., {Alexander}, D.~M., {Fine}, S., {et~al.} 2013, \mnras, 433, 622, \dodoi{10.1093/mnras/stt751}

\bibitem[{{Nesvadba} {et~al.}(2017){Nesvadba}, {De Breuck}, {Lehnert}, {Best}, \& {Collet}}]{nesvadba17}
{Nesvadba}, N.~P.~H., {De Breuck}, C., {Lehnert}, M.~D., {Best}, P.~N., \& {Collet}, C. 2017, \aap, 599, A123, \dodoi{10.1051/0004-6361/201528040}

\bibitem[{{Nesvadba} {et~al.}(2007){Nesvadba}, {Lehnert}, {De Breuck}, {Gilbert}, \& {van Breugel}}]{nesvadba07}
{Nesvadba}, N.~P.~H., {Lehnert}, M.~D., {De Breuck}, C., {Gilbert}, A., \& {van Breugel}, W. 2007, \aap, 475, 145, \dodoi{10.1051/0004-6361:20078175}

\bibitem[{{Nesvadba} {et~al.}(2008){Nesvadba}, {Lehnert}, {De Breuck}, {Gilbert}, \& {van Breugel}}]{nesvadba08}
{Nesvadba}, N.~P.~H., {Lehnert}, M.~D., {De Breuck}, C., {Gilbert}, A.~M., \& {van Breugel}, W. 2008, \aap, 491, 407, \dodoi{10.1051/0004-6361:200810346}

\bibitem[{{Nesvadba} {et~al.}(2006){Nesvadba}, {Lehnert}, {Eisenhauer}, {Gilbert}, {Tecza}, \& {Abuter}}]{nesvadba06}
{Nesvadba}, N.~P.~H., {Lehnert}, M.~D., {Eisenhauer}, F., {et~al.} 2006, \apj, 650, 693, \dodoi{10.1086/507266}

\bibitem[{{O'Dea}(1998)}]{odea98}
{O'Dea}, C.~P. 1998, \pasp, 110, 493, \dodoi{10.1086/316162}

\bibitem[{{O'Dea} \& {Saikia}(2021)}]{odea21}
{O'Dea}, C.~P., \& {Saikia}, D.~J. 2021, \aapr, 29, 3, \dodoi{10.1007/s00159-021-00131-w}

\bibitem[{{Oke} \& {Gunn}(1983)}]{oke83}
{Oke}, J.~B., \& {Gunn}, J.~E. 1983, \apj, 266, 713, \dodoi{10.1086/160817}

\bibitem[{{Pentericci} {et~al.}(2000){Pentericci}, {Van Reeven}, {Carilli}, {R{\"o}ttgering}, \& {Miley}}]{pentericci00}
{Pentericci}, L., {Van Reeven}, W., {Carilli}, C.~L., {R{\"o}ttgering}, H.~J.~A., \& {Miley}, G.~K. 2000, \aaps, 145, 121, \dodoi{10.1051/aas:2000104}

\bibitem[{{Perna} {et~al.}(2023){Perna}, {Arribas}, {Marshall}, {D'Eugenio}, {{\"U}bler}, {Bunker}, {Charlot}, {Carniani}, {Jakobsen}, {Maiolino}, {Rodr{\'\i}guez Del Pino}, {Willott}, {B{\"o}ker}, {Circosta}, {Cresci}, {Curti}, {Husemann}, {Kumari}, {Lamperti}, {P{\'e}rez-Gonz{\'a}lez}, \& {Scholtz}}]{perna23}
{Perna}, M., {Arribas}, S., {Marshall}, M., {et~al.} 2023, \aap, 679, A89, \dodoi{10.1051/0004-6361/202346649}

\bibitem[{{Rafferty} {et~al.}(2006){Rafferty}, {McNamara}, {Nulsen}, \& {Wise}}]{rafferty06}
{Rafferty}, D.~A., {McNamara}, B.~R., {Nulsen}, P.~E.~J., \& {Wise}, M.~W. 2006, \apj, 652, 216, \dodoi{10.1086/507672}

\bibitem[{{Revalski} {et~al.}(2018){Revalski}, {Crenshaw}, {Kraemer}, {Fischer}, {Schmitt}, \& {Machuca}}]{revalski18}
{Revalski}, M., {Crenshaw}, D.~M., {Kraemer}, S.~B., {et~al.} 2018, \apj, 856, 46, \dodoi{10.3847/1538-4357/aab107}

\bibitem[{{Revalski} {et~al.}(2021){Revalski}, {Meena}, {Martinez}, {Polack}, {Crenshaw}, {Kraemer}, {Collins}, {Fischer}, {Schmitt}, {Schmidt}, {Maksym}, \& {Rafelski}}]{revalski21}
{Revalski}, M., {Meena}, B., {Martinez}, F., {et~al.} 2021, \apj, 910, 139, \dodoi{10.3847/1538-4357/abdcad}

\bibitem[{{Roy} {et~al.}(2018){Roy}, {Bundy}, {Cheung}, {Rujopakarn}, {Cappellari}, {Belfiore}, {Yan}, {Heckman}, {Bershady}, {Greene}, {Westfall}, {Drory}, {Rubin}, {Law}, {Zhang}, {Gelfand}, {Bizyaev}, {Wake}, {Masters}, {Thomas}, {Li}, \& {Riffel}}]{roy18}
{Roy}, N., {Bundy}, K., {Cheung}, E., {et~al.} 2018, \apj, 869, 117, \dodoi{10.3847/1538-4357/aaee72}

\bibitem[{{Roy} {et~al.}(2021{\natexlab{a}}){Roy}, {Bundy}, {Nevin}, {Belfiore}, {Yan}, {Campbell}, {Riffel}, {Riffel}, {Bershady}, {Westfall}, {Drory}, \& {Zhang}}]{roy21a}
{Roy}, N., {Bundy}, K., {Nevin}, R., {et~al.} 2021{\natexlab{a}}, \apj, 913, 33, \dodoi{10.3847/1538-4357/abf1e6}

\bibitem[{{Roy} {et~al.}(2021{\natexlab{b}}){Roy}, {Moravec}, {Bundy}, {Hardcastle}, {G{\"u}rkan}, {Diego Baldi}, {Leslie}, {Masters}, {Gelfand}, {Riffel}, {Riffel}, {Mingo Fernandez}, \& {Drabent}}]{roy21c}
{Roy}, N., {Moravec}, E., {Bundy}, K., {et~al.} 2021{\natexlab{b}}, \apj, 922, 230, \dodoi{10.3847/1538-4357/ac24a0}

\bibitem[{{Roy} {et~al.}(2024){Roy}, {Heckman}, {Overzier}, {Saxena}, {Duncan}, {Miley}, {Villar Mart{\'\i}n}, {Gab{\'a}nyi}, {Aydar}, {Bosman}, {Rottgering}, {Pentericci}, {Onoue}, \& {Reynaldi}}]{roy24}
{Roy}, N., {Heckman}, T., {Overzier}, R., {et~al.} 2024, \apj, 970, 69, \dodoi{10.3847/1538-4357/ad4bda}

\bibitem[{{Ruschel-Dutra} {et~al.}(2021){Ruschel-Dutra}, {Storchi-Bergmann}, {Schnorr-M{\"u}ller}, {Riffel}, {Dall'Agnol de Oliveira}, {Lena}, {Robinson}, {Nagar}, \& {Elvis}}]{ruschel-dutra21}
{Ruschel-Dutra}, D., {Storchi-Bergmann}, T., {Schnorr-M{\"u}ller}, A., {et~al.} 2021, \mnras, 507, 74, \dodoi{10.1093/mnras/stab2058}

\bibitem[{{Sanders} {et~al.}(2016){Sanders}, {Shapley}, {Kriek}, {Reddy}, {Freeman}, {Coil}, {Siana}, {Mobasher}, {Shivaei}, {Price}, \& {de Groot}}]{sanders16}
{Sanders}, R.~L., {Shapley}, A.~E., {Kriek}, M., {et~al.} 2016, \apj, 816, 23, \dodoi{10.3847/0004-637X/816/1/23}

\bibitem[{{Saxena} {et~al.}(2024){Saxena}, {Overzier}, {Villar-Mart{\'\i}n}, {Heckman}, {Roy}, {Duncan}, {R{\"o}ttgering}, {Miley}, {Aydar}, {Best}, {Bosman}, {Cameron}, {{\'E}va Gab{\'a}nyi}, {Humphrey}, {Morais}, {Onoue}, {Pentericci}, {Reynaldi}, \& {Venemans}}]{saxena24}
{Saxena}, A., {Overzier}, R.~A., {Villar-Mart{\'\i}n}, M., {et~al.} 2024, arXiv e-prints, arXiv:2401.12199, \dodoi{10.48550/arXiv.2401.12199}

\bibitem[{{Seymour} {et~al.}(2007){Seymour}, {Stern}, {De Breuck}, {Vernet}, {Rettura}, {Dickinson}, {Dey}, {Eisenhardt}, {Fosbury}, {Lacy}, {McCarthy}, {Miley}, {Rocca-Volmerange}, {R{\"o}ttgering}, {Stanford}, {Teplitz}, {van Breugel}, \& {Zirm}}]{seymour07}
{Seymour}, N., {Stern}, D., {De Breuck}, C., {et~al.} 2007, \apjs, 171, 353, \dodoi{10.1086/517887}

\bibitem[{{Shimizu} {et~al.}(2019){Shimizu}, {Davies}, {Lutz}, {Burtscher}, {Lin}, {Baron}, {Davies}, {Genzel}, {Hicks}, {Koss}, {Maciejewski}, {M{\"u}ller-S{\'a}nchez}, {Orban de Xivry}, {Price}, {Ricci}, {Riffel}, {Riffel}, {Rosario}, {Schartmann}, {Schnorr-M{\"u}ller}, {Sternberg}, {Sturm}, {Storchi-Bergmann}, {Tacconi}, \& {Veilleux}}]{shimizu19}
{Shimizu}, T.~T., {Davies}, R.~I., {Lutz}, D., {et~al.} 2019, \mnras, 490, 5860, \dodoi{10.1093/mnras/stz2802}

\bibitem[{{Smail} \& {Blundell}(2013)}]{smail13}
{Smail}, I., \& {Blundell}, K.~M. 2013, \mnras, 434, 3246, \dodoi{10.1093/mnras/stt1240}

\bibitem[{{Smail} {et~al.}(2012){Smail}, {Blundell}, {Lehmer}, \& {Alexander}}]{smail12}
{Smail}, I., {Blundell}, K.~M., {Lehmer}, B.~D., \& {Alexander}, D.~M. 2012, \apj, 760, 132, \dodoi{10.1088/0004-637X/760/2/132}

\bibitem[{{Speagle}(2020)}]{speagle20}
{Speagle}, J.~S. 2020, \mnras, 493, 3132, \dodoi{10.1093/mnras/staa278}

\bibitem[{{Speranza} {et~al.}(2021){Speranza}, {Balmaverde}, {Capetti}, {Massaro}, {Tremblay}, {Marconi}, {Venturi}, {Chiaberge}, {Baldi}, {Baum}, {Grandi}, {Meyer}, {O'Dea}, {Sparks}, {Terrazas}, \& {Torresi}}]{speranza21}
{Speranza}, G., {Balmaverde}, B., {Capetti}, A., {et~al.} 2021, \aap, 653, A150, \dodoi{10.1051/0004-6361/202140686}

\bibitem[{{Swinbank} {et~al.}(2015){Swinbank}, {Vernet}, {Smail}, {De Breuck}, {Bacon}, {Contini}, {Richard}, {R{\"o}ttgering}, {Urrutia}, \& {Venemans}}]{swinbank15}
{Swinbank}, A.~M., {Vernet}, J.~D.~R., {Smail}, I., {et~al.} 2015, \mnras, 449, 1298, \dodoi{10.1093/mnras/stv366}

\bibitem[{{{\"U}bler} {et~al.}(2023){{\"U}bler}, {Maiolino}, {Curtis-Lake}, {P{\'e}rez-Gonz{\'a}lez}, {Curti}, {Perna}, {Arribas}, {Charlot}, {Marshall}, {D'Eugenio}, {Scholtz}, {Bunker}, {Carniani}, {Ferruit}, {Jakobsen}, {Rix}, {Rodr{\'\i}guez Del Pino}, {Willott}, {Boeker}, {Cresci}, {Jones}, {Kumari}, \& {Rawle}}]{ubler23}
{{\"U}bler}, H., {Maiolino}, R., {Curtis-Lake}, E., {et~al.} 2023, \aap, 677, A145, \dodoi{10.1051/0004-6361/202346137}

\bibitem[{{Ulivi} {et~al.}(2024){Ulivi}, {Venturi}, {Cresci}, {Marconi}, {Marconcini}, {Amiri}, {Belfiore}, {Bertola}, {Carniani}, {D'Amato}, {Di Teodoro}, {Ginolfi}, {Girdhar}, {Harrison}, {Maiolino}, {Mannucci}, {Mingozzi}, {Perna}, {Scialpi}, {Tomicic}, {Tozzi}, \& {Treister}}]{ulivi24}
{Ulivi}, L., {Venturi}, G., {Cresci}, G., {et~al.} 2024, \aap, 685, A122, \dodoi{10.1051/0004-6361/202347436}

\bibitem[{{van Ojik} {et~al.}(1997){van Ojik}, {Roettgering}, {Miley}, \& {Hunstead}}]{van97}
{van Ojik}, R., {Roettgering}, H.~J.~A., {Miley}, G.~K., \& {Hunstead}, R.~W. 1997, \aap, 317, 358, \dodoi{10.48550/arXiv.astro-ph/9608092}

\bibitem[{{Vayner} {et~al.}(2023){Vayner}, {Zakamska}, {Ishikawa}, {Sankar}, {Wylezalek}, {Rupke}, {Veilleux}, {Bertemes}, {Barrera-Ballesteros}, {Chen}, {Diachenko}, {Goulding}, {Greene}, {Hainline}, {Hamann}, {Heckman}, {Johnson}, {Lim}, {Liu}, {Lutz}, {Lutzgendorf}, {Mainieri}, {McCrory}, {Murphree}, {Nesvadba}, {Ogle}, {Sturm}, \& {Whitesell}}]{vayner23}
{Vayner}, A., {Zakamska}, N.~L., {Ishikawa}, Y., {et~al.} 2023, arXiv e-prints, arXiv:2307.13751, \dodoi{10.48550/arXiv.2307.13751}

\bibitem[{{Veilleux} {et~al.}(2020){Veilleux}, {Maiolino}, {Bolatto}, \& {Aalto}}]{veilleux20}
{Veilleux}, S., {Maiolino}, R., {Bolatto}, A.~D., \& {Aalto}, S. 2020, \aapr, 28, 2, \dodoi{10.1007/s00159-019-0121-9}

\bibitem[{{Veilleux} {et~al.}(2023){Veilleux}, {Liu}, {Vayner}, {Wylezalek}, {Rupke}, {Zakamska}, {Ishikawa}, {Bertemes}, {Barrera-Ballesteros}, {Chen}, {Diachenko}, {Goulding}, {Greene}, {Hainline}, {Hamann}, {Heckman}, {Johnson}, {Grace Lim}, {Lutz}, {L{\"u}tzgendorf}, {Mainieri}, {Maiolino}, {McCrory}, {Murphree}, {Nesvadba}, {Ogle}, {Sankar}, {Sturm}, \& {Whitesell}}]{veilleux23}
{Veilleux}, S., {Liu}, W., {Vayner}, A., {et~al.} 2023, \apj, 953, 56, \dodoi{10.3847/1538-4357/ace10f}

\bibitem[{{Venturi} {et~al.}(2021){Venturi}, {Cresci}, {Marconi}, {Mingozzi}, {Nardini}, {Carniani}, {Mannucci}, {Marasco}, {Maiolino}, {Perna}, {Treister}, {Bland-Hawthorn}, \& {Gallimore}}]{venturi21}
{Venturi}, G., {Cresci}, G., {Marconi}, A., {et~al.} 2021, \aap, 648, A17, \dodoi{10.1051/0004-6361/202039869}

\bibitem[{{Villar Mart{\'\i}n} {et~al.}(2021){Villar Mart{\'\i}n}, {Emonts}, {Cabrera Lavers}, {Bellocchi}, {Alonso Herrero}, {Humphrey}, {Dall'Agnol de Oliveira}, \& {Storchi-Bergmann}}]{villar-martin21}
{Villar Mart{\'\i}n}, M., {Emonts}, B.~H.~C., {Cabrera Lavers}, A., {et~al.} 2021, \aap, 650, A84, \dodoi{10.1051/0004-6361/202039642}

\bibitem[{Virtanen {et~al.}(2020)Virtanen, Gommers, Oliphant, Haberland, Reddy, Cournapeau, Burovski, Peterson, Weckesser, Bright, {van der Walt}, Brett, Wilson, Millman, Mayorov, Nelson, Jones, Kern, Larson, Carey, Polat, Feng, Moore, {VanderPlas}, Laxalde, Perktold, Cimrman, Henriksen, Quintero, Harris, Archibald, Ribeiro, Pedregosa, {van Mulbregt}, \& {SciPy 1.0 Contributors}}]{scipy20}
Virtanen, P., Gommers, R., Oliphant, T.~E., {et~al.} 2020, Nature Methods, 17, 261, \dodoi{10.1038/s41592-019-0686-2}

\bibitem[{{Wagner} \& {Bicknell}(2011)}]{wagner11}
{Wagner}, A.~Y., \& {Bicknell}, G.~V. 2011, \apj, 728, 29, \dodoi{10.1088/0004-637X/728/1/29}

\bibitem[{{Wagner} {et~al.}(2012){Wagner}, {Bicknell}, \& {Umemura}}]{wagner12}
{Wagner}, A.~Y., {Bicknell}, G.~V., \& {Umemura}, M. 2012, \apj, 757, 136, \dodoi{10.1088/0004-637X/757/2/136}

\bibitem[{{Wang} {et~al.}(2024){Wang}, {Wylezalek}, {De Breuck}, {Vernet}, {Rupke}, {Zakamska}, {Vayner}, {Lehnert}, {Nesvadba}, \& {Stern}}]{wang24}
{Wang}, W., {Wylezalek}, D., {De Breuck}, C., {et~al.} 2024, \aap, 683, A169, \dodoi{10.1051/0004-6361/202348531}

\bibitem[{{Wang} {et~al.}(2025){Wang}, {De Breuck}, {Wylezalek}, {Vernet}, {Lehnert}, {Stern}, {Rupke}, {Nesvadba}, {Vayner}, {Zakamska}, {Lin}, {Kukreti}, {Dall'Agnol de Oliveira}, \& {Groth}}]{wang25}
{Wang}, W., {De Breuck}, C., {Wylezalek}, D., {et~al.} 2025, \aap, 696, A88, \dodoi{10.1051/0004-6361/202553668}

\bibitem[{{Xu} {et~al.}(2019){Xu}, {Arav}, {Miller}, \& {Benn}}]{xu19}
{Xu}, X., {Arav}, N., {Miller}, T., \& {Benn}, C. 2019, \apj, 876, 105, \dodoi{10.3847/1538-4357/ab164e}

\bibitem[{{Xu} {et~al.}(2020){Xu}, {Arav}, {Miller}, {Kriss}, \& {Plesha}}]{xu20}
{Xu}, X., {Arav}, N., {Miller}, T., {Kriss}, G.~A., \& {Plesha}, R. 2020, \apjs, 247, 42, \dodoi{10.3847/1538-4365/ab5f68}

\end{thebibliography}
\bibliographystyle{aasjournal}



\end{document}